\documentclass{article}
\usepackage[spanish,english,es-tabla]{babel}
\usepackage[utf8]{inputenc}
\usepackage[T1]{fontenc}
\usepackage{lmodern}     
\usepackage[protrusion=true, expansion=true]{microtype}  
\usepackage{amsmath,amssymb} 
\usepackage{tabularx}    
\usepackage{multicol}    
\usepackage{multirow}    
\usepackage{longtable}   
\usepackage{booktabs}    
\usepackage{ltablex}
\usepackage{setspace} 
\usepackage{tikz}                
\usepackage{graphicx,graphics}   
\usepackage{subfigure}           
\usepackage[export]{adjustbox}
\usepackage{ragged2e} 
\usepackage{blindtext}
\usepackage[acronym,nonumberlist,nomain]{glossaries}  
\usepackage{hyperref}
\usepackage[a4paper]{geometry}
\usepackage{parskip}     
\usepackage{fancyhdr}    
\usepackage{enumitem}
\usepackage{xcolor}
\usepackage{emptypage}   
\usepackage[small,hang,bf]{caption}    
\usepackage{natbib}
\usepackage{csquotes}
\usetikzlibrary{arrows.meta}
\usepackage{algpseudocode}
\usepackage{pgfplots}
\usepackage{wasysym}
\pgfplotsset{compat=1.18}
\usepackage{changes}
\usepackage[noblocks]{authblk}

\DeclareGraphicsExtensions{.pdf,.png,.jpg}

\usepackage{xstring} 

\newcommand{\figdir}{} 

\newif\ifuseflattened
\useflattenedtrue   

\ifuseflattened
   \usepackage{xparse}
   \ExplSyntaxOn
   \NewDocumentCommand{\figpath}{m O{}}{
      \StrSubstitute{#1}{/}{-F-}[\l_tmpa_str]
      \includegraphics[#2]{\figdir\l_tmpa_str}
   }
   \ExplSyntaxOff
\else
   \usepackage{xparse}
   \ExplSyntaxOn
   \NewDocumentCommand{\figpath}{m O{}}{
      \includegraphics[#2]{#1}
   }
   \ExplSyntaxOff
\fi

\title{A Phase-Field Approach to Fracture and Fatigue Analysis: Bridging Theory and Simulation}
\author[1,2]{M. Castillón}
\author[1,2]{I. Romero}
\author[3,2]{J. Segurado}
\affil[1]{Dept. of Mechanical Engineering, Universidad Politécnica de Madrid, Spain}
\affil[2]{IMDEA Materials Institute, Getafe, Madrid, Spain}
\affil[3]{Dept. of Materials Sciences, Universidad Politécnica de Madrid, Spain}

\begin{document}
\maketitle

\begin{abstract} 
This article presents a novel, robust and efficient framework for fatigue crack-propagation that combines the principles of Linear Elastic Fracture Mechanics (LEFM) with phase-field fracture (PFF). Contrary to cycle-by-cycle PFF approaches, this work relies on a single simulation and uses standard crack propagation models such as Paris' law for the material response, simplifying its parametrization.

The core of the methodology is the numerical evaluation of the derivative of a specimen's compliance with respect to the crack area. To retrieve this compliance the framework relies on a PFF-FEM simulation, controlled imposing a monotonic crack growth. This control of the loading process is done by a new crack-control scheme which allows to robustly trace the complete equilibrium path of a crack, capturing complex instabilities. The specimen's compliance obtained from the PFF simulation enables the integration of Paris' law to predict fatigue life. 

The proposed methodology is first validated through a series of benchmarks with analytical solutions to demonstrate its accuracy. The framework is then applied to more complex geometries where the crack path is unknown, showing a very good agreement with experimental results of both crack paths and fatigue life.

\end{abstract}

\newcommand{\Cparis}{C_{\text{Paris}}}
\newcommand{\nparis}{n}
\newcommand{\Ccompliance}{C}
\newcommand{\SpecimenWidth}{W}
\newcommand{\Nax}{N_a(\boldsymbol x)}
\newcommand{\Nbx}{N_b(\boldsymbol x)}

\section{Introduction}
\label{sec:Introduction}
%

Fracture mechanics and fatigue analysis are two fundamental theories for understanding the failure of materials and structures. Fracture mechanics provides a theoretical framework to study the conditions for quasistatic crack propagation, while fatigue analysis focuses on the behavior of materials under cyclic loading conditions. 

Among the different models available to simulate failure, the phase-field fracture model (PFF) has captured significant attention in the last decade. The theoretical foundation for this method was laid by Francfort and Marigo~\cite{phase_field_FrancfortMarigo1998}, who first proposed a variational formulation of brittle fracture, revisiting Griffith's theory as an energy minimization problem. This variational framework was subsequently regularized and developed into the modern phase-field method through the pioneering work of researchers like Bourdin et al.~\cite{phase_field_Bourdin2000} and Miehe et al.~\cite{phase_field_Miehe2010}, both of whom introduced key concepts such as the diffusive crack representation and robust algorithmic implementations. In this approach, cracks are represented implicitly by a continuous phase-field variable, eliminating the need for explicit crack tracking. This implicit formulation greatly simplifies numerical implementation and naturally accommodates complex crack phenomena such as branching, merging, and multiple crack interactions.

The extensions of phase-field fracture to model fatigue crack propagation are relatively recent \cite{phase_field_Carrara2020, phase_field_fatigue_Seles, phase_field_fatigue_Golahmar, phase_field_fatigue_Schreiber, phase_field_fatigue_Mesgarnejad}. A common ingredient of most of the models proposed, as Carrara et al.~\cite{phase_field_Carrara2020}, is to introduce a fatigue degradation function that reduces the fracture toughness according to a cumulative history variable, where the simulation progress is tracked by a pseudotime. This history variable typically monitors quantities such as accumulated plastic strain or energy dissipation, effectively capturing the material’s fatigue history. In these models, fatigue crack growth is simulated by incrementally applying cyclic loads and updating the history variable at each cycle. From a fundamental viewpoint, this background is very interesting since the parametrization of the history dependence of the damage is independent of the classical Paris law parameters and therefore it allows the application of non-uniform cycles and loading conditions during propagation. Moreover, it has been shown that this law can be recovered as a result of the cycle-by-cycle simulations. However, from a practical viewpoint, Paris law parameters have been experimentally determined for many materials and their use in a linear fatigue simulation is very convenient to avoid expensive ad-hoc experimental campaigns and also for comparability with the results of other techniques. Another limitation of this cycle-by-cycle modeling approach is its computational cost, especially for high-cycle fatigue scenarios, as it requires resolving each individual load cycle within the phase-field simulation. This is particularly prohibitive for high-cycle fatigue scenarios, and the computational burden becomes even more significant in three-dimensional simulations, where the number of degrees of freedom and the complexity of crack evolution increase substantially. Although some techniques have been proposed to accelerate these simulations, such as cycle-jumping or cycle-skipping strategies~\cite{phase_field_fatigue_Heinzmann,phase_field_fatigue_Kristensen}, the cost is still inherently high since its foundation is following the loading history, therefore requiring a full set of simulations during loading.

In classical linear elastic fatigue frameworks, crack propagation is typically modeled using Paris-type laws, which provide the crack growth rate and direction as a function of the cyclic stress intensity factors at the crack tip. These stress intensity factors are well defined within linear elastic fracture mechanics (LEFM) theory and can be derived as a function of the specimen's stiffness for a given crack length. While analytical expressions exist for simple cases, most modern fatigue simulation frameworks rely on the numerical evaluation of these factors using the J-integral within the finite element method (FEM). To simulate the complete process of crack growth, numerical techniques such as the Extended Finite Element Method (X-FEM)~\cite{introduction_x_fem} or remeshing strategies~\cite{introduction_remeshing} are commonly employed. These methods model crack propagation incrementally, either by enriching the finite element space (as in X-FEM) or by explicitly modifying the mesh to accommodate the advancing crack. Criteria such as the Maximum Tangential Stress~\cite{introduction_maximum_tangential_stress_criterion} and the Maximum Energy Release Rate~\cite{introduction_maximum_energy_release_rate_criterion} are often used to predict the direction of crack propagation. However, both approaches are computationally intensive, as they require frequent updates to the mesh connectivity and repeated assembly and factorization of the stiffness matrix at each propagation step.

To overcome the challenges identified thus far in the models reviewed, a new method is proposed to model fatigue. The method is based on obtaining the compliance rate with respect to the crack area using the phase-field fracture (PFF) model, combined with novel energy-controlled techniques. In this form, the effective stress intensity factor is obtained at each stage of crack propagation and is used to integrate a Paris type law to provide the crack path evolution with the corresponding number of cycles at stage in a single simulation. With our modeling approach it is not necessary to simulate cycle by cycle propagation, allowing to reach cases with millions of steps without increasing the computational cost. Moreover, contrary to X-FEM or re-meshing techniques, the use of PFF eliminates the need of recomputing and factorizing the stiffness matrix at each crack propagation step also strongly reducing the computational cost.

The key point of this approach is the development of a novel energy-controlled equilibrium path solver, formulated in both variational and non-variational forms. These solvers robustly trace the complete equilibrium path --- including snap-back and snap-through instabilities --- with a single Lagrange multiplier as an additional variable, resulting in a simple and efficient implementation.
Alternative equilibrium path tracking methods have been proposed in the literature, such as crack-length control techniques \cite{phase_field_snap_pedro}, arc-length methods integrated with fracture energy control \cite{phase_field_snap_Ritukesh}, and arc-length control for staggered phase-field schemes \cite{phase_field_snap_Zambrano}. While these approaches also address the challenge of tracing unstable equilibrium paths, the method presented here stands out for its simplicity: only one Lagrange multiplier is added to the system, and the implementation is straightforward within a monolithic framework. This enables efficient and robust simulation of quasi-static crack growth and provides direct access to the rate of compliance with respect to the crack area, which is essential for subsequent fracture and fatigue analysis.

This article is organized as follows. Section~\ref{sec:theoretical_background_lefm} reviews the theoretical foundations of LEFM and Paris' law. Section~\ref{sec:phase_field_models} introduces the phase-field fracture method, detailing the variational and non-variational energy-controlled solvers and explaining the computation of key quantities like the compliance derivative ($\mathrm{d}C/\mathrm{d}a$). Section~\ref{sec:numerical_aspects} details the numerical implementation, including the adaptive solution strategy and methods for correcting the overestimation of the crack area and its resulting effect on the fracture toughness. Section~\ref{sec:validation} provides a comprehensive validation of the framework. It begins with the three-point bending test (Section~\ref{example:three_point_bending_test_specimen}), where the equivalence of the proposed solvers is validated and the results are compared against analytical solutions. It then analyzes the center-cracked specimen (Section~\ref{example:center_cracked_test_specimen}) to investigate numerical aspects such as the length scale parameter and crack measurement corrections, comparing results with LEFM. Once the approach is validated, Section~\ref{example:compact_tension_test_specimen} demonstrates the framework's capabilities on a complex example, analyzing the unknown crack paths in a modified compact tension specimen with multiple holes. Finally, Section~\ref{sec:conclusions} summarizes the findings and outlines future research and Section~\ref{sec:code_availability} presents the open-source software and code repositories.

\section{Theoretical Background and Key Concepts in Linear Elastic Fracture Mechanics}
\label{sec:theoretical_background_lefm}

This section establishes the theoretical foundation of linear elastic fracture mechanics (LEFM) that underpins the proposed methodology. While these concepts are well-established, a succinct presentation is essential as this work demonstrates that combining fundamental fracture mechanics principles with energy-controlled phase-field models can achieve remarkable accuracy and computational efficiency.

The stress intensity factor, $K$, is a fundamental parameter in LEFM that quantifies the intensity of the stress field in the vicinity of a crack tip. For a given specimen geometry and loading configuration, it is expressed as:
\begin{equation}
    K = \frac{P}{B\sqrt{\SpecimenWidth}} \, f(a/\SpecimenWidth),
    \label{eq:stress_intensity_factor}
\end{equation}
where $P$ is the applied force, $B$ is the specimen thickness, $\SpecimenWidth$ is a characteristic dimension (typically the specimen width), $a$ is the crack length, and $f(a/\SpecimenWidth)$ is a dimensionless geometry factor that depends on the specimen configuration and loading conditions. 

Fracture is predicted to occur when the stress intensity factor reaches a critical value, the fracture toughness $K_c$, which is an intrinsic material property. The critical force $P_c$ required to initiate fracture is determined by setting $K = K_c$:
\begin{equation}
   P_c = \frac{B\sqrt{\SpecimenWidth} \, K_c}{f(a/\SpecimenWidth)}.
   \label{eq:critical_force_geometry_factor}
\end{equation}

In cases involving mixed-mode fracture, the critical parameter is the equivalent stress intensity factor, $K_{\text{eq}}$, which accounts for the combined effects of all active fracture modes. Specifically, $K_{\text{eq}}$ is a function of the individual mode contributions.

The precise form of $K_{\text{eq}} = f(K_I, K_{II}, K_{III})$ depends on the material and loading conditions, and it enables a unified treatment of fracture processes where multiple modes are present simultaneously.

Under cyclic loading conditions, crack growth is commonly described by Paris' law, which relates the crack growth rate per cycle, $\mathrm{d}a/\mathrm{d}N$, to the equivalent stress intensity factor range $\Delta K_{\text{eq}}$:
\begin{equation}
   \frac{\mathrm{d}a}{\mathrm{d}N} = \Cparis (\Delta K_{\text{eq}})^{\nparis},
   \label{eq:paris_law}
\end{equation}
where $\Cparis$ and $\nparis$ are empirical material constants determined experimentally. The equivalent stress intensity factor range accounts for all active fracture modes and is calculated from the cyclic loading conditions. For simple loading cases, this reduces to the familiar form:
\begin{equation}
    \Delta K = \frac{\Delta P}{B\sqrt{\SpecimenWidth}} \, f({a}/{\SpecimenWidth}),
    \label{eq:delta_K}
\end{equation}
where $\Delta P = P_{\max} - P_{\min}$ is the cyclic force range.
The fatigue life of a component can be calculated by integrating Paris' law employing Eqs.~\eqref{eq:paris_law} and~\eqref{eq:delta_K}:
\begin{equation}
   N_f = N_i + \frac{1}{\Cparis \left(\frac{\Delta P}{B\sqrt{\SpecimenWidth}}\right)^{\nparis}} \int_{a_i}^{a_f} \frac{\mathrm{d}a}{[f(a/\SpecimenWidth)]^{\nparis}} \ ,
   \label{eq:total_number_of_cycles_geometry}
\end{equation}
where $N_i$ and $N_f$ are the initial and final cycle numbers, and $a_i$ and $a_f$ are the corresponding crack lengths. This equation enables fatigue life calculations based on material properties ($\Cparis$, $\nparis$), applied loading ($\Delta P$), specimen geometry ($B$, $W$), and calculations from LEFM ($f(a/W)$).

An alternative, yet equivalent, formulation in LEFM is based on the energy release rate, $G$, which represents the energy dissipated per unit of newly created crack surface area. For a body under an applied force $P$, the energy release rate is related to the compliance derivative as:
\begin{equation}
   G = \frac{P^2}{2B} \frac{\mathrm{d}\Ccompliance}{\mathrm{d}a},
   \label{eq:energy_release_rate_compliance}
\end{equation}
where $\Ccompliance$ is the structural compliance and $a$ is the crack length. For linear elastic materials, the energy release rate and equivalent stress intensity factor are related through:
\begin{equation}
   G = \frac{K_{\text{eq}}^2}{E'},
   \label{eq:relation_K_G}
\end{equation}
where $E'$ is the effective Young's modulus, defined as $E' = E$ for plane stress conditions and $E' = E/(1-\nu^2)$ for plane strain conditions, with $E$ being Young's modulus and $\nu$ Poisson's ratio. Notably, the energy release rate $G$ intrinsically incorporates the effects of all fracture modes, as it is defined in terms of the equivalent stress intensity factor $K_{\text{eq}}$, which aggregates the contributions from Modes I, II, and III. This ensures that $G$ provides a unified energetic criterion for crack propagation regardless of the mode mixity.

The relationship between the applied force $P$ and the energy release rate $G$ can be expressed by rearranging Eq.~\eqref{eq:energy_release_rate_compliance}:
\begin{equation}
   P = \sqrt{\frac{2B G}{\mathrm{d}\Ccompliance/\mathrm{d}a}}\ .
   \label{eq:force_from_G}
\end{equation}
Fracture is predicted to occur when the energy release rate $G$ reaches its critical value $G_c$. The critical force, $P_c$, is therefore determined by substituting $G = G_c$ into Eq.~\eqref{eq:force_from_G}. This expression provides a method for fracture prediction based on the rate of compliance with
respect to the crack area, which can be determined experimentally or numerically.

The energetic framework can be extended to fatigue analysis. Using the relationship $\Delta K = \sqrt{E'} \ \Delta\sqrt{G}$ into Paris' Law (Eq.~\eqref{eq:paris_law}), we obtain an expression for crack growth in terms of the energy release rate range, $\Delta\sqrt{G}$:
\begin{equation}
   \frac{da}{dN} = \Cparis \left( \sqrt{E'} \Delta \sqrt{G} \right)^\nparis.
   \label{eq:paris_law_G}
\end{equation}
Integrating from the initial crack length $a_i$ to the final crack length $a_f$ gives the fatigue life:
\begin{equation}
    N_f = N_i + \frac{1}{\Cparis (E'/(2B))^{\nparis/2} (\Delta P)^{\nparis}} \int_{a_i}^{a_f} \frac{\mathrm{d}a}{(\mathrm{d}\Ccompliance/\mathrm{d}a)^{\nparis/2}}.
    \label{eq:total_number_of_cycles_dCda}
\end{equation}
This equation is equivalent to Eq.~\eqref{eq:total_number_of_cycles_geometry} and demonstrates that fatigue life can be predicted directly from compliance evolution.

It is important to note that while Eq.~\eqref{eq:total_number_of_cycles_dCda} contains the effective Young's modulus $E'$, the fatigue behavior described by Paris' law depends fundamentally on the material constants $\Cparis$ and $\nparis$. The term $E'$ appears because compliance $\Ccompliance$ is inversely proportional to $E'$. The derivative can be expressed as:
\begin{equation}
    \frac{\mathrm{d}\Ccompliance}{\mathrm{d}a} = \frac{1}{E'} \frac{\mathrm{d}\bar{\Ccompliance}}{\mathrm{d}a},
    \label{eq:normalized_compliance}
\end{equation}
where $\bar{\Ccompliance}$ is a normalized compliance depending only on geometry. Substituting this into Eq.~\eqref{eq:total_number_of_cycles_dCda}, one confirms that the $E'$ terms cancel out and that fatigue life prediction is independent of elastic modulus, consistent with the physical nature of Paris' law.

The direct relationship between the geometry factor and compliance derivative can be established by combining Eqs.~\eqref{eq:stress_intensity_factor}, \eqref{eq:energy_release_rate_compliance}, and \eqref{eq:relation_K_G}:
\begin{equation}
   f({a}/{\SpecimenWidth}) = \sqrt{\frac{E' B \SpecimenWidth}{2} \frac{\mathrm{d}\Ccompliance}{\mathrm{d}a}}.
   \label{eq:geometry_factor_dCda}
\end{equation}
This equivalence confirms that, within the framework of linear elastic fracture mechanics, both stress-based and energy-based approaches are consistent and interchangeable. The choice between them depends on whether the geometry factor or compliance evolution is more convenient to work with, for a given problem. For complex geometries or unknown crack paths, calculating compliance evolution numerically via phase-field simulations offers a powerful and versatile method for fracture and fatigue analysis in the linear elastic regime.

\section{Energy-Controlled Phase-Field Fracture Solvers}
\label{sec:phase_field_models}
This work employs a phase-field approach to model fracture, based on the variational framework of \cite{phase_field_FrancfortMarigo1998} and the regularization scheme of \cite{phase_field_Miehe2010}. This method represents cracks implicitly using a continuous scalar field, $\phi$, which varies smoothly from $\phi=0$ (intact material) to $\phi=1$ (fully fractured). This approach eliminates the need for explicit crack tracking and naturally handles complex crack phenomena such as branching and merging.

The evolution of the system is governed by the minimization of a total energy functional, $\mathcal{E}$, which combines the elastic strain energy and the energy required to create new crack surfaces:
\begin{equation}
\mathcal{E}(\boldsymbol{u}, \phi) = \int_\Omega\left[ g(\phi) \psi_a(\boldsymbol{\epsilon}(\boldsymbol{u})) + \psi_b(\boldsymbol{\epsilon}(\boldsymbol{u}))\right] \,\mathrm{d}\Omega + G_c \int_\Omega \gamma(\phi, \nabla\phi) \,\mathrm{d}\Omega - \mathcal{E}_{\text{ext}}[\boldsymbol{u}].
\label{eq:phase_field_fracture_functional}
\end{equation}
Here, $\boldsymbol{u}$ is the displacement field, and the strain tensor is defined as the symmetric part of the displacement gradient: $\boldsymbol{\epsilon}(\boldsymbol{u}) = \frac{1}{2}(\nabla \boldsymbol{u} + \nabla \boldsymbol{u}^T)$. The functions $\psi_a$ and $\psi_b$ represent the split of elastic energy $\psi(\boldsymbol{\epsilon}) = \frac{1}{2}\lambda (\text{tr}(\boldsymbol{\epsilon}))^2 + \mu \boldsymbol{\epsilon}:\boldsymbol{\epsilon}$, where $\lambda$ and $\mu$ are the Lamé parameters, in a part $\psi_a$ which can lead to a crack growth and which is affected by the phase-field variable, and other one which does not contribute to generate damage. Several energy decomposition strategies have been proposed in the literature, such as the volumetric-deviatoric split~\cite{phase_field_Amor2009} and the spectral decomposition~\cite{phase_field_Miehe2010}. The degradation function $g(\phi) = (1 - \phi)^2$ reduces the material's stiffness as damage accumulates. The crack surface density functional, $\gamma(\phi, \nabla\phi) = \frac{1}{2l}\phi^2 + \frac{l}{2} |\nabla \phi|^2$, approximates the total crack surface energy, $G_c$ is the material's critical energy release rate and $l$ is the length scale parameter controlling the width of the diffuse crack region. The term $\mathcal{E}_{\text{ext}}$ represents the potential energy of the external forces.
\begin{equation}
\mathcal{E}_{\text{ext}}[\boldsymbol{u}] = \int_\Omega \boldsymbol{f} \cdot \boldsymbol{u} \,\mathrm{d}\Omega + \int_{\partial_N\Omega} \boldsymbol{t} \cdot \boldsymbol{u} \,\mathrm{d}S.
\end{equation}
where $\boldsymbol{f}$ represents the body forces and $\boldsymbol{t}$ are the prescribed tractions on the Neumann boundary $\partial_N\Omega$. Since body forces are not considered in this work, the corresponding term is omitted in the subsequent equations.

The weak form of the governing equations is obtained by taking the first variation of the functional in Eq.~\eqref{eq:phase_field_fracture_functional} with respect to $\boldsymbol{u}$ and $\phi$ and setting it to zero, which yields:
\begin{equation}
   \int_\Omega [g(\phi)\boldsymbol{\sigma}_a(\boldsymbol{\epsilon}(\boldsymbol{u})) + \boldsymbol{\sigma}_b(\boldsymbol{\epsilon}(\boldsymbol{u}))]:\boldsymbol{\epsilon}(\delta \boldsymbol{u}) \,\mathrm{d}\Omega -  \int_{\partial_N\Omega} \boldsymbol{t} \cdot \delta\boldsymbol{u} \,\mathrm{d}S  = {0} \ ,
\end{equation}
\begin{equation}
   \int_\Omega g'(\phi) \delta\phi \, \psi_a(\boldsymbol{\epsilon}(\boldsymbol{u})) \,\mathrm{d}\Omega + G_c \int_\Omega \left( \frac{1}{l} \phi \delta\phi  + l \nabla\phi \cdot \nabla \delta \phi \right) \,\mathrm{d}\Omega = 0\ ,
\end{equation}
where $\boldsymbol{\sigma}_a$ and $\boldsymbol{\sigma}_b$ are the stress tensors corresponding to the chosen energy decomposition,  and $\delta\boldsymbol{u}, \delta\phi$ are, respectively, the displacement and phase field variations.  Solving this coupled system enables the prediction of crack initiation and propagation without requiring prior knowledge of the crack path.

Traditional solution strategies for phase-field fracture, such as displacement or force control, often fail to capture the complete equilibrium path when the structural response exhibits instabilities such as snap-back or snap-through behavior. In these cases, displacement-controlled solvers may jump between stable equilibrium points, missing critical portions of the response, while force-controlled schemes may be unable to trace the path entirely.

To overcome these limitations, this section introduces two novel energy-controlled schemes that build upon the crack growth control technique developed in \cite{phase_field_snap_pedro} for FFT-based micromechanical solvers, adapting it for general macroscopic boundary value problems. These methods are designed to robustly trace the complete equilibrium path during crack propagation by controlling the simulation with a monotonically increasing energy-like parameter, which ensures stable and continuous progress even through complex instabilities. This allows for an accurate characterization of the system's response throughout the entire fracture process. A key advantage is that a single simulation provides the rate of compliance change with respect to crack area, $\mathrm{d}C/\mathrm{d}a$, for all stages of crack evolution, which is essential for both fracture and fatigue analysis within the proposed methodology.

\subsection{Variational scheme}
\label{subsec:variational_scheme}
The first approach is a variational scheme in which the simulation is driven by an energetic control function, $\tau(t)$. This is accomplished by introducing a constraint that equates $\tau(t)$ to a weighted combination of the crack surface energy and the work done by external forces. A Lagrange multiplier, $\lambda$, is employed to enforce this constraint. The constraint equation is defined as:
\newcommand{\cphi}{c_1}
\newcommand{\cu}{c_2}
\begin{equation}
   \cphi \int_\Omega \left( \frac{1}{2l} \phi^2 + \frac{l}{2} |\nabla \phi|^2 \right) \,\mathrm{d}\Omega 
   + \cu \int_{\partial_N\Omega} \boldsymbol{t} \cdot \boldsymbol{u} \,\mathrm{d}S = \tau(t).
\end{equation}
Here, $\cphi$ and $\cu$ are numerical parameters introduced to improve solver convergence and ensure dimensional consistency in the constraint. Their values can be chosen to appropriately scale the contributions of the crack surface energy and the external work, respectively. For example, if $\cphi$ is assigned the same units as the critical energy release rate $G_c$ (energy per unit length), then $\cu$ becomes dimensionless, and $\tau(t)$ has units of energy. Alternatively, if $\cphi$ is dimensionless, then $\cu$ must have units of compliance (length per force), and $\tau(t)$ will have units of area, in this case the algorithm can be referred to as a crack area controlled algorithm. In practice, these parameters are selected to balance the terms in the constraint and facilitate numerical stability.
 
In this formulation, a force is applied in a prescribed direction $\boldsymbol{n}$, while the actual load magnitude is determined as part of the solution through the Lagrange multiplier. This multiplier emerges naturally from the constrained optimization, ensuring that the equilibrium path is traced by enforcing the energy balance. Concurrently, the displacement field evolves to satisfy the equilibrium equations under the imposed energy constraint. The augmented functional for this constrained system, $V$, is defined as:
\begin{align}
    V(\boldsymbol{u}, \phi, \lambda) &= \int_\Omega [g(\phi)\psi_a(\boldsymbol{\epsilon}(\boldsymbol{u})) + \psi_b(\boldsymbol{\epsilon}(\boldsymbol{u}))] \,\mathrm{d}\Omega \nonumber \\
    &\quad + G_c \int_\Omega \left( \frac{1}{2l} \phi^2 + \frac{l}{2} |\nabla \phi|^2 \right) \,\mathrm{d}\Omega 
    - \int_{\partial_N\Omega} \boldsymbol{t} \cdot \boldsymbol{u} \,\mathrm{d}S \nonumber \\
    &\quad + \lambda \left[ \cphi \int_\Omega \left( \frac{1}{2l} \phi^2 + \frac{l}{2} |\nabla \phi|^2 \right) \,\mathrm{d}\Omega 
    + \cu \int_{\partial_N\Omega} \boldsymbol{t} \cdot \boldsymbol{u} \,\mathrm{d}S - \tau(t) \right].
    \label{eq:variational_augmented_functional}
\end{align}
The equilibrium equations are obtained by enforcing stationarity, $\delta V = 0$. The resulting weak form is:
\begin{equation}
    \int_\Omega [g(\phi)\boldsymbol{\sigma}_a(\boldsymbol{\epsilon}(\boldsymbol{u})) + \boldsymbol{\sigma}_b(\boldsymbol{\epsilon}(\boldsymbol{u}))]:\boldsymbol{\epsilon}(\delta \boldsymbol{u}) \,\mathrm{d}\Omega 
    - (1 - \lambda \cu) \int_{\partial_N\Omega} \boldsymbol{t} \cdot \delta\boldsymbol{u} \,\mathrm{d}S = {0},
    \label{eq:variational_weak_1}
\end{equation}
\begin{equation}
    \int_\Omega g'(\phi) \delta\phi \, \psi_a(\boldsymbol{\epsilon}(\boldsymbol{u})) \,\mathrm{d}\Omega 
    + (G_c + \lambda \cphi) \int_\Omega \left( \frac{1}{l} \phi \delta\phi + l \nabla\phi \cdot \nabla \delta \phi \right) \,\mathrm{d}\Omega = 0,
    \label{eq:variational_weak_2}
\end{equation}
\begin{equation}
    \delta \lambda \left[ \cphi \int_\Omega \left( \frac{1}{2l} \phi^2 + \frac{l}{2} |\nabla \phi|^2 \right) \,\mathrm{d}\Omega 
    +  \cu \int_{\partial_N\Omega} \boldsymbol{t} \cdot \boldsymbol{u} \,\mathrm{d}S - \tau(t) \right] = 0,
    \label{eq:variational_weak_3}
\end{equation}
where $\delta\boldsymbol{u},\delta\phi,\delta\lambda$ are, respectively, the variations of the displacement field, the phase field, and the Lagrange multiplier.
From the solution $(\boldsymbol{u}, \phi, \lambda)$, the Lagrange multiplier provides key physical insights: the effective applied force magnitude is scaled by the factor $(1 - \lambda \cu)$, and the effective critical energy release rate becomes $G_c^{\text{eff}} = (G_c + \lambda \cphi)$.

A fundamental characteristic of the variational scheme is that introducing the Lagrange multiplier replaces the toughness $G_c$ with a modified value $G_c^{\text{eff}}$. As a result, the equations do not directly trace the physical response for a constant material toughness $G_c$. Nevertheless, a precise mathematical relationship, detailed in Appendix~\ref{sec:scaling_relationships}, shows the relation between two problems with different critical energy release rates.

Based on this, equivalence with a problem that has a constant $G_c$ can be established by introducing a scaling factor, $\alpha$.
\begin{equation}
    \alpha = \sqrt{\frac{G_c}{G_c^{\text{eff}}}} = \frac{1}{\sqrt{1 + \frac{\lambda \cphi}{G_c}}}\ .
    \label{eq:correction_factor_variational}
\end{equation}
The physical quantities for constant $G_c$ can then be recovered from the variational simulation results using the following transformation equations:
\begin{equation}
   P_{G_c} = \alpha \cdot P_{\text{variational}}
   \label{eq:force_correction_variational}
\end{equation}
\begin{equation}
   u_{G_c} = \alpha \cdot u_{\text{variational}}
   \label{eq:displacement_correction_variational}
\end{equation}
\begin{equation}
   \psi_{G_c} = \alpha^2 \cdot \psi_{\text{variational}}
   \label{eq:energy_correction_variational}
\end{equation}
Here, $P_{\text{variational}}$, $u_{\text{variational}}$, and $\psi_{\text{variational}}$ are the reaction force, displacement, and strain energy density obtained from the variational solution, while $P_{G_c}$, $u_{G_c}$, and $\psi_{G_c}$ are the corresponding quantities for a solution with a constant $G_c$. These transformations provide a systematic approach to reconstruct the physical quantities. The factor $\alpha$ depends on the instantaneous value of the Lagrange multiplier $\lambda$ and ensures that the recovered response corresponds to the true material behavior.

\subsection{Non-variational scheme}
\label{subsec:non_variational_scheme}
As an alternative approach, we propose a non-variational energy-controlled scheme that directly computes the physical equilibrium path for an effective constant $G_c^{\text{eff}}=G_c$ without requiring post-processing transformations. This formulation is achieved by modifying the weak form equations to ensure that the critical energy release rate, $G_c$, remains constant throughout the simulation and is not influenced by the Lagrange multiplier $\lambda$.

The governing equations for this non-variational scheme are formulated as follows:
\begin{equation}
    \int_\Omega [g(\phi)\boldsymbol{\sigma}_a(\boldsymbol{\epsilon}(\boldsymbol{u})) + \boldsymbol{\sigma}_b(\boldsymbol{\epsilon}(\boldsymbol{u}))]:\boldsymbol{\epsilon}(\delta \boldsymbol{u}) \,\mathrm{d}\Omega 
    - (1 - \lambda\cu) \int_{\partial_N\Omega} \boldsymbol{t} \cdot \delta\boldsymbol{u} \,\mathrm{d}S = {0}, \label{eq:non_variational_momentum}
\end{equation}
\begin{equation}
    \int_\Omega g'(\phi) \delta\phi \, \psi_a(\boldsymbol{\epsilon}(\boldsymbol{u})) \,\mathrm{d}\Omega 
    + G_c \int_\Omega \left( \frac{1}{l} \phi \delta\phi + l \nabla\phi \cdot \nabla \delta \phi \right) \,\mathrm{d}\Omega = 0, \label{eq:non_variational_phase_field}
\end{equation}
\begin{equation}
    \delta \lambda \left[ \cphi \int_\Omega \left( \frac{1}{2l} \phi^2 + \frac{l}{2} |\nabla \phi|^2 \right) \,\mathrm{d}\Omega 
    +  \cu \int_{\partial_N\Omega} \boldsymbol{t} \cdot \boldsymbol{u} \,\mathrm{d}S - \tau(t) \right] = 0.
    \label{eq:non_variational_constraint}
\end{equation}

The fundamental distinction of this formulation lies in the treatment of the scalar $\lambda$, which — strictly speaking — is not a Lagrange multiplier but plays a similar role to $\lambda$ in Eqs.~\eqref{eq:variational_weak_1}-\eqref{eq:variational_weak_3}. This key difference ensures that the material toughness $G_c$ remains constant throughout the simulation, thereby enabling the solver to directly trace the physical force-displacement curve without requiring post-processing corrections.

Despite their fundamentally different mathematical formulations, both energy-controlled solvers—the variational (with post-processing corrections) and non-variational schemes—are mathematically equivalent in that they trace the same physical equilibrium path. The choice between them depends on computational preferences. In both energy-controlled approaches, the energy-based control parameter ensures a robust tracing of the complete equilibrium path, including complex instabilities such as snap-back and snap-through behavior.

\subsection{Computation of Compliance and its Derivative}
\label{subsec:computation_of_compliance}
A key advantage of the proposed energy-controlled framework is the ability to directly compute essential LEFM parameters from phase-field simulation results. The structural compliance, as a function of crack area $a$, is defined as the ratio of the displacement at the loading point, $u$, to the corresponding reaction force, $P$:
\begin{equation}
   C = \frac{u}{P}.
\end{equation}
This compliance can also be calculated from the total elastic energy of the body, $\mathcal{U}$, using $\mathcal{U} = \frac{1}{2}Pu$:
\begin{equation}
   C = \frac{2\mathcal{U}}{P^2} = \frac{2}{P^2} \int_\Omega \left[ g(\phi) \psi_a(\boldsymbol{\epsilon}(\boldsymbol{u})) + \psi_b(\boldsymbol{\epsilon}(\boldsymbol{u})) \right] \,\mathrm{d}\Omega\ .
   \label{eq:compliance_definition}
\end{equation}
Similarly, an energetically consistent displacement, $u_{\text{energy}}$, is:
\begin{equation}
   u_\mathrm{energy} = \frac{2\mathcal{U}}{P} = \frac{2}{P} \int_\Omega \left[ g(\phi) \psi_a(\boldsymbol{\epsilon}(\boldsymbol{u})) + \psi_b(\boldsymbol{\epsilon}(\boldsymbol{u})) \right] \,\mathrm{d}\Omega\ .
   \label{eq:energy_displacement}
\end{equation}
This energy-based calculation provides a robust measure of displacement that is consistent with the variational principles of the model.

For the fatigue analysis methodology proposed in this work, it is critical to obtain the rate of change of compliance with respect to crack area, $\mathrm{d}C/\mathrm{d}a$. A significant benefit of our approach is that this quantity can be computed directly without resorting to numerical differentiation, thus avoiding the inaccuracies associated with finite difference approximations. The calculation differs slightly depending on the solver used.

For the non-variational solver, the energy release rate is always equal to the material's fracture toughness, $G_c$. By rearranging the fundamental relationship from LEFM (Eq.~\eqref{eq:energy_release_rate_compliance}), the compliance derivative is computed directly as:
\begin{equation}
   \frac{\mathrm{d}C}{\mathrm{d}a} = \frac{2BG_c}{P^2}\ ,
   \label{eq:dC_da_direct_Gc}
\end{equation}
where $P$ is the reaction force obtained from the simulation.

For the variational solver, to ensure the result corresponds to the true material behavior, it is crucial to use the corrected physical force, $P_{G_c}$, as defined in Eq.~\eqref{eq:force_correction_variational}. This ensures that the computed compliance evolution is consistent with a constant fracture toughness $G_c$.

\section{Numerical implementation}
\label{sec:numerical_aspects}
The governing equations are discretized using the finite element method. Both the displacement field $\boldsymbol u$ and the phase-field $\phi$ are approximated within a standard Galerkin framework using the same set of nodal shape functions $N_a$. The function space for the unknowns is defined as:
\begin{gather}
   V = \left\{ \nu: \Omega \rightarrow \mathbb{R},\ \nu(\boldsymbol x) = \sum_{a=1}^{n_{\text{node}}} \Nax\, \nu_a \right\}
\end{gather}
The continuous fields and their variations are then expressed as:
\begin{gather}
   \boldsymbol u_h(\boldsymbol x) = \sum_{a=1}^{n_{\text{node}}} \Nax \boldsymbol u_a, \quad \delta \boldsymbol u_h(\boldsymbol x) = \sum_{a=1}^{n_{\text{node}}} \Nax \delta \boldsymbol u_a\ , \\
   \phi_h(\boldsymbol x) = \sum_{a=1}^{n_{\text{node}}} \Nax \phi_a, \quad \delta \phi_h(\boldsymbol x) = \sum_{a=1}^{n_{\text{node}}} \Nax \delta \phi_a\ .
\end{gather}

Substituting these approximations into the weak form equations yields a system of non-linear algebraic equations. The arbitrariness of the variations $\delta \boldsymbol u_a$ and $\delta \phi_a$ requires the residual vectors at each node, $\boldsymbol{R}^{\boldsymbol u}_a$ and $R^{\phi}_a$, to be zero. These are defined as:
\begin{equation}
\label{eq-rurphi}
        \boldsymbol{R}^{\phi}_a = 
        \underset{e}{\mathbf{A}} \left( R^{\phi e}_a \right) = \boldsymbol{0} \quad \text{and} \quad 
        \boldsymbol{R}^{\boldsymbol u}_a = \underset{e}{\mathbf{A}} \left( R^{\boldsymbol u e}_a \right) = \boldsymbol{0}
\end{equation}
where $\mathbf{A}$ is the assembly operator. The elemental residuals and tangent stiffness matrices are shown in appendix~\ref{sec:appendix_fem}.

The nonlinear system of Eqs.~\eqref{eq-rurphi} may be solved using an iterative method such as Newton-Raphson's; in this case, the Lagrange multiplier $\lambda$ would be treated as an additional global unknown.

Regarding the irreversibility behavior of crack growth, in both solvers, the Lagrange multiplier ensures that $\tau(t)$ increases monotonically over time, guaranteeing that the weighted sum of the two energy quantities in the constraint always increases. This condition does not prevent scenarios where, for example, in specimens with multiple cracks, one crack decreases in favor of increasing another, or where the applied energy can transfer from one part to another. However, regarding the solution strategy and implementation, this behavior does not influence any of the analyses performed.

\subsection{Adaptive Solution Strategy}
\label{subsec:adaptive_solution_strategy}
The non-linear system is solved monolithically with Newton-Raphson. An adaptive control-step algorithm adjusts $\Delta\tau$ based on convergence: if the solver converges quickly, $\Delta\tau$ increases; if not, it decreases. Failed steps are reverted and retried with a smaller $\Delta\tau$. The adaptive control-step algorithm used in this work is summarized in Algorithm~\ref{alg:adaptive_control_step}.

\begin{figure}[h!]
   \centering
   \begin{minipage}{0.95\linewidth}
      \begin{algorithmic}[1]
      \State \textbf{Initialize:} solution state $(\boldsymbol{u}_c, \phi_c, \lambda_c)$, control parameter $\tau$, step size $\Delta\tau$
      \State \textbf{Set:} min/max step sizes $\Delta\tau_{\min}, \Delta\tau_{\max}$, final control value $\tau_{\text{final}}$
      \State \textbf{Set:} golden ratio $\approx 0.618$ for step size adaptation
      \While{$\gamma < \gamma_{\text{final}}$}
         \State $\tau \gets \tau + \Delta\tau$ \Comment{Attempt a new step}
         \State $k, \text{converged} \gets \text{SolveNonlinearSystem}(\tau)$
         \If{\text{converged}}
            \State \textit{// Step accepted}
            \State UpdateState $(\boldsymbol{u}_c, \phi_c, \lambda_c) \gets (\boldsymbol{u}, \phi, \lambda)$
            \State SaveResults()
            \If{$k \le 2$}
               \State $\Delta\tau \gets \min(\Delta\tau \times (1 + \text{golden\_ratio}), \Delta\tau_{\max})$ \Comment{Increase step size}
            \EndIf
            \State $\text{step} \gets \text{step} + 1$
         \Else
            \State \textit{// Step rejected}
            \State $\tau \gets \tau - \Delta\tau$ \Comment{Revert control parameter}
            \State RevertState $(\boldsymbol{u}, \phi, \lambda) \gets (\boldsymbol{u}_c, \phi_c, \lambda_c)$
            \State $\Delta\tau \gets \max(\Delta\tau \times \text{golden\_ratio}, \Delta\tau_{\min})$ \Comment{Reduce step size}
            \If{$\Delta\tau \le \Delta\tau_{\min}$}
               \State \textbf{break} \Comment{Stop if step size is too small}
            \EndIf
         \EndIf
      \EndWhile
      \end{algorithmic}
   \end{minipage}
   \caption{Adaptive control-step algorithm for energy-controlled phase-field fracture simulation. The step size $\Delta\tau$ is adjusted based on convergence performance, with fast convergence ($\leq2$ iterations) triggering step size increases and failed convergence causing step size reduction using the golden ratio.}
   \label{alg:adaptive_control_step}
\end{figure}

\subsection{Correction of Simulated Crack Area and Physical Quantities}
\label{subsec:crack_area_measurement}

Accurate determination of crack surface area is crucial for reliable fatigue and fracture analysis. In the phase-field model, crack area is typically estimated by integrating the crack surface density functional, $\gamma(\phi, \nabla\phi)$, over the computational domain. However, this integral tends to systematically overestimate the true physical crack area.

To address this limitation, several correction methods have been proposed in the literature. In general, the corrected crack area can be expressed as:
\begin{equation}
   \gamma_\mathrm{phys} = \frac{\gamma_\mathrm{sim}}{\mathcal{F}_\mathrm{corr}}
   \label{eq:crack_area_correction}
\end{equation}
where $\gamma_\mathrm{phys}$ is the corrected crack area, $\gamma_\mathrm{sim}$ is the area obtained from the phase-field simulation, and $\mathcal{F}_\mathrm{corr}$ is a correction factor that compensates for the overestimation caused by the diffuse crack representation.

One classical approach, introduced by Francfort and Marigo~\cite{phase_field_FrancfortMarigo1998}, applies a correction factor based on mesh resolution:
\begin{equation}
   \mathcal{F}_\mathrm{corr} = 1 + \frac{h}{2l}
\end{equation}
where $h$ is the characteristic mesh size and $l$ is the phase-field length scale parameter. This correction remains constant for a given mesh and length scale.

An alternative, is the skeletonization or thresholding technique~\cite{phase_field_skeleton}. In this approach, crack area is measured directly from the finite element mesh by identifying regions where the phase-field variable exceeds a specified threshold (typically $\phi > 0.9$). The skeletonization algorithm then extracts the central path of the crack, representing it as a sequence of points with unit pixel length. To improve accuracy, especially for cracks propagating diagonally or with complex geometry, a spline is fitted through these points, and the crack length is calculated from the spline curve. This procedure accounts for sub-pixel crack propagation and avoids overestimation due to pixelation effects. Unlike the mesh-based correction, the correction factor in this method varies throughout the simulation and is computed at each step as:
\begin{equation}
   \mathcal{F}_\mathrm{corr} = \frac{\gamma_\mathrm{sim}}{\gamma_\mathrm{measured}}
\end{equation}
where $\gamma_\mathrm{measured}$ is the crack area obtained through the thresholding and skeletonization procedure. Details of the skeletonization algorithm are provided in the appendix \ref{sec:appendix_crack_measurement}.

\begin{table}[h!]
   \centering
   \begin{tabular}{ll}
      \toprule
      \textbf{Correction Method} & \textbf{Correction Factor} \\
      \midrule
      Francfort-Marigo~\cite{phase_field_FrancfortMarigo1998} & $1 + \dfrac{h}{2l}$ \\
      Skeleton/threshold method~\cite{phase_field_skeleton} & $\frac{\gamma_\mathrm{sim}}{\gamma_\mathrm{measured}}$ \\
      \bottomrule
   \end{tabular}
   \caption{Summary of methods for determining the crack area correction factor.}
   \label{tab:crack_area_corrections}
\end{table}

Once the correction factor $\mathcal{F}_\mathrm{corr}$ is determined, it is necessary to transform the simulated results into physically meaningful quantities. This ensures that the corrected values are consistent with the principles of LEFM. The basis for this transformation is the assumption that the structural compliance, a global property, is accurately captured by the simulation, i.e., $C_\mathrm{phys} = C_\mathrm{sim}$.

The relationship between the physical compliance derivative and the simulated one is established through the chain rule, using the corrected crack area $a_\mathrm{phys} = a_\mathrm{sim} / \mathcal{F}_\mathrm{corr}$:
\begin{equation}
   \frac{\mathrm{d}C}{\mathrm{d}a_\mathrm{phys}} = \frac{\mathrm{d}C}{\mathrm{d}a_\mathrm{sim}} \frac{\mathrm{d}a_\mathrm{sim}}{\mathrm{d}a_\mathrm{phys}} = \frac{\mathrm{d}C}{\mathrm{d}a_\mathrm{sim}} \mathcal{F}_\mathrm{corr}\ .
   \label{eq:corrected_dC_da}
\end{equation}
This direct computation approach does not require numerical differentiation, avoiding introducing the correction factor in the numerical derivative, and eviting the noise from the skeletonization-based crack measurements.

It should be noted that when this term is corrected, Eq.~\eqref{eq:dC_da_direct_Gc} becomes physically inconsistent. To restore the validity of this equation, the right-hand side must also be multiplied by the correction factor. The correction factor affects not only the surface area but also the energy release rate. Therefore, the effective energy release rate incorporates this correction:
\begin{equation}
   G_\mathrm{eff} = G_\mathrm{sim}\mathcal{F}_\mathrm{corr}
   \label{eq:corrected_G}
\end{equation}

As shown previously, the energy release rate does not affect specimen compliance. However, as demonstrated in Appendix \ref{sec:scaling_relationships}, displacement and force values are scaled by a factor, so in this case, the effective displacement and force must also be corrected. All quantities of interest, as shown in Table~\ref{tab:corrected_quantities}, are corrected by the factor $\mathcal{F}_\mathrm{corr}$. Note that depending on the type of correction factor, it can be either a constant or an array that affects each simulation step with a different correction factor. The correction factor is applied to the quantities listed in Table~\ref{tab:corrected_quantities}.
   \begin{table}[h!]
      \centering
      \renewcommand{\arraystretch}{1.5}
      \setlength{\tabcolsep}{12pt}
      \begin{tabular}{ll}
          \toprule
          \textbf{Physical Quantity} & \textbf{Transformation} \\ 
          \midrule
          Crack Area & $a_\mathrm{phys} = a_\mathrm{sim} / \mathcal{F}_\mathrm{corr}$ \\
          Force & $P_\mathrm{phys} = P_\mathrm{sim} / \sqrt{\mathcal{F}_\mathrm{corr}}$ \\
          Displacement & $u_\mathrm{phys} = u_\mathrm{sim} / \sqrt{\mathcal{F}_\mathrm{corr}}$ \\
          Compliance Derivative & $\frac{\mathrm{d}C}{\mathrm{d}a_\mathrm{phys}} = \frac{\mathrm{d}C}{\mathrm{d}a_\mathrm{sim}} \cdot \mathcal{F}_\mathrm{corr}$ \\
          Compliance & $C_\mathrm{phys} = C_\mathrm{sim}$ \\ 
          Critical Energy Release Rate & $G_\mathrm{phys} = G_\mathrm{sim} \cdot \mathcal{F}_\mathrm{corr}$ \\
          \bottomrule
      \end{tabular}
      \caption{Transformations from simulated to physical quantities using the correction factor $\mathcal{F}_\mathrm{corr}$.}
      \label{tab:corrected_quantities}
   \end{table}

\section{Validation}
\label{sec:validation} 
In this section different aspects involved in the methodology proposed will be analyzed and validated against closed form expressions and analytical results. For computational efficiency, several examples will take advantage of symmetry in the simulations. However, it is important to note that all graphs and theoretical solutions presented for comparison purposes will always refer to the complete model to ensure a clear and consistent validation.

\subsection{Energy-Controlled Schemes Validation}
\label{example:three_point_bending_test_specimen}
%
%
%
%
In this section the proposed energy-controlled techniques are validated by comparing the variational and non-variational phase-field solvers with each other and with the analytical solution from linear elastic fracture mechanics (LEFM) for a classical three-point bending test. The simulations use identical material parameters, mesh resolution, and boundary conditions to ensure a direct and fair comparison.

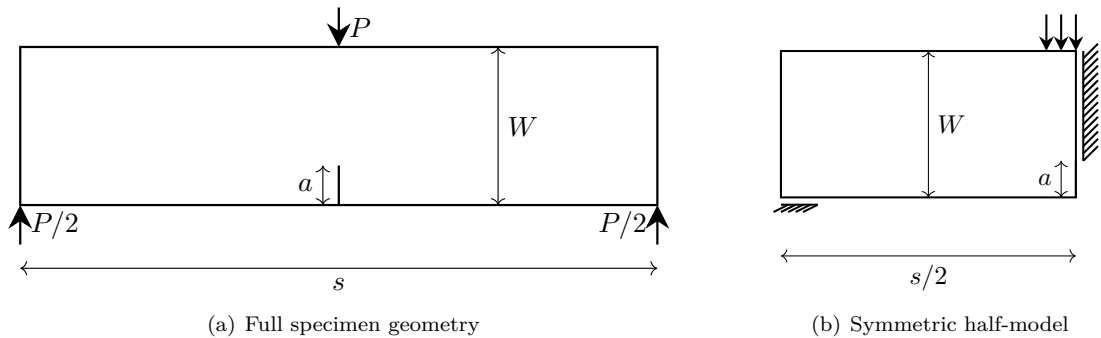
\begin{figure}[h!]
   \centering
   \subfigure[Full specimen geometry]{
       \resizebox{0.6\textwidth}{!}{%
         \begin{tikzpicture}[scale=1.0]
             \draw[thick] (0,0) rectangle (8,2);
             \draw[thick] (4,0) -- (4.0,0.5);
             \draw[<->] (3.8,0) -- (3.8,0.5) node[midway, left] {$a$};

             \draw[-{Stealth[length=3mm, width=3mm]}, thick] (8,-0.5) -- (8,0) node[midway, left] {$P/2$};

             \draw[-{Stealth[length=3mm, width=3mm]}, thick] (0,-0.5) -- (0,0) node[midway, right] {$P/2$};

             \draw[-{Stealth[length=3mm, width=3mm]}, thick] (4,2.5) -- (4,2) node[midway, right] {$P$};
             
             \draw[<->] (0, -0.8) -- (8, -0.8) node[midway, below] {$s$};
             \draw[<->] (6.0,0) -- (6.0,2) node[midway, right] {$\SpecimenWidth$};
         \end{tikzpicture}
      }
      \label{fig:three_point_full_geometry}
   }
   \hfill
   \subfigure[Symmetric half-model]{
      \resizebox{0.3\textwidth}{!}{%
         \begin{tikzpicture}[scale=1.0]
         \draw[thick] (0,0) rectangle (4,2);
         \draw[thick] (4,0) -- (4.0,0.5);
         \draw[<->] (3.8,0) -- (3.8,0.5) node[midway, left] {$a$};
         
         \foreach \x in {3.6, 3.8, ..., 4.0}
         {
             \draw[-{Stealth[length=2mm, width=2mm]}, thick] (\x,2.5) -- (\x,2.0);
         }
         
         \draw[thick] (0.0,-0.1) -- (0.5,-0.1);
         \foreach \x in {0.1, 0.2, ..., 0.5}
         {
             \draw[thick] (\x,-0.1) -- ++(-0.2,-0.1);
         }
         
         \draw[thick] (4.1,0.5) -- (4.1,2);
         \foreach \y in {0.5, 0.6,..., 2} {
            \draw[thick] (4.1,\y) -- ++(0.2, 0.2);
         }

         \draw[<->] (0, -0.8) -- (4, -0.8) node[midway, below] {$s/2$};
         \draw[<->] (2.0,0) -- (2.0,2) node[midway, right] {$\SpecimenWidth$};
         \end{tikzpicture}
      }
      \label{fig:three_point_half_geometry}
   }
   \caption{Geometry and boundary conditions for the three-point bending test. (a) Complete specimen geometry showing loads, supports, and dimensions. (b) Symmetric half-model used for computational efficiency, with symmetry boundary conditions applied to the right edge.}
   \label{fig:three_point_specimen_geometry}
\end{figure}

The three-point bending specimen consists of a rectangular plate with a centrally located notch, supported at both ends, as shown in Figure~\ref{fig:three_point_full_geometry}. According to LEFM theory~\cite{lefm_Tada}, the stress intensity factor for a specimen with a span-to-width ratio of $s/\SpecimenWidth = 8$ is given by Eq.~\eqref{eq:stress_intensity_factor}, where the geometric factor $f(a/\SpecimenWidth)$ is defined in Eq.~\eqref{eq:geometry_factor_three_point_bending} with an accuracy of $0.2\%$ for $a/\SpecimenWidth \leq 0.6$.
\begin{equation}
   f(a/\SpecimenWidth) = \frac{3 \sqrt{\pi} (s/\SpecimenWidth) \sqrt{a/\SpecimenWidth}}{2} 
   \left[ 1.106 - 1.552 \left(\frac{a}{\SpecimenWidth}\right) + 7.71 \left(\frac{a}{\SpecimenWidth}\right)^2 
   - 13.53 \left(\frac{a}{\SpecimenWidth}\right)^3 + 14.23 \left(\frac{a}{\SpecimenWidth}\right)^4 \right]
   \label{eq:geometry_factor_three_point_bending}
\end{equation}
The specimen's compliance, $C(a)$, can be determined by combining Eqs.~\eqref{eq:energy_release_rate_compliance} and \eqref{eq:relation_K_G}. Integrating the resulting expression with respect to the crack length yields the compliance for a specimen of thickness $B$:
\begin{equation}
   C(a) = \frac{C_{a_0}}{B} \left[1 + \frac{1}{C_{a_0}} \frac{2}{E' B \SpecimenWidth} \int_{a_0}^{a} f \left(\frac{a}{\SpecimenWidth} \right)^2 \, da \right],  
   \label{eq:compliance_three_point_bending} 
\end{equation}
where $C_{a_0}$ is the initial compliance per unit thickness, for a crack length $a_0$ and a specimen under plain strain conditions and unit thickness $B=1\,\text{mm}$. In the case of an uncracked specimen ($a_0=0\,\text{mm}$), the initial compliance can be calculated using beam theory as
\begin{equation}
   C_0 = \frac{s^3}{48 E' I},
\end{equation}
where $I = B \SpecimenWidth^3 / 12$ is the area moment of inertia of the beam's cross-section.

The critical force $P_c$ that the specimen can withstand before fracture can be determined either from the critical stress intensity factor $K_{Ic}$ using Eq.~\eqref{eq:critical_force_geometry_factor} or from the critical energy release rate considering $G_c$ in the Eq.~\eqref{eq:force_from_G}. With the compliance $C(a)$ and critical force $P_c(a)$ known, the full force-displacement equilibrium path, including snap-back, can be constructed. The force-displacement curve initially follows a linear behavior governed by the compliance $C(a_0)$ until the critical force $P_c(a_0)$ is reached. Subsequently, the system follows the equilibrium path defined by $u(a) = C(a) P_c(a)$ as the crack propagates.

For the numerical validation, a specimen with a span of $s = 8.0$~mm, a width of $\SpecimenWidth = 1.0$~mm, and a thickness of $B = 1.0$~mm was modeled, matching the $s/\SpecimenWidth = 8$ ratio. Figure~\ref{fig:three_point_half_geometry} shows the symmetric half-model used to improve computational efficiency. The boundary conditions consist of a fixed vertical displacement over a small area ($A_{\text{surface}} = 0.075$~mm$^2$) at the lower-left support and an applied downward vertical force over an equal area at the top center. The symmetry plane is constrained horizontally. Although to perfectly match the problem with the analytical solution, a contact problem for the support and application of the force must be considered, this configuration provides a good approximation due to the large span-to-width ratio, and the initial stiffness of the FEM model matches the analytical solution, due to the specific surface value considered. For other configurations, the initial compliance from the FEM model should be used for accurate comparison.

All simulations were performed with an initial crack of $a_0 = 0.2$~mm on a regular $400 \times 100$ grid of square elements with a size of $h=0.01$~mm. The material properties were set to $G_c = 0.0005$~kN/mm, $E = 20.8$~kN/mm$^2$, and $\nu = 0.3$. The phase-field length scale was $l = 0.03$~mm, resulting in a ratio of $l/h = 3$. To prevent damage under compression at the loading surface and supports, an anisotropic formulation with spectral energy decomposition was used. The applied traction vector was $\boldsymbol{t} = (0, -1/A_{\text{surface}})$. For the variational solver, the control parameters were $\cphi = 35.0$~kN/mm and $\cu = 1.0$, while for the non-variational solver, they were $\cphi = 1.0$~kN/mm and $\cu = 1.0$. An initial $\Delta\tau = 0.001$~kN$\cdot$mm was used for both solvers.

\begin{figure}[h!]
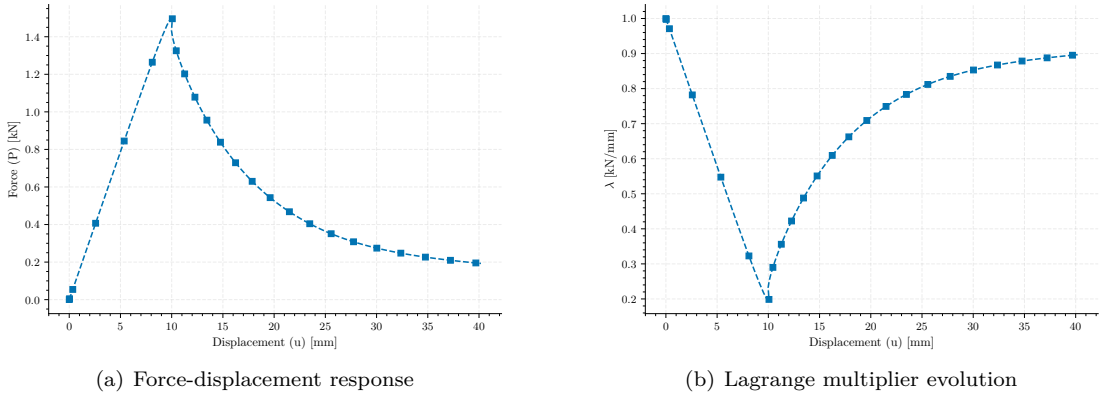

    \centering
    \subfigure[Force-displacement response]{
        \figpath{examples/Compare_Three_Point/results_three_point/variational_displacement_vs_force}[width=0.45\textwidth]
        \label{fig:variational_results_force_displacement}
    }
    \hfill
    \subfigure[Lagrange multiplier evolution]{
        \figpath{examples/Compare_Three_Point/results_three_point/variational_displacement_vs_lambda}[width=0.45\textwidth]
        \label{fig:variational_results_lagrange_displacement}
    }
    \caption{Results from the variational energy-controlled solver for the three-point bending test: (a) force-displacement curve showing the complete equilibrium path; (b) evolution of the Lagrange multiplier with displacement.}
    \label{fig:variational_results}
\end{figure}

For the variational solver, the resulting force-displacement and Lagrange multiplier-displacement curves are shown in Figures~\ref{fig:variational_results_force_displacement} and~\ref{fig:variational_results_lagrange_displacement}, respectively. It should be noted that this solution does not correspond to a constant $G_c$, as equilibrium is achieved by modifying the effective critical energy release rate, as explained in Section~\ref{subsec:variational_scheme}. To recover the solution for a constant $G_c$, Eqs.~\eqref{eq:force_correction_variational} and~\eqref{eq:displacement_correction_variational} must be applied.

\begin{figure}[h!]
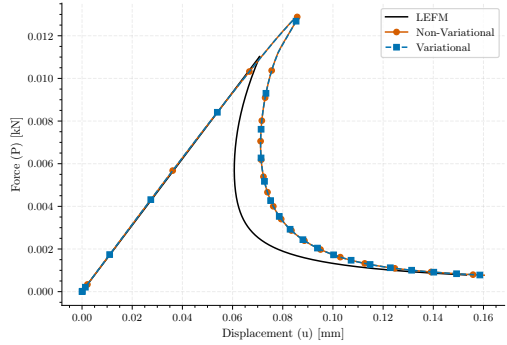
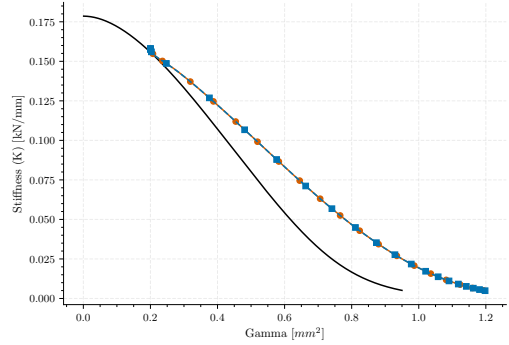

   \centering
   \subfigure[Force-displacement curves for both solvers compared with LEFM theory.]{
      \figpath{examples/Compare_Three_Point/results_three_point/compare_displacement_vs_force}[width=0.45\textwidth]
      \label{fig:comparation_three_point_reaction_force_vs_displacement}
   }
   \hfill
   \subfigure[Specimen stiffness as a function of crack length for both solvers and LEFM.]{
      \figpath{examples/Compare_Three_Point/results_three_point/compare_gamma_vs_stiffness}[width=0.45\textwidth]
      \label{fig:comparation_three_point_reaction_gamma_vs_stiffness}
   }
   \caption{Comparison of the energy-controlled phase-field solvers with the analytical LEFM solution for the three-point bending test. The phase-field results show an overestimation of both the peak force (a) and the crack length (b) compared to the LEFM predictions.}
   \label{fig:comparation_three_point}
\end{figure}

Figure~\ref{fig:comparation_three_point_reaction_force_vs_displacement} compares the force-displacement curves from both solvers with the LEFM solution. The non-variational solver directly computes the physical response for a constant $G_c$, while the variational solver's output yields identical results after applying the necessary correction. Both solvers show excellent agreement with each other, confirming their equivalence. However, the simulated peak force for the phase-field solvers is higher than the LEFM prediction. This is a known artifact attributed to the diffuse crack representation in phase-field models, which leads to an effective overestimation of the critical energy release rate, as explained in Section~\ref{subsec:crack_area_measurement}. A related phenomenon is observed in Figure~\ref{fig:comparation_three_point_reaction_gamma_vs_stiffness}, which shows that the crack length measured via the $\gamma$ functional is also overestimated. This causes the stiffness degradation curve to deviate from the LEFM solution.

\begin{figure}[h!]
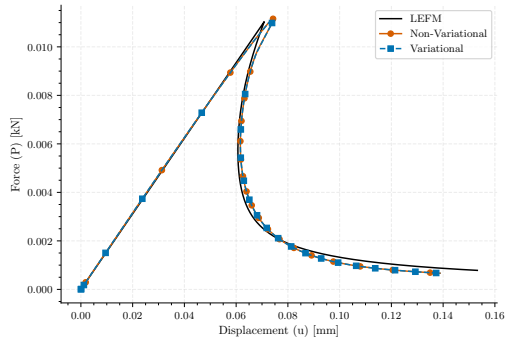
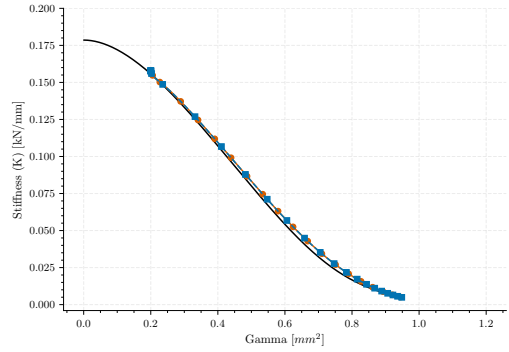

    \centering
    \subfigure[Corrected force-displacement curves showing improved agreement with LEFM.]{
        \figpath{examples/Compare_Three_Point/results_three_point/compare_displacement_vs_force_corrected_Gc}[width=0.45\textwidth]
        \label{fig:comparation_three_point_reaction_force_vs_displacement_corrected}
    }
    \hfill
    \subfigure[Corrected stiffness versus crack length relationship.]{
        \figpath{examples/Compare_Three_Point/results_three_point/compare_gamma_vs_stiffness_corrected_Gc}[width=0.45\textwidth]
        \label{fig:comparation_three_point_reaction_gamma_vs_stiffness_corrected}
    }
   \caption{Validation of the correction methodology for the three-point bending test. By applying the correction factor $\mathcal{F}_\mathrm{corr} = 1 + h/l$, the force-displacement (a) and stiffness-crack length (b) curves show excellent agreement with LEFM predictions.}
    \label{fig:comparation_three_point_corrected}
\end{figure}

To address this, the scaling relationships from Section~\ref{subsec:crack_area_measurement} are applied. In particular, the correction factor of $\mathcal{F}_\mathrm{corr} = 1 + 2h/(2l)$ proposed in \cite{phase_field_FrancfortMarigo1998} is used due to the regular shape and size of the elements. This correction is applied to the crack area, the force and displacement. It is important to emphasize that the factor of 2 multiplying the element size $h$ is specifically introduced because of the symmetry boundary condition used in the simulation, as referenced in~\cite{phase_field_effective_Gc_factor_2}. The corrected results are presented in Figures~\ref{fig:comparation_three_point_reaction_force_vs_displacement_corrected} and~\ref{fig:comparation_three_point_reaction_gamma_vs_stiffness_corrected}. It can be observed that both force-displacement curve and stiffness perfectly match the analytical LEFM expressions after the rescaling to correct the overestimation of the crack length.

In summary, this benchmark demonstrates that both energy-controlled schemes are equivalent and capable of robustly tracing the complete equilibrium path, including snap-back behavior. The validation confirms the accuracy of the proposed methodology and its potential for reliable fracture analysis, if the overestimation of the crack length and the critical energy release rate are taken properly into account.
\subsection{Validation of fatigue crack propagation: The center cracked test specimen}
\label{example:center_cracked_test_specimen}
%
%
%
%
%
%
%
This section evaluates the method's ability to predict the critical force and fatigue life for a center-cracked tension specimen. Similar to the three-point bending test, this example is particularly relevant because the crack path and geometric factor are well-known, allowing for a direct comparison with analytical solutions. First, the analytical LEFM solution is presented. Subsequently, to validate the LEFM solution numerically, a series of pure linear elastic simulations are performed by explicitly defining the crack in the mesh. Finally, a comprehensive analysis of the phase-field method is conducted. This includes investigating the influence of the length scale parameter $l$, the resulting compliance, and the correction factor used to address the overestimation of crack length. A fatigue analysis is then performed using Paris' law to demonstrate the framework's ability to predict fatigue life.

\begin{figure}[h!]
   \centering
   \subfigure[Center-cracked tension specimen geometry and loading.]{
   \begin{tikzpicture}[scale=1.0]
      \draw[thick] (0,0) rectangle (4,8);   
       \draw[thick] (1,4) -- ++(2,0); 
       \draw[<->] (1, 4.2) -- (2.0, 4.2) node[midway, above]{$a$};
       \draw[<->] (2, 4.2) -- (3, 4.2) node[midway, above]{$a$};
       \draw[<->] (0, 2.2) -- (2, 2.2) node[midway, above]{$\SpecimenWidth$};
       \draw[<->] (2, 2.2) -- (4, 2.2) node[midway, above]{$\SpecimenWidth$};
       \draw[<->] (4.5,0) -- (4.5,4) node[midway, right]{$H$};
       \draw[<->] (4.5,4) -- (4.5,8) node[midway, right]{$H$};
       \foreach \x in {0.0, 0.2, ..., 4.0}
           {
           \draw[-{Stealth[length=2mm, width=2mm]}, thick] (\x,8) -- (\x,8.5);
           }
        \node[anchor=center] at (2.0, 8.75) {$\boldsymbol \sigma$};
       \foreach \x in {0.0, 0.2, ..., 4.0}
           {
           \draw[-{Stealth[length=2mm, width=2mm]}, thick] (\x,0) -- (\x,-0.5);
           }
        \node[anchor=center] at (2.0, -0.75) {$\boldsymbol \sigma$};
   \end{tikzpicture}
   \label{fig:lefm_center_cracked_speciment}
   }
   \quad 
   \subfigure[Quarter-symmetry FEM model with boundary conditions.]{
      \begin{tikzpicture}[scale=1.0]
         \draw[thick] (2,4) rectangle (4,8);
         \draw[<->] (2, 4.2) -- (3, 4.2) node[midway, above] {$a$};
         \draw[<->] (2, 6.2) -- (4, 6.2) node[midway, above] {$\SpecimenWidth$};
         \draw[<->] (4.5,4) -- (4.5,8) node[midway, right] {$H$};
         \foreach \x in {2.0, 2.2, ..., 4.0}
         {
             \draw[-{Stealth[length=2mm, width=2mm]}, thick] (\x,8) -- (\x,8.5);
         }
         \draw[thick] (4.0,3.85) -- (3.0,3.85);
         \foreach \x in {3.1, 3.2, ..., 4.0}
         {
             \draw[thick] (\x,3.85) -- ++(-0.2,-0.2); 
         }
         \draw[thick] (1.8,4) -- (1.8,8);
         \foreach \y in {4.1, 4.2,..., 8} {
            \draw[thick] (1.8,\y) -- ++(-0.2,-0.2); 
         }
      \end{tikzpicture}
      \label{fig:fem_quarter_center_cracked_speciment}
   }
   \caption{Schematic of the center-cracked tension test specimen. (a) Full geometry, showing dimensions and loading configuration. (b) Quarter-symmetry finite element model, illustrating the applied boundary conditions for efficient simulation.}
   \label{fig:center_cracked_specimen_fig}
\end{figure}
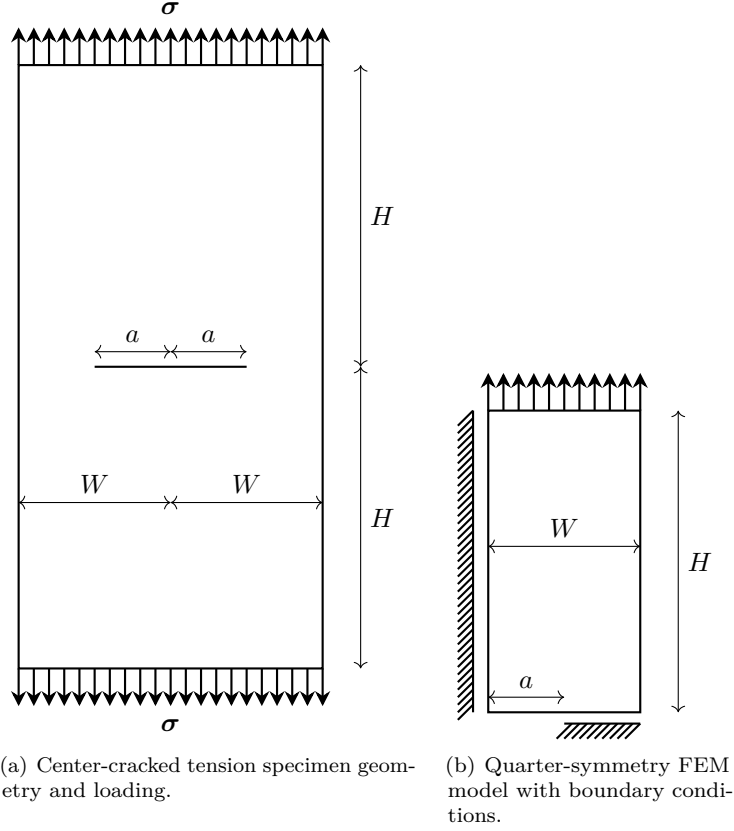

The center-cracked tension specimen, shown in Figure~\ref{fig:lefm_center_cracked_speciment}, is a rectangular plate with a centrally located crack subjected to uniform tensile loading. The specimen has a total width of $2\SpecimenWidth = 2.0\,\text{mm}$, a total height of $2H = 6.0\,\text{mm}$, and an initial crack of total length $2a$. According to \cite{lefm_Tada}, for the chosen aspect ratio of $H/\SpecimenWidth = 3$, the plate can be approximated as an infinite strip, which implies that the test yields equivalent results under both force- and displacement-controlled loading. The material properties are provided in Table~\ref{tab:lefm_material_properties_center_cracked_speciment}.

\begin{table}[h!]
   \centering
   \renewcommand{\arraystretch}{1.5} 
   \setlength{\tabcolsep}{12pt}      
   \begin{tabular}{lcc}
       \toprule
       \textbf{Parameter} & \textbf{Value} & \textbf{Units} \\ 
       \midrule
       Young's Modulus ($E$)                & 210       & $\text{kN/mm}^2$ \\ 
       Poisson's Ratio ($\nu$)              & 0.3       & --               \\ 
       Critical Energy Release Rate ($G_c$) & 0.0027    & $\text{kN/mm}$   \\
       Paris Exponent ($\nparis$)           & 2.5       & -- \\
       Fatigue Constant ($\Cparis$)         & $10^{-9}$ & $\frac{mm^{\frac{2+3\nparis}{2}}}{cycle \ kN^{\nparis}}$   \\
       \bottomrule
   \end{tabular}
   \caption{Material parameters for the center-cracked tension test specimen.}
   \label{tab:lefm_material_properties_center_cracked_speciment}
\end{table}

According to LEFM theory \cite{lefm_Tada}, the stress intensity factor for this specimen is given by Eq.~\eqref{eq:stress_intensity_factor}, where the geometric factor $f(a/\SpecimenWidth)$ is given in Eq.~\eqref{eq:geometry_factor_center_cracked}. This expression provides an accuracy within $0.1\%$ for any value of $a/\SpecimenWidth$.
\begin{equation}
   f(a/\SpecimenWidth)= \sqrt{\frac{\pi a}{4 \SpecimenWidth} \sec\left(\frac{\pi a}{2 \SpecimenWidth}\right)} \left[ 1.0 - 0.025\left(\frac{a}{\SpecimenWidth}\right)^2 + 0.06\left(\frac{a}{\SpecimenWidth}\right)^4 \right]
   \label{eq:geometry_factor_center_cracked}
\end{equation}

From this expression, the critical force $P_c$ required for fracture can be determined by combining Eq.~\eqref{eq:relation_K_G} and the critical stress intensity factor $K_{Ic}$:
\begin{equation}
   P_c = \frac{B \sqrt{\SpecimenWidth} \sqrt{E' G_c}}{f(a/\SpecimenWidth)}.
\end{equation}

Additionally, the specimen's compliance can be obtained by integrating the relationship between the energy release rate and the compliance derivative (Eqs.~\eqref{eq:energy_release_rate_compliance} and \eqref{eq:relation_K_G}):
\begin{equation}
   C(a) = \frac{C_{a_0}}{B} \left[1+ \frac{1}{C_{a_0}}\frac{4}{E' B \SpecimenWidth}  \int_{a_0}^{a} f \left(\frac{a}{\SpecimenWidth} \right)^2 \, da \right],   
\end{equation}
where $C_{a_0}$ is the initial compliance per unit thickness, for a crack length $a_0$ and a specimen under plain strain conditions and unit thickness $B=1\,\text{mm}$. For an uncracked specimen ($a_0=0\,\text{mm}$), the initial compliance is given by the theoretical expression $C_0 = \frac{H}{E' \SpecimenWidth}$.

\begin{figure}[h!]
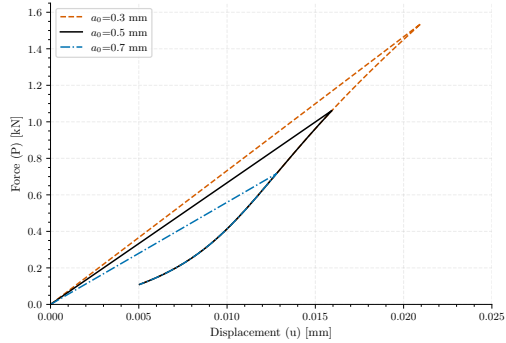
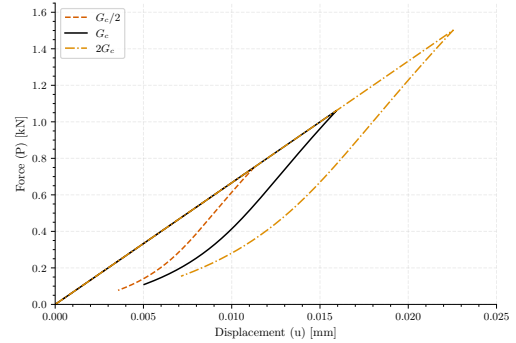

    \centering
    \subfigure[Effect of initial crack length $a_0$ on force-displacement response.]{
        \figpath{examples/LEFM/results_central_cracked/force_vs_displacement}[width=0.45\textwidth]
        \label{fig:lefm_force_displacement_center_cracked}
    }
    \hfill
    \subfigure[Effect of critical energy release rate $G_c$ on force-displacement response.]{
        \figpath{examples/LEFM/results_central_cracked/compare_gc_force_vs_displacement}[width=0.45\textwidth]
        \label{fig:lefm_influence_gc}
    }
    \caption{LEFM analytical force-displacement curves for the center-cracked specimen. (a) Shows that a larger initial crack length $a_0$ leads to lower initial stiffness, but the equilibrium paths converge upon crack propagation. (b) Illustrates that a higher $G_c$ increases the peak force but does not affect the specimen's stiffness.}
    \label{fig:compare_gc}
\end{figure}

By knowing the compliance $C(a)$ and critical force $P_c(a)$ as functions of crack length, the force-displacement equilibrium path ($G=G_c$) can be constructed for any initial crack length.
Figure~\ref{fig:lefm_force_displacement_center_cracked} shows the force-displacement response calculated with LEFM theory for different initial crack lengths. While a larger initial crack reduces specimen stiffness, the equilibrium paths converge to a single trajectory once propagation begins. Figure~\ref{fig:lefm_influence_gc} illustrates the effect of the critical energy release rate, $G_c$, for a fixed initial crack length of $a_0 = 0.5$~mm. As detailed in Appendix~\ref{sec:scaling_relationships}, a higher $G_c$ increases the critical force but does not alter the post-peak curve, since stiffness is independent of fracture toughness.

\subsubsection{FEM Elasticity Validation}
\label{subsec:central_cracked_validation_fem_elasticity}
To asses the validity of the LEFM expressions of stiffness and critical force in the case considered here, these values were compared with the results of an elastic finite element model of the specimen under both force- and displacement-controlled loading. The quarter-symmetry model shown in Figure~\ref{fig:fem_quarter_center_cracked_speciment} was used to improve computational efficiency. The crack was explicitly defined in the mesh, and appropriate boundary conditions were applied: the bottom edge was constrained in the vertical direction, except for the region representing the initial crack, while the left edge was fixed in the horizontal direction to enforce symmetry. Simulations were performed for crack lengths ranging from $a=0.0$ to $0.95\,\text{mm}$. The mesh consisted of quadrilateral elements with a size of $h = 0.005\,\text{mm}$.

\begin{figure}[h!]
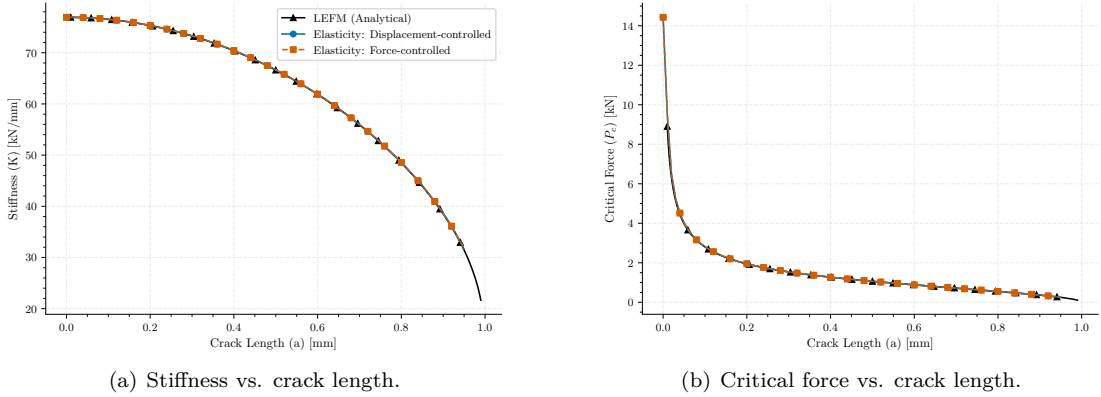

    \centering
    \subfigure[Stiffness vs. crack length.]{
        \figpath{examples/Compare_Central_Cracked/results_central_cracked/stiffness_vs_crack_length}[width=0.45\textwidth]
        \label{fig:compare_central_cracked_elasticity_lefm_stiffness}
    }
    \hfill
    \subfigure[Critical force vs. crack length.]{
        \figpath{examples/Compare_Central_Cracked/results_central_cracked/critical_force_vs_crack_length}[width=0.45\textwidth]
        \label{fig:compare_central_cracked_elasticity_critical_force}
    }
    \caption{Validation of stiffness and critical force for the center-cracked specimen. The results from both displacement- and force-controlled FEM simulations show excellent agreement with the analytical LEFM solutions, confirming the accuracy of the numerical model.}
    \label{fig:compare_central_cracked_elasticity_lefm}
\end{figure}

Figure~\ref{fig:compare_central_cracked_elasticity_lefm_stiffness} compares the specimen stiffness as a function of crack length from FEM simulations (under both force and displacement control) with the LEFM analytical solution. Figure~\ref{fig:compare_central_cracked_elasticity_critical_force} shows the corresponding critical force. For the FEM results, the compliance derivative, $dC/da$, was obtained by numerical differentiation. The excellent agreement between the numerical and analytical predictions for both stiffness and critical force confirms the reliability of the FEM approach.

\subsubsection{Phase-Field: Length Scale Influence}
Next, the phase-field approach is applied to the center-cracked specimen to investigate the influence of the length scale parameter $l$ and the initial crack length $a_0$. A series of simulations were performed with three initial crack lengths ($a_0 = 0.3, 0.5, 0.7$ mm) and two length scale parameters ($l_1 = 0.025$ mm and $l_2 = 0.0025$ mm), as summarized in Table~\ref{tab:phase_field_parameters_center_cracked_speciment}. The material properties are identical to those in the LEFM analysis (Table~\ref{tab:lefm_material_properties_center_cracked_speciment}).

The mesh size $h$ was refined proportionally to the length scale parameter $l$, ensuring a consistent resolution ratio of $l/h = 2.5$ across all simulations. For these analyses, the non-variational solver was employed without energy decomposition, adopting an isotropic formulation.

\begin{table}[h!]
   \centering
   \renewcommand{\arraystretch}{1.3}
   \setlength{\tabcolsep}{12pt}
   \begin{tabular}{cccc}
       \toprule
       \textbf{Simulation} & \textbf{Initial Crack Length ($a_0$)} & \textbf{Length Scale ($l$)} & \textbf{Mesh Size ($h$)} \\
       \midrule
       1 & 0.3 mm & 0.025 mm & 0.01 mm \\
       2 & 0.3 mm & 0.0025 mm & 0.001 mm \\
       \midrule
       3 & 0.5 mm & 0.025 mm & 0.01 mm \\
       4 & 0.5 mm & 0.0025 mm & 0.001 mm \\
       \midrule
       5 & 0.7 mm & 0.025 mm & 0.01 mm \\
       6 & 0.7 mm & 0.0025 mm & 0.001 mm \\
       \bottomrule
   \end{tabular}
   \caption{Phase-field simulations for the center-cracked tension test specimen. Three different initial crack lengths are considered, each with two different length scale parameters. The mesh size is refined proportionally to the length scale parameter to ensure accurate resolution of the diffuse crack interface.}
   \label{tab:phase_field_parameters_center_cracked_speciment}
\end{table}

\begin{figure}[h!]
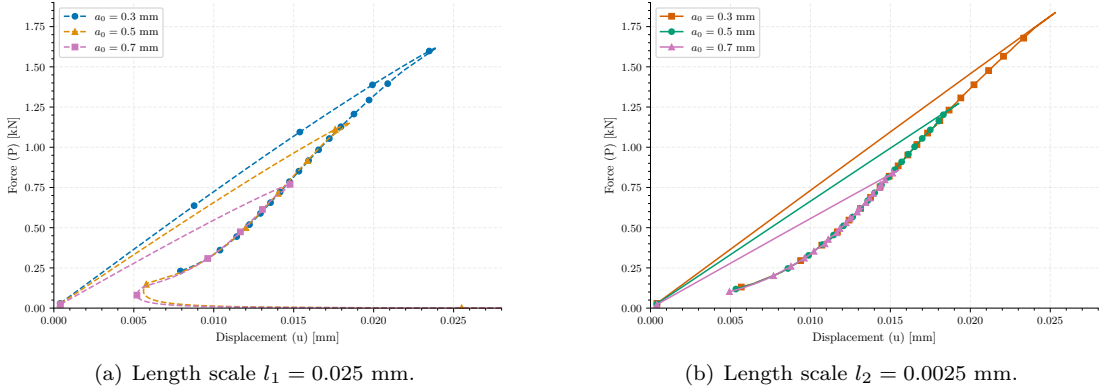

   \centering
   \subfigure[Length scale $l_1 = 0.025$ mm.]{
      \figpath{examples/Compare_Central_Cracked/results_compare_lenght_scale/compare_all_l1_u_vs_force}[width=0.45\textwidth]
      \label{fig:ppf_l1_a0}
   }
   \hfill
   \subfigure[Length scale $l_2 = 0.0025$ mm.]{
      \figpath{examples/Compare_Central_Cracked/results_compare_lenght_scale/compare_all_l2_u_vs_force}[width=0.45\textwidth]
      \label{fig:ppf_l2_a0}
   }
\caption{Force-displacement curves for three initial crack lengths under two phase-field length scales: (a) $l_1 = 0.025$ mm and (b) $l_2 = 0.0025$ mm. For each length scale, the curves converge after crack propagation, indicating that the equilibrium path becomes independent of the initial crack length.}
   \label{fig:ppf_l_a0}
\end{figure}

Figure~\ref{fig:ppf_l1_a0} and Figure~\ref{fig:ppf_l2_a0} show the force-displacement curves for the larger ($l_1$) and smaller ($l_2$) length scales, respectively. For a given length scale, the equilibrium paths converge after sufficient crack propagation, becoming independent of the initial crack length $a_0$. It is also observed that simulations with a smaller length scale parameter ($l_2$) yield a higher peak force.

Furthermore, for the larger length scale ($l_1$), the initial part of the force-displacement curve deviates from the linear elastic response, indicating damage initiation before the peak load. This effect is diminished for the smaller length scale ($l_2$), which aligns more closely with the expected elastic behavior.

\subsubsection{Phase-Field: Correction and Length Scale Effect}

To facilitate a direct and faithful comparison with the LEFM solution, the results from the phase-field simulations have been adjusted. As detailed in Section~\ref{subsec:crack_area_measurement}, raw phase-field models tend to overestimate the critical force. To counteract this, a constant Bourdin correction factor has been applied to the simulated force and displacement values. This straightforward scaling brings the numerical results into much closer alignment with the analytical predictions.

\begin{figure}[h!]
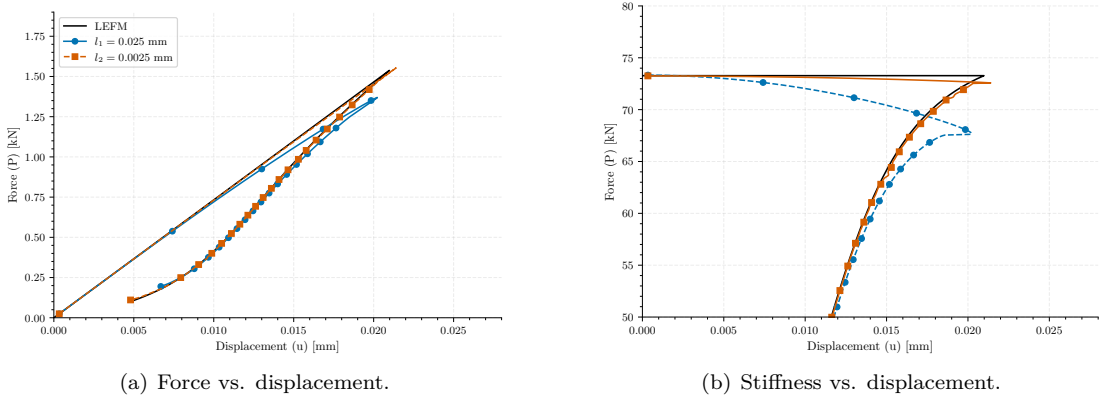

   \centering
   \subfigure[Force vs. displacement.]{
      \figpath{examples/Compare_Central_Cracked/results_compare_lenght_scale/compare_u_vs_force_l1_l2_lefm_corrected_gc}[width=0.45\textwidth]
      \label{fig:ppf_l1_l2_force_u}
   }
   \hfill
   \subfigure[Stiffness vs. displacement.]{
      \figpath{examples/Compare_Central_Cracked/results_compare_lenght_scale/compare_displacement_vs_stiffness_l1_l2_lefm_corrected_gc}[width=0.45\textwidth]
      \label{fig:ppf_l1_l2_stiffness_disp}
   }
\caption{Comparison of force-displacement (a) and stiffness-displacement (b) curves for an initial crack of $a_0 = 0.7$ mm. The results for two phase-field length scales are compared with the LEFM solution. The simulation with the smaller length scale ($l_2$) shows better agreement with the linear elastic behavior predicted by LEFM. The Bourdin correction has been applied to the phase-field results.}
   \label{fig:ppf_l1_l2_general}
\end{figure}

Figures~\ref{fig:ppf_l1_l2_force_u} and~\ref{fig:ppf_l1_l2_stiffness_disp} show the corrected force-displacement and stiffness-displacement curves, respectively, for an initial crack of $a_0 = 0.3$ mm, comparing two length scales ($l_1 > l_2$) against the LEFM solution. The corrected force-displacement curves show excellent agreement with LEFM. However, the stiffness plots reveal that the phase-field simulations still exhibit a gradual reduction prior to fracture, a sign of premature damage accumulation not present in the ideal LEFM case. This effect is more pronounced for the larger length scale ($l_1$), confirming that the simulation with the smaller length scale ($l_2$) more accurately captures the expected linear elastic behavior.

A critical aspect of the analysis is the measurement of crack length and its impact on key fracture metrics. To evaluate this, we compare the analytical LEFM solution with phase-field results (for a simulation with $a_0=0.3$~mm and $l=0.0025$~mm) using three different crack length definitions: direct integration of the $\gamma$ functional, the Bourdin correction, and the skeletonization measurement technique.

\begin{figure}[h!]
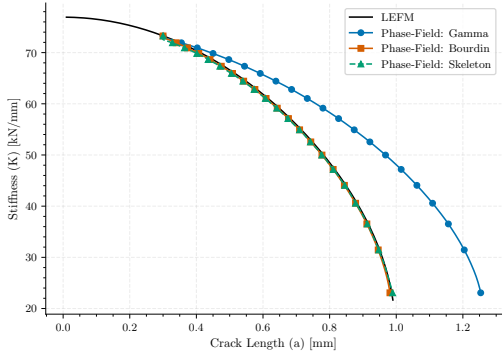
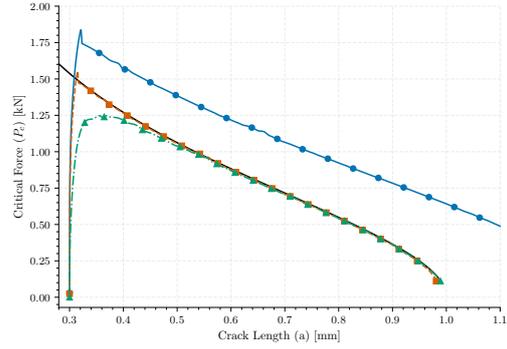

    \centering
    \subfigure[Stiffness vs. crack length.]{
        \figpath{examples/Compare_Central_Cracked/results_compare_lenght_scale/gamma_compare_vs_k}[width=0.45\textwidth]
        \label{fig:central_cracked_gamma_vs_k_correction}
    }
    \hfill
    \subfigure[Critical force vs. crack length.]{
        \figpath{examples/Compare_Central_Cracked/results_compare_lenght_scale/critical_force_vs_crack_length_pff}[width=0.45\textwidth]
        \label{fig:central_cracked_gamma_vs_critical_force_correction}
    }
    \caption{Validation of stiffness and critical force for the center-cracked specimen. The results from both displacement- and force-controlled FEM simulations show excellent agreement with the analytical LEFM solutions, confirming the accuracy of the numerical model.}
    \label{fig:compare_central_cracked_elasticity_lefm_gamma}
\end{figure}

Figure~\ref{fig:central_cracked_gamma_vs_k_correction} illustrates the resulting relationship between stiffness and crack length, while Figure~\ref{fig:central_cracked_gamma_vs_critical_force_correction} shows the corresponding critical force. Together, these figures demonstrate the performance of each correction method in aligning the phase-field predictions with the analytical solution.

\subsubsection{Fatigue Life Prediction}

A fatigue analysis was conducted to predict the crack propagation life of the center-cracked specimen. The analysis employed the Paris' law parameters specified in Table~\ref{tab:lefm_material_properties_center_cracked_speciment} and subjected the specimen to a cyclic load range of $\Delta P = 33$~kN (from $P_{\text{min}} = 0$~kN to $P_{\text{max}} = 33$~kN). The simulation tracked crack growth from an initial length of $a=0.4$~mm until complete failure.

\begin{figure}[h!]
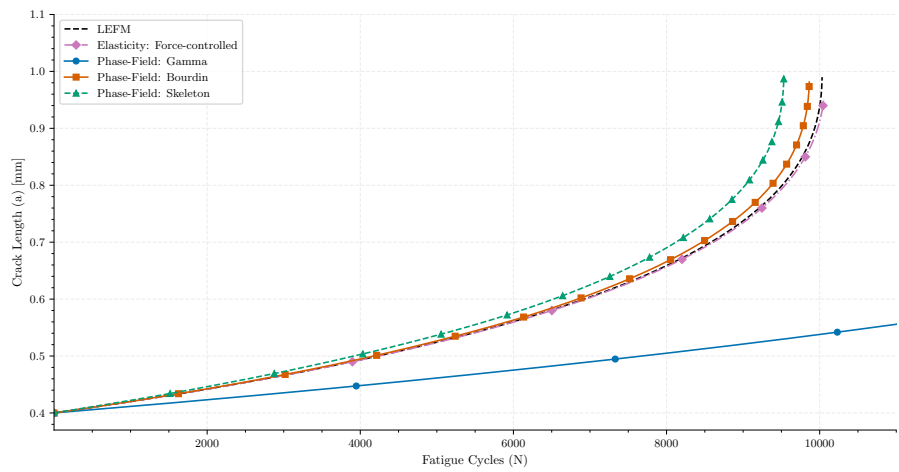

   \centering
   \figpath{examples/Compare_Central_Cracked/results_compare_fatigue/compare_paris_law_phase_field.pdf}[width=0.8\textwidth]
   \caption{Comparison of fatigue life predictions for the center-cracked specimen. The plot shows the number of cycles to failure as a function of crack length, comparing the phase-field model (with and without correction), the elastic simulation, and the analytical LEFM solution.}
   \label{fig:central_cracked_paris_law_lefm_elasticity}
\end{figure}

Figure~\ref{fig:central_cracked_paris_law_lefm_elasticity} presents the results, plotting the number of cycles to failure against the corresponding crack length. The figure provides a comprehensive comparison of fatigue life predictions obtained from various approaches. It includes the phase-field simulation results with $l=0.0025$~mm and $a_0=0.3$~mm (referring to simulation 2 in Table~\ref{tab:phase_field_parameters_center_cracked_speciment}), considering the fatigue analysis starting from $a=0.4$~mm. The results are shown for crack lengths measured directly via gamma integration, those corrected using the Bourdin gamma correction, and those adjusted with the skeleton correction. Additionally, the figure displays the results from the elasticity FEM analysis under force controlled conditions, where the crack was explicitly imposed through boundary conditions as explained in \ref{subsec:central_cracked_validation_fem_elasticity}, and the predictions derived from the analytical LEFM solution. This comparative assessment highlights the accuracy and differences among the methods.

The results indicate that the FEM simulation with a explicit crack representation in the mesh geometry aligns most closely with the LEFM reference solution. This suggests that the most accurate predictions from the phase-field model are achieved when its behavior approaches that of the pure elastic case. Notably, without applying a correction for the inherent overestimation of crack length in the phase-field model, the predicted number of cycles is significantly overestimated. However, when correction factors derived from skeletonization or Bourdin's method are applied, the phase-field predictions show excellent agreement with the theoretical solution, validating the approach proposed in the case of a straight crack path.

\clearpage
\section{Results and validation on complex geometries}
\label{example:compact_tension_test_specimen}
This section aims at evaluating the capability of the proposed methodology to predict fatigue life in complex scenarios with unknown crack paths. To this aim modified compact tension (CT) specimens with different initial cracks and containing internal voids are analyzed, serving as a challenging benchmark due to their geometric complexity and non-trivial crack propagation. The results of the proposed modeling approach for curvilinear paths  will be compared with the results of the experimental campaign performed by Wagner et. al. on the same modified CT specimens \cite{example_Wagner2019_article,example_Wagner2018_phd_thesis}.

\begin{figure}[h!]
   \centering
   \begin{tikzpicture}[scale=0.2]  
      \def\b{40}
      \def\hg{0.6*\b}         
      \def\a{0.2*\b}          
      \def\hone{0.275*\b}     
      \def\c{0.25*\b}         
      \def\D{0.25*\b}         
      \def\H{2.0}             
      
      \def\Da{0.2*\b}         
      \def\xa{0.45*\b}        
      \def\ha{0.25*\b}        
      
      \def\Db{0.1*\b}         
      \def\xb{0.75*\b}        
      \def\hb{0}              
      
      \def\Dc{0.2*\b}         
      \def\xc{0.625*\b}       
      \def\hc{-0.25*\b}       
      
      \def\f{0.06375*\b}      
      \def\j{0.08875*\b}      
      \def\k{0.04250*\b}      
      \def\l{0.00375*\b}      
      \def\g{\a - \l/0.57735}   
      
      \draw[very thick, black] (-\c,-\hg) -- (\b,-\hg);
      \draw[very thick, black] ( \b,-\hg) -- (\b, \hg);
      \draw[very thick, black] (-\c, \hg) -- (\b, \hg);

      \draw[very thick, black] (-\c,\hg) -- (-\c,\H+\f);
      \draw[very thick, black] (-\c,\H-\f) -- (-\c,-\hg);
      
      \draw[very thick, black] (0,\hone) circle (\D/2);
      \draw[thick, black] ({-0.5}, \hone) -- ({0.5}, \hone);
      \draw[thick, black] (0, {\hone-0.5}) -- (0, {\hone+0.5});
      
      \draw[very thick, black] (0,-\hone) circle (\D/2);
      \draw[thick, black] ({-0.5}, {-\hone}) -- ({0.5}, {-\hone});
      \draw[thick, black] (0, {-\hone-0.5}) -- (0, {-\hone+0.5});
      
      \draw[very thick, black] (\xa,\ha) circle (\Da/2);
      \draw[thick, black] ({\xa-0.5}, \ha) -- ({\xa+0.5}, \ha) node[right] {a};
      \draw[thick, black] (\xa, {\ha-0.5}) -- (\xa, {\ha+0.5});
      \draw[very thick, black] (\xb,\hb) circle (\Db/2) node[left] {b};
      \draw[thick, black] ({\xb-0.5}, \hb) -- ({\xb+0.5}, \hb);
      \draw[thick, black] (\xb, {\hb-0.5}) -- (\xb, {\hb+0.5});

      \draw[very thick, black] (\xc,\hc) circle (\Dc/2);
      \draw[thick, black] ({\xc-0.5}, \hc) -- ({\xc+0.5}, \hc) node[right] {c};
      \draw[thick, black] (\xc, {\hc-0.5}) -- (\xc, {\hc+0.5});
      \draw[<->, blue] (0,\H+1.5) -- (\a, \H+1.5) node[midway, above, blue] {$a$};
      \draw[<->, blue] (-\c,-\hg-3) -- (0,-\hg-3) node[midway, below, blue] {$D_{X}$};
      \draw[<->, blue] (0,-\hg-3) -- (\b,-\hg-3) node[midway, below, blue] {$\SpecimenWidth$};
      
      \draw[<->, blue] (\b+2, 0 ) -- (\b+2, \hg) node[midway, right, blue] {$h=D_{bY}$};
      \draw[<->, blue] (\b+2, 0 ) -- (\b+2,-\hg) node[midway, right, blue] {$h$};

      \draw[<->, blue] (-\c-4, \hg) -- (-\c-4, \H) node[midway, left, blue] {$H$};
      \draw[dashed, blue] (-\c-4, \H) -- (\a, \H);
      \draw[dashed, blue] (-\c-4, \hg) -- (-\c, \hg);
      \draw[<-, blue] ({\D/2*cos(45)}, {\hone + \D/2*sin(45)}) -- ({\D/2*cos(45) + 0.5}, {\hone + \D/2*sin(45) + 0.5}) -- ({\D/2*cos(45) + 1.5}, {\hone + \D/2*sin(45) + 0.5}) node[right, blue] {$\diameter D$};
      
      \draw[<-, blue] ({\D/2*cos(45)}, {-\hone + \D/2*sin(45)}) -- ({\D/2*cos(45) + 0.5}, {-\hone + \D/2*sin(45) + 0.5}) -- ({\D/2*cos(45) + 1.5}, {-\hone + \D/2*sin(45) + 0.5}) node[right, blue] {$\diameter D$};

      \draw[<-, blue] ({\xa + \Da/2*cos(45)}, {\ha + \Da/2*sin(45)}) -- ({\xa + \Da/2*cos(45) + 0.5}, {\ha + \Da/2*sin(45) + 0.5}) -- ({\xa + \Da/2*cos(45) + 1.5}, {\ha + \Da/2*sin(45) + 0.5}) node[right, blue] {$\diameter D_a$};

      \draw[<-, blue] ({\xb - \Db/2*cos(45)}, {\hb + \Db/2*sin(45)}) -- ({\xb - \Db/2*cos(45) - 0.5}, {\hb + \Db/2*sin(45) + 0.5}) -- ({\xb - \Db/2*cos(45) - 1.5}, {\hb + \Db/2*sin(45) + 0.5}) node[left, blue] {$\diameter D_b$};

      \draw[<-, blue] ({\xc + \Dc/2*cos(45)}, {\hc + \Dc/2*sin(45)}) -- ({\xc + \Dc/2*cos(45) + 0.5}, {\hc + \Dc/2*sin(45) + 0.5}) -- ({\xc + \Dc/2*cos(45) + 1.5}, {\hc + \Dc/2*sin(45) + 0.5}) node[right, blue] {$\diameter D_c$};
      
      \draw[<->, blue] (-\D/2-1.25,\hone) -- (-\D/2-1.25,\hg) node[midway, left, blue] {$D_Y$};
      \draw[<->, blue] (-\D/2-1.0,-\hg) -- (-\D/2-1.0,-\hone) node[midway, left, blue] {$D_Y$};
      
      \draw[dashed, blue] (0, -\hg-3) -- (0, \hg);
      \draw[dashed, blue] (-\D/2-1.0,-\hone) -- (0,-\hone);
      \draw[dashed, blue] (-\D/2-1.0, \hone) -- (0, \hone);

      \draw[<->, blue] (\xa-\Da/2-1.25,\ha) -- (\xa-\Da/2-1.25,\hg) node[midway, left, blue] {$D_{aY}$};
      \draw[<->, blue] (\xa,\ha+\Da/2+1.2) -- (\b,\ha+\Da/2+1.2) node[midway, above, blue] {$D_{aX}$};

      \draw[dashed, blue] (\xa,\ha+\Da/2+1.2) -- (\xa,\ha);
      \draw[dashed, blue] (\xa-\Da/2-1.25,\ha) -- (\xa,\ha);

      \draw[<->, blue] (\xb,\hb+\Db/2+1.2) -- (\b,\hb+\Db/2+1.2) node[midway, above, blue] {$D_{bX}$};
      \draw[dashed, blue] (\xb,\hb+\Db/2+1.2) -- (\xb,\hb);
      \draw[dashed, blue] (\b+2, 0) -- (\xb,\hb);

      \draw[<->, blue] (\xc-\Dc/2-1.25,\hc) -- (\xc-\Dc/2-1.25,-\hg) node[midway, left, blue] {$D_{cY}$};
      \draw[<->, blue] (\xc,\hc-\Dc/2-1.2) -- (\b,\hc-\Dc/2-1.2) node[midway, below, blue] {$D_{cX}$};

      \draw[dashed, blue] (\xc-\Dc/2-1.25,\hc) -- (\xc,\hc);
      \draw[dashed, blue] (\xc,\hc-\Dc/2-1.2) -- (\xc,\hc);

      \def\a{0.2*\b}          
      \def\c{0.25*\b}         

      \def\f{0.06375*\b}      
      \def\j{0.08875*\b}      
      \def\k{0.04250*\b}      
      \def\l{0.015*\b}      
      \def\g{\a - \l/0.57735}      

      \draw[thick, black] (-\c,\H+\f) -- (-\c+\k,\H+\j);
      \draw[thick, black] (-\c+\k,\H+\j) -- (-\c+\k,\H+\l);
      \draw[thick, black] (-\c+\k,\H+\l) -- (\g,\H+\l);
      \draw[thick, black] (\g,\H+\l) -- (\a,\H);
      \draw[thick, black] (\a,\H) -- (\g,\H-\l);
      \draw[thick, black] (\g,\H-\l) -- (-\c+\k,\H-\l);
      \draw[thick, black] (-\c+\k,\H-\l) -- (-\c+\k,\H-\j);
      \draw[thick, black] (-\c+\k,\H-\j) -- (-\c,\H-\f);

      \draw[dashed, blue] (-\c-4,\H-\j) -- (-\c+\k,\H-\j);

      \draw[dashed, blue] (-\c-2,\H-\f) -- (-\c+\k,\H-\f);
      
      \draw[<->, blue] (-\c-2,\H) -- (-\c-2,\H-\f) node[midway, left, blue] {$f$};
      \draw[<->, blue] (-\c-4,\H) -- (-\c-4,\H-\j) node[midway, left, blue] {$j$};

      \draw[<-, blue] (-2,\H-\l) -- (-2,\H-\f);
      \draw[<-, blue] (-2,\H+\l) -- (-2,\H+\f) node[midway, left, blue] {$2e$};

      \draw[<->, blue] (-\c,\H-\j-1.5) -- (-\c+\k,\H-\j-1.5) node[midway, below, blue] {$k$};
      \draw[dashed, blue] (-\c+\k,\H-\j-1.5) -- (-\c+\k,\H-\j);
      \draw[dashed, blue] (-\c,\H-\j-1.5) -- (-\c,\H-\f);

      \draw[thick] (-0.5*\c-0.5, \H-\l) -- (-0.5*\c+0.5, \H-\l);
      \draw[thick] (-0.5*\c-0.5, \H+\l) -- (-0.5*\c+0.5, \H+\l);

      \draw[thin, blue, dashed] (\a,\H) -- ({\a+2.6*cos(30)},{\H+2.6*sin(30)});
      \draw[thin, blue, dashed] (\a,\H) -- ({\a+2.6*cos(30)},{\H-2.6*sin(30)});

      \draw[<->, thick, blue] ({\a+2.0*cos(30)},{\H+2.0*sin(30)}) arc[start angle=30, end angle=-30, radius=2.0] node[midway, right, blue] {$60^\circ$};

   \end{tikzpicture}
   \caption{Compact tension specimen geometry with modified design featuring three circular holes. All dimensions are proportional to the base width $\SpecimenWidth = 40$ mm. The specimen includes a notched crack region positioned at distance $H$ from the top edge and three circular holes (labeled $a$, $b$, and $c$) that disrupt the specimen symmetry and influence crack propagation patterns.}
   \label{fig:compact_specimen_Hxx_geometry}
\end{figure}
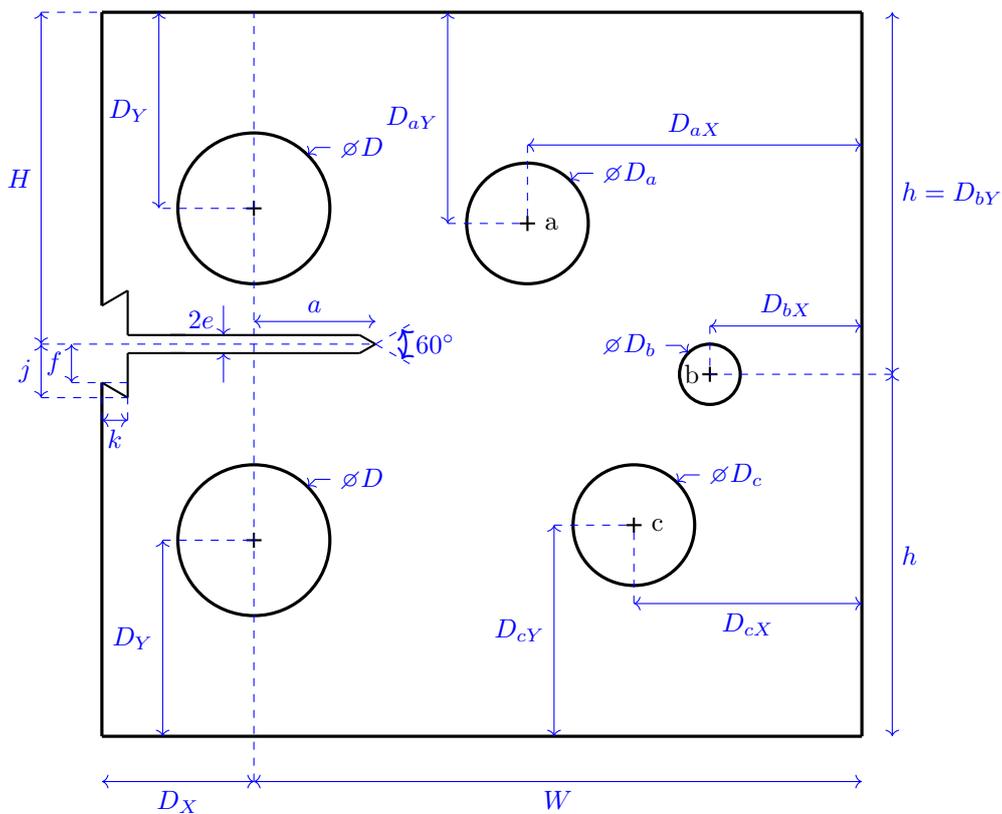

The specimen geometry is based on the ASTM-E647 standard and modified by introducing three additional circular holes and varying the initial crack tip position, denoted by the parameter $H$. The overall geometry is illustrated in Figure~\ref{fig:compact_specimen_Hxx_geometry}. All dimensions, proportional to the specimen width $\SpecimenWidth = 40$~mm, are listed in Table~\ref{tab:wagner_specimen_parameters}. The material is a Haynes 230 nickel-based superalloy and the properties used for this material are given in Table~\ref{tab:lefm_material_properties_compact}. Among these values, the elastic parameters $E,\nu$ were characterized in the publications of the experimental campaign \cite{example_Wagner2019_article,example_Wagner2018_phd_thesis}. On the contrary, fatigue related parameters were not reported and are taken from literature or estimated. Paris' law exponent is assumed to be independent on the particular material condition and is taken from \cite{example_Paris_material}. On the other hand, the prefactor has been adjusted such that the fatigue life of a simple specimen matches approximately the lifes reported in \cite{example_Wagner2019_article,example_Wagner2018_phd_thesis}. It is important to note that the value obtained is in the order of magnitude of the ones provided in other studies \cite{example_Paris_material}.
Finally, the critical energy released rate ($G_c$) is obtained from \cite{example_Gc}, although this parameter does not influence at all the fatigue life predictions in our framework.

The geometric complexity, particularly the additional holes, induces curvilinear crack growth, posing a significant challenge for traditional fracture models. In contrast, the phase-field method is well-suited to this problem, as it naturally predicts complex crack trajectories without remeshing. This study investigates the influence of the initial crack position through a parametric analysis of multiple configurations, each with a different value of $H$, as specified in Table~\ref{tab:compact_specimen_simulation_configurations}. The proposed energy-controlled approach enables the direct computation of the rate of change of the compliance with respect to crack length, $\mathrm{d}C/\mathrm{d}a$, providing the necessary inputs for the subsequent fatigue life analysis using Paris' law. The objective is to demonstrate the model's ability to capture how small variations in geometry can significantly alter crack paths and fatigue life predictions.

Once the compliance-crack length relationship is established for each specimen, the fatigue life is predicted by integrating Paris' law. For the standard compact tension specimen (without holes), the analytical LEFM solution is used as a reference, since closed-form expressions for the geometry factor and compliance are available. For the modified specimens with additional holes—where analytical solutions are not possible due to the complex crack paths—the fatigue life predictions from the numerical phase-field simulations are directly compared with experimental data reported in \cite{example_Wagner2018_phd_thesis}.

To ensure a meaningful comparison with the experimental data, the numerical fatigue analysis meticulously replicates the experimental loading protocol. The experiments involved first the generation of an initial pre-crack of approximately $3.5$~mm, followed by a fatigue test where the Stress Intensity Factor (SIF) range increased exponentially. Following the ASTM-E647 standard, the variable-amplitude loading is defined by:
\begin{equation}
   \Delta K(a) = (1-R) K_0 e^{\theta(a-a_i)}
\end{equation} 
where the initial SIF is $K_0 = 0.474$~kN/mm$^{3/2}$, the load ratio is $R=0.1$, the scaling parameter is $\theta = 0.04$~mm$^{-1}$, and $a_i$ is the initial crack length including the pre-crack.

Since the standard fatigue life calculation (Eq.~\eqref{eq:total_number_of_cycles_dCda}) was derived for constant amplitude loading, it must be adapted to the variable loading applied here. The resulting expression for the number of cycles to failure, $N_f$, is given by:
\begin{equation}
    N_f = N_i + \frac{1}{\Cparis (E'/(2B))^{\nparis/2} ((1-R) P_0)^{\nparis}} \int_{a_i}^{a_f} \frac{\mathrm{d}a}{(\mathrm{d}\Ccompliance/\mathrm{d}a)^{\nparis/2} \ e^{n \theta(a-a_i)}}.
    \label{eq:total_number_of_cycles_dCda_experiment}
\end{equation}
Here, the initial force, $P_0$, is calculated from the initial stress intensity factor, $K_0$, by first converting it to the energy release rate, $G$, using Eq.~\eqref{eq:relation_K_G}, and then applying Eq.~\eqref{eq:force_from_G}. This calculation uses the compliance derivative, $\mathrm{d}C_i/\mathrm{d}a_i$, corresponding to the initial crack length, $a_i$, which includes the pre-crack. The fatigue analysis employs the Paris law constant, $\Cparis$, and exponent, $\nparis$, from Table~\ref{tab:lefm_material_properties_compact}. It is important to note that while the load ratio, $R$, influences the crack growth rate, the crack path is primarily determined by the stress state at peak load.

\begin{table}[h!]
   \centering
   \renewcommand{\arraystretch}{1.5}
   \setlength{\tabcolsep}{10pt}
   \begin{tabular}{lcl}
      \toprule
      \textbf{Parameter} & \textbf{Value} & \textbf{Description} \\
      \midrule
      $\SpecimenWidth$ & $40$ mm & Specimen width (reference dimension) \\
      $H$ & Variable & Initial crack tip vertical position \\
      $L$   & $0.6\,\SpecimenWidth = 24$ mm & Half specimen height \\
      $a$ & $0.2\,\SpecimenWidth = 8$ mm & Initial crack length \\
      $D$ & $0.25\,\SpecimenWidth = 10$ mm & Loading hole diameter \\
      $D_X$ & $0.25\,\SpecimenWidth = 10$ mm & Notch horizontal offset \\
      $D_Y$ & $0.325\,\SpecimenWidth = 13$ mm & Notch mouth vertical offset \\
      \midrule
      $\diameter D_{a}$ & $0.2\,\SpecimenWidth = 8$ mm & Diameter of hole $a$ \\
      $\diameter D_{b}$ & $0.1\,\SpecimenWidth = 4$ mm & Diameter of hole $b$ \\
      $\diameter D_{c}$ & $0.2\,\SpecimenWidth = 8$ mm & Diameter of hole $c$ \\
      \midrule
      $D_{aX}$ & $0.55\,\SpecimenWidth = 22$ mm & Horizontal position of hole $a$ (from right edge) \\
      $D_{aY}$ & $0.35\,\SpecimenWidth = 14$ mm & Vertical position of hole $a$ (from specimen top edge) \\
      $D_{bX}$ & $0.25\,\SpecimenWidth = 10$ mm & Horizontal position of hole $b$ (from right edge) \\
      $D_{bY}$ & $0.6\,\SpecimenWidth = 24$ mm & Vertical position of hole $b$ (from specimen top edge) \\
      $D_{cX}$ & $0.375\,\SpecimenWidth = 15$ mm & Horizontal position of hole $c$ (from right edge) \\
      $D_{cY}$ & $0.35\,\SpecimenWidth = 14$ mm & Vertical position of hole $c$ (from specimen bottom edge) \\
      \midrule
      $f$ & $0.06375\,\SpecimenWidth = 2.55$ mm & Vertical half-width of the crack mouth opening \\
      $j$ & $0.08875\,\SpecimenWidth = 3.55$ mm & Vertical half-width of the notch throat section \\
      $k$ & $0.04250\,\SpecimenWidth = 1.70$ mm & Horizontal depth of the notch throat indentation \\
      $e$ & $0.00375\,\SpecimenWidth = 0.15$ mm & Half-height of the crack tip region \\
      \midrule
      $B$ & $0.08\,\SpecimenWidth = 3.2$ mm & Specimen thickness  \\
      \bottomrule
   \end{tabular}
   \caption{Geometric parameters for the compact tension specimen with additional holes and notched crack geometry. All dimensions are given as functions of the reference width $\SpecimenWidth = 40$ mm. The table summarizes the main specimen dimensions, hole locations and sizes, and crack notch geometric parameters used in the simulations.}
   \label{tab:wagner_specimen_parameters}
\end{table}

\begin{table}[h!]
   \centering
   \renewcommand{\arraystretch}{1.5} 
   \setlength{\tabcolsep}{12pt}      
   \begin{tabular}{lcc}
       \toprule
       \textbf{Specimen} & \textbf{H (Crack Position)} & \textbf{Holes Considered} \\ 
       \midrule
       Specimen 1 & $H = 0.60 \SpecimenWidth = 24.0$ mm   & No holes   \\
       Specimen 2 & $H = 0.56 \SpecimenWidth = 22.4$ mm   & With holes \\
       Specimen 3 & $H = 0.58 \SpecimenWidth = 23.2$ mm   & With holes \\
       Specimen 4 & $H = 0.64 \SpecimenWidth = 25.6$ mm   & With holes \\
       \bottomrule
   \end{tabular}
    \caption{Specimen configurations with different initial crack positions $H$ and hole presence. The first specimen represents the standard compact tension specimen with the crack positioned in the middle and no holes, while specimens 2-4 include three additional holes and have the crack positioned off-center.}
   \label{tab:compact_specimen_simulation_configurations}
\end{table}

\begin{table}[h!]
   \centering
   \renewcommand{\arraystretch}{1.5} 
   \setlength{\tabcolsep}{12pt}      
   \begin{tabular}{lcc}
       \toprule
       \textbf{Parameter} & \textbf{Value} & \textbf{Units} \\ 
       \midrule
       Young modulus ($E$)                 & $211$                       & $\text{kN/mm}^2$ \\ 
       Poisson ratio ($\nu$)               & $0.3$                       & --               \\ 
       Critial energy release rate ($G_c$) & $0.073$                     & $\text{kN/mm}$   \\
       Paris exponent ($\nparis$)          & $2.08$                      & -- \\
       Fatigue constant ($\Cparis$)        & $2.129 \times 10^{-5}$   & $\frac{mm^{\frac{2+3\nparis}{2}}}{cycle \ kN^{\nparis}}$   \\
       \bottomrule
   \end{tabular}
   \caption{Material properties for the Haynes 230 nickel-based superalloy. Elastic constants taken from \cite{example_Wagner2018_phd_thesis}. The Paris law exponent $n$ is obtained from \cite{example_Paris_material}, while the Paris constant $\Cparis$ is calibrated from experiments. The critical energy release rate $G_c$, which does not influence fatigue life in our model, is taken from \cite{example_Gc}.}
   \label{tab:lefm_material_properties_compact}
\end{table}



\subsection{Standard Compact Tension Test Specimen}
First, as a baseline for comparison, the standard CT specimen without additional holes is analyzed. This corresponds to Specimen 1 in Table~\ref{tab:compact_specimen_simulation_configurations}, where the initial crack is centered ($H=L$), resulting in a symmetric configuration. For this case, the theoretical crack path is straight, and an analytical expression for the geometry factor $f(a/\SpecimenWidth)$ is available in the literature~\cite{lefm_Tada, lefm_Anderson2005}. The stress intensity factor is given by Eq.~\eqref{eq:stress_intensity_factor}, with the geometry factor for the standard CT specimen defined as:
\begin{equation}
   f(a/\SpecimenWidth) = \frac{2 + \frac{a}{\SpecimenWidth}}{\left(1-\frac{a}{\SpecimenWidth}\right)^{\frac{3}{2}}}\left[0.886 + 4.64 \left(\frac{a}{\SpecimenWidth}\right) - 13.32 \left(\frac{a}{\SpecimenWidth}\right)^2 + 14.72 \left(\frac{a}{\SpecimenWidth}\right)^3 - 5.60 \left(\frac{a}{\SpecimenWidth}\right)^4\right]
   \label{eq:geometry_factor_compact_tension}
\end{equation}
which is accurate to within $0.5\%$ for $a/\SpecimenWidth > 0.2$.

Following the approach from the previous section, the specimen's compliance is obtained using LEFM with the expression:
\begin{equation}
   C(a) = \frac{C_{a_0}}{B} \left[ 1 + \frac{1}{C_{a_0}} \frac{2}{E'\SpecimenWidth}  \int_{a_0}^{a}  f\left(\frac{a}{\SpecimenWidth}\right)^2 \, da \right],
\end{equation}
where $C_{a_0}$ is the initial compliance per unit thickness, for a crack length $a_0$ and a specimen under plain strain conditions and unit thickness $B=1\,\text{mm}$. Since an analytical expression for $C_{a_0}$ is not available, it is calculated numerically via an elastic finite element analysis.

\begin{figure}[h!]
   \centering

\subfigure[Theoretical crack path based on geometry and loading conditions]{
      \resizebox{0.45\textwidth}{0.3\textheight}{%
         \begin{tikzpicture}[scale=0.15]  
      \def\b{40}
      \def\hg{0.6*\b}         
      \def\a{0.2*\b}          
      \def\hone{0.275*\b}     
      \def\c{0.25*\b}         
      \def\D{0.25*\b}         
      \def\H{0.0}             
      
      \def\Da{0.2*\b}         
      \def\xa{0.45*\b}        
      \def\ha{0.25*\b}        
      
      \def\Db{0.1*\b}         
      \def\xb{0.75*\b}        
      \def\hb{0}              
      
      \def\Dc{0.2*\b}         
      \def\xc{0.625*\b}       
      \def\hc{-0.25*\b}       
      
      \def\f{0.06375*\b}      
      \def\j{0.08875*\b}      
      \def\k{0.04250*\b}      
      \def\l{0.00375*\b}      
      \def\g{\a - \l/0.57735}           
      
      
      \draw[thick] (-\c,-\hg) -- (\b,-\hg);
      \draw[thick] ( \b,-\hg) -- (\b, \hg);
      \draw[thick] (-\c, \hg) -- (\b, \hg);

      \draw[thick] (-\c,\hg) -- (-\c,\H+\f);
      \draw[thick] (-\c,\H+\f) -- (-\c+\k,\H+\j);
      \draw[thick] (-\c+\k,\H+\j) -- (-\c+\k,\H+\l);
      \draw[thick] (-\c+\k,\H+\l) -- (\g,\H+\l);
      \draw[thick] (\g,\H+\l) -- (\a,\H);
      \draw[thick] (\a,\H) -- (\g,\H-\l);
      \draw[thick] (\g,\H-\l) -- (-\c+\k,\H-\l);
      \draw[thick] (-\c+\k,\H-\l) -- (-\c+\k,\H-\j);
      \draw[thick] (-\c+\k,\H-\j) -- (-\c,\H-\f);
      \draw[thick] (-\c,\H-\f) -- (-\c,-\hg);
      
      \draw[thick] (0,\hone) circle (\D/2);
      \draw[thick] (0,-\hone) circle (\D/2);
      \foreach \angle in {180, 190, ..., 360}
      {
          \draw[thick, blue] ({0 + (\D/2)*cos(\angle)}, {-\hone + (\D/2)*sin(\angle)}) 
            -- ++({-(0.8/40)*\b*cos(\angle)}, {-(0.8/40)*\b*sin(\angle)});
      }
      \node[anchor=north, blue] at (0, -\hone - \D/2 - 1) {$u_x = u_y = 0$};
      
      \foreach \angle in {180, 190, ..., 360}
      {
          \draw[thick, blue] ({0 + (\D/2)*cos(\angle)}, {-\hone + (\D/2)*sin(\angle)}) 
               -- ++({-(0.8/40)*\b*cos(\angle)}, {-(0.8/40)*\b*sin(\angle)});
      }

      \foreach \angle in {10, 30,..., 170}
      {
          \draw[->, thick, blue] ({0 + (\D/2)*cos(\angle)}, {\hone + (\D/2)*sin(\angle)}) 
               -- ++({0}, {(2.4/40)*\b});
      }

   \end{tikzpicture}
      }
      \label{fig:compact_specimen_a8_boundary}
   }
   \hfill
   \subfigure[Theoretical crack path based on geometry and loading conditions]{
      \resizebox{0.45\textwidth}{0.3\textheight}{%
         \begin{tikzpicture}[scale=0.15]  
      \def\b{40}
      \def\hg{0.6*\b}         
      \def\a{0.75*\b}          
      \def\hone{0.275*\b}     
      \def\c{0.25*\b}         
      \def\D{0.25*\b}         
      \def\H{0.0}             
      
      \def\Da{0.2*\b}         
      \def\xa{0.45*\b}        
      \def\ha{0.25*\b}        
      
      \def\Db{0.1*\b}         
      \def\xb{0.75*\b}        
      \def\hb{0}              
      
      \def\Dc{0.2*\b}         
      \def\xc{0.625*\b}       
      \def\hc{-0.25*\b}       
      
      \def\f{0.06375*\b}      
      \def\j{0.08875*\b}      
      \def\k{0.04250*\b}      
      \def\l{0.00375*\b}      
      \def\g{\a - \l/0.57735}           
      
      
      \draw[thick] (-\c,-\hg) -- (\b,-\hg);
      \draw[thick] ( \b,-\hg) -- (\b, \hg);
      \draw[thick] (-\c, \hg) -- (\b, \hg);

      \draw[thick] (-\c,\hg) -- (-\c,\H+\f);
      \draw[thick] (-\c,\H+\f) -- (-\c+\k,\H+\j);
      \draw[thick] (-\c+\k,\H+\j) -- (-\c+\k,\H+\l);
      \draw[thick] (-\c+\k,\H+\l) -- (\g,\H+\l);
      \draw[thick] (\g,\H+\l) -- (\a,\H);
      \draw[thick] (\a,\H) -- (\g,\H-\l);
      \draw[thick] (\g,\H-\l) -- (-\c+\k,\H-\l);
      \draw[thick] (-\c+\k,\H-\l) -- (-\c+\k,\H-\j);
      \draw[thick] (-\c+\k,\H-\j) -- (-\c,\H-\f);
      \draw[thick] (-\c,\H-\f) -- (-\c,-\hg);
      
      \draw[thick] (0,\hone) circle (\D/2);
      \draw[thick] (0,-\hone) circle (\D/2);
      \foreach \angle in {180, 190, ..., 360}
      {
          \draw[thick, blue] ({0 + (\D/2)*cos(\angle)}, {-\hone + (\D/2)*sin(\angle)}) 
            -- ++({-(0.8/40)*\b*cos(\angle)}, {-(0.8/40)*\b*sin(\angle)});
      }
      \node[anchor=north, blue] at (0, -\hone - \D/2 - 1) {$u_x = u_y = 0$};
      
      \foreach \angle in {180, 190, ..., 360}
      {
          \draw[thick, blue] ({0 + (\D/2)*cos(\angle)}, {-\hone + (\D/2)*sin(\angle)}) 
               -- ++({-(0.8/40)*\b*cos(\angle)}, {-(0.8/40)*\b*sin(\angle)});
      }

      \foreach \angle in {10, 30,..., 170}
      {
          \draw[->, thick, blue] ({0 + (\D/2)*cos(\angle)}, {\hone + (\D/2)*sin(\angle)}) 
               -- ++({0}, {(2.4/40)*\b});
      }

   \end{tikzpicture}
      }
      \label{fig:compact_specimen_a30_boundary}
   }
    \caption{Boundary conditions for the compact tension specimen (Specimen 1). The lower half of the bottom loading hole is fixed in all directions ($u_x = u_y = 0$), while a distributed vertical force is applied to the upper half of the top loading hole.}
   \label{fig:compact_specimen_boundary_conditions}
\end{figure}
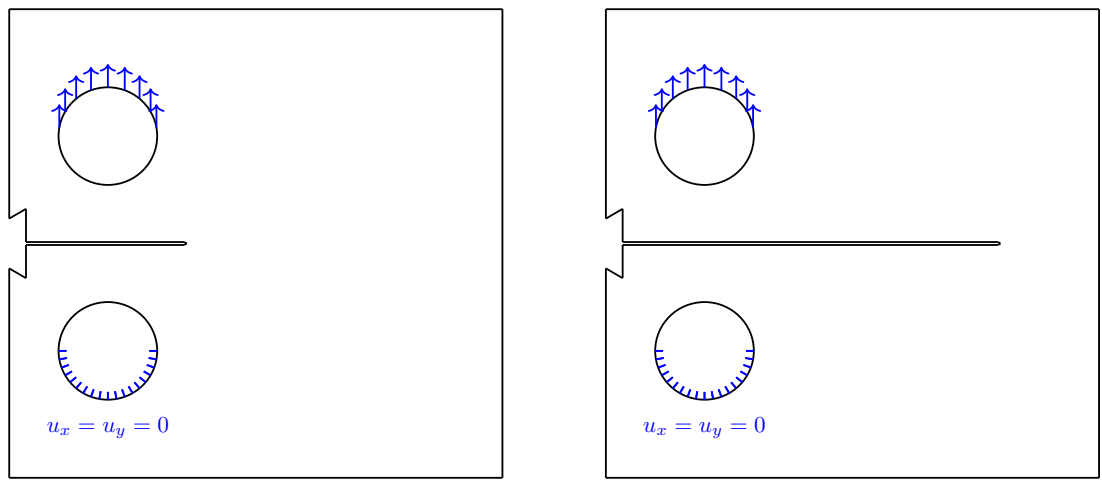

Guided by the analysis in~\cite{example_boundary_conditions}, the boundary conditions were chosen to ensure close agreement with established LEFM solutions. As illustrated in Figure~\ref{fig:compact_specimen_boundary_conditions}, these conditions involve constraining the lower half of the bottom loading hole in all directions and applying a distributed vertical force to the upper half of the top loading hole. To validate the chosen boundary conditions, a series of linear elastic finite element analyses were performed. This involved generating multiple meshes, each with an explicitly defined crack of length $a$, ranging from an initial length of $a_0 = 0.2\SpecimenWidth = 8.0$~mm up to $0.95\SpecimenWidth = 38.0$~mm in increments of $\Delta a = 0.5$~mm. For each mesh, a single force-controlled linear elastic analysis was conducted to compute the structural compliance. For instance, Figure~\ref{fig:compact_specimen_a8_boundary} depicts the schematic with an initial crack length of $a_{0}=8.0$~mm, whereas Figure~\ref{fig:compact_specimen_a30_boundary} presents the configuration for a crack length of $a=30.0$~mm. This process established a numerical compliance-crack length relationship, $C(a)$, which was then compared with the analytical LEFM solution to verify the accuracy of the simulation setup.

The linear elastic finite element model employs quadrilateral elements with a characteristic size of $h = 0.001\SpecimenWidth = 0.04$~mm in the central region, ensuring consistency with the mesh refinement used in subsequent phase-field simulations. The Young's modulus $E$ and Poisson's ratio $\nu$ for these simulations are specified in Table~\ref{tab:lefm_material_properties_compact}. 

For the phase-field simulation a length scale parameter of $l = 0.0025\SpecimenWidth = 0.1$~mm is chosen, resulting in a ratio of $l/h = 2.5$.  The simulation is performed with the same boundary conditions as the linear elastic simulations. An isotropic damage PFF model is used, and the non-variational phase-field solver is employed with constants $\cphi = 1.0$~kN/mm and $\cu = 1.0$ (dimensionless). The applied traction force is $\boldsymbol{t} = (0, 1/\text{surface})$~kN, where "surface" is the area over which the force is applied (the upper half of the top circle).

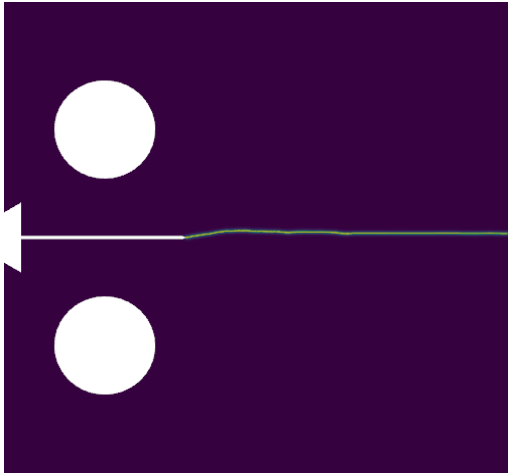
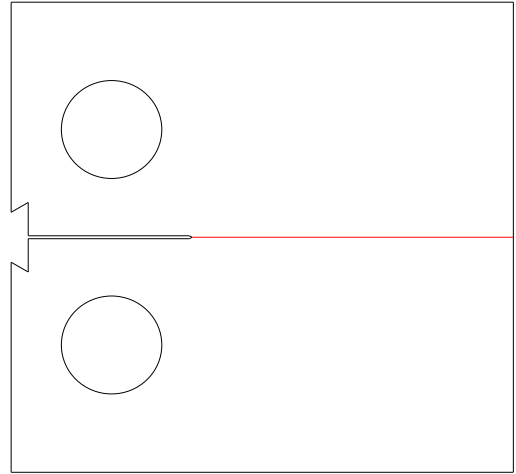
\begin{figure}[h!]
   \centering

   \subfigure[Final phase-field simulation results showing the crack path at the last step of the simulation]{
      \figpath{examples/Phase_Field_Compact_Specimen/results_specimen_1_H00/paraview_phi}[width=0.45\textwidth, height=0.3\textheight, keepaspectratio=false]
      \label{fig:simulation_1_results}
   }
   \hfill
   \subfigure[Theoretical crack path based on geometry and loading conditions]{
      \resizebox{0.45\textwidth}{0.3\textheight}{%
         \begin{tikzpicture}[scale=1.0]  
         \def\b{40}
         \def\hg{0.6*\b}         
         \def\a{0.2*\b}          
         \def\hone{0.275*\b}     
         \def\c{0.25*\b}         
         \def\D{0.25*\b}         
         \def\H{0.0}             
         
         \def\Da{0.2*\b}         
         \def\xa{0.45*\b}        
         \def\ha{0.25*\b}        
         
         \def\Db{0.1*\b}         
         \def\xb{0.75*\b}        
         \def\hb{0}              
         
         \def\Dc{0.2*\b}         
         \def\xc{0.625*\b}       
         \def\hc{-0.25*\b}       
         
         \def\f{0.06375*\b}      
         \def\j{0.08875*\b}      
         \def\k{0.04250*\b}      
         \def\l{0.00375*\b}      
         \def\g{\a - \l/0.57735}            
               
         \draw[thick] (-\c,-\hg) -- (\b,-\hg);
         \draw[thick] ( \b,-\hg) -- (\b, \hg);
         \draw[thick] (-\c, \hg) -- (\b, \hg);

         \draw[thick] (-\c,\hg) -- (-\c,\H+\f);
         \draw[thick] (-\c,\H+\f) -- (-\c+\k,\H+\j);
         \draw[thick] (-\c+\k,\H+\j) -- (-\c+\k,\H+\l);
         \draw[thick] (-\c+\k,\H+\l) -- (\g,\H+\l);
         \draw[thick] (\g,\H+\l) -- (\a,\H);
         \draw[thick] (\a,\H) -- (\g,\H-\l);
         \draw[thick] (\g,\H-\l) -- (-\c+\k,\H-\l);
         \draw[thick] (-\c+\k,\H-\l) -- (-\c+\k,\H-\j);
         \draw[thick] (-\c+\k,\H-\j) -- (-\c,\H-\f);
         \draw[thick] (-\c,\H-\f) -- (-\c,-\hg);
         
         \draw[thick] (0,\hone) circle (\D/2);
         \draw[thick] (0,-\hone) circle (\D/2);

          \draw[thick, red, line width=2pt] (8, 0) -- (40, 0);
         \end{tikzpicture}%
      }
      \label{fig:simulation_1_geometry}
   }
   \caption{Comparison of phase-field simulation results with theoretical predictions for the standard compact tension (CT) specimen without additional holes. (a) The phase-field simulation accurately captures the straight crack propagation path. (b) Theoretical crack path based on the specimen's geometry and loading conditions.}
   \label{fig:simulation_no_holes_complete}
\end{figure}

Figure~\ref{fig:simulation_no_holes_complete} compares the final crack path predicted by the phase-field simulation with the theoretical straight path. The left panel shows the phase-field variable $\phi$ at the final step, while the right panel shows the theoretical crack path. The simulation accurately captures the straight crack propagation, demonstrating the method's ability to predict crack evolution without prior knowledge of the path. A slight upward deviation at the beginning of the crack path is observed, which can be attributed to mesh effects.

\begin{figure}[h!]
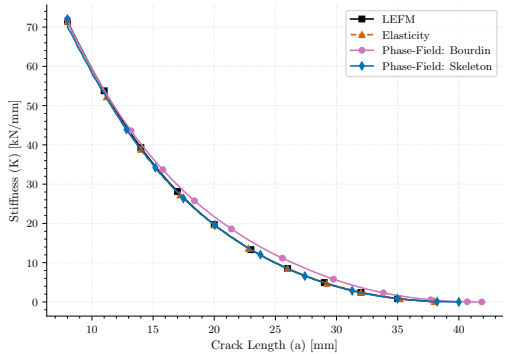
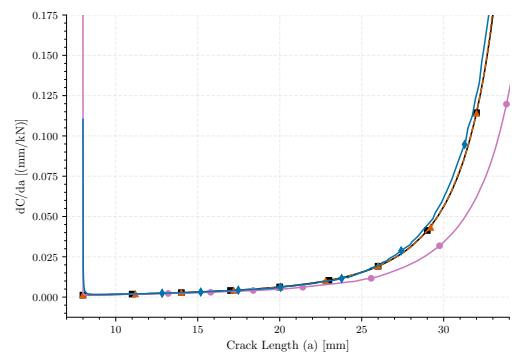

   \centering
   \subfigure[Stiffness versus crack length, validating the phase-field simulation against analytical and numerical linear elastic solutions.]{
      \figpath{examples/Compare_Compact_Specimen/compact_tension/compare_stiffness}[width=0.45\textwidth]
      \label{fig:compact_tension_stiffness_crack_lenght}
   }
   \hfill
   \subfigure[Compliance derivative ($\mathrm{d}C/\mathrm{d}a$) as a function of crack length, comparing phase-field simulation results with analytical and numerical solutions.]{
      \figpath{examples/Compare_Compact_Specimen/compact_tension/dCda_vs_crack_length}[width=0.45\textwidth]
      \label{fig:compact_tension_dCda}
   }
   \caption{Validation results for the standard compact tension specimen: (a) Stiffness versus crack length, validating the phase-field simulation against analytical and numerical linear elastic solutions; (b) Compliance derivative ($\mathrm{d}C/\mathrm{d}a$) as a function of crack length, demonstrating the phase-field method's accuracy in capturing compliance evolution.}
   \label{fig:compact_tension_stiffness_crack_length}
\end{figure}

Figure~\ref{fig:compact_tension_stiffness_crack_lenght} compares the stiffness results from the phase-field simulation, using crack lengths measured by both the Bourdin and skeleton methods, against the analytical LEFM solution and linear elastic finite element simulations. Similarly, Figure~\ref{fig:compact_tension_dCda} highlights the relationship between the rate of change of compliance with respect to crack area ($\mathrm{d}C/\mathrm{d}a$) and the crack length, a key parameter for fatigue analysis.

From these results, it is concluded that for an unstructured mesh where the crack does not propagate exactly along a symmetry plane, the skeleton crack length measurement provides results closer to the theoretical solution than the Bourdin correction. Therefore, the skeleton algorithm is used for all subsequent analyses. The elastic simulation also shows good correlation with the theoretical one, despite minor differences.

\begin{figure}[h!]
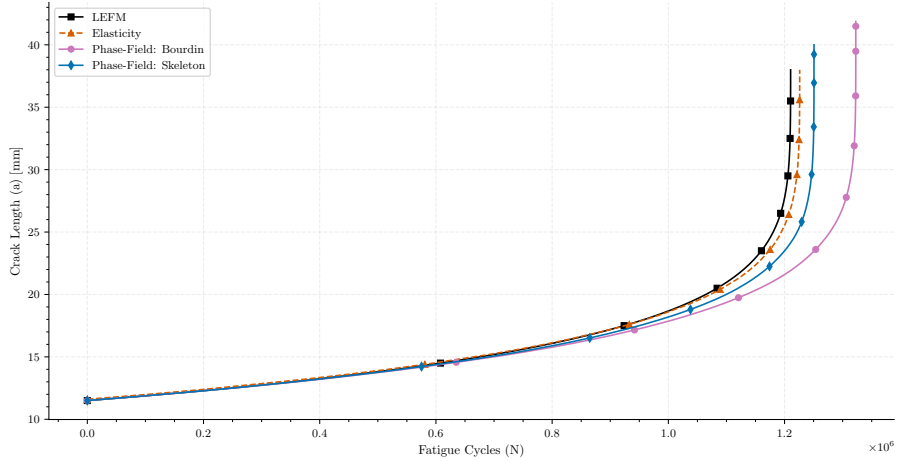

   \centering
   \figpath{examples/Compare_Compact_Specimen/compact_tension/cycles_vs_crack_length}[width=0.8\textwidth]
   \caption{Fatigue life prediction using Paris' law, highlighting the phase-field method's accuracy in fatigue analysis.}
   \label{fig:compact_tension_fatigue_life}
\end{figure}

With the compliance derivative, $\mathrm{d}C/\mathrm{d}a$, established for each method (Figure~\ref{fig:compact_tension_dCda}), the fatigue life is calculated using Eq.~\eqref{eq:total_number_of_cycles_dCda_experiment}. The analysis begins from an initial crack length of $a_i=11.5$~mm to ensure consistency with the experimental data presented later.

Figure~\ref{fig:compact_tension_fatigue_life} compares the fatigue life predictions from the theoretical LEFM solution, the elasticity simulation, and the phase-field model (with both Bourdin and skeleton corrections). While the underlying stiffness and $\mathrm{d}C/\mathrm{d}a$ curves show strong agreement, the resulting fatigue life predictions exhibit more pronounced differences. These discrepancies are primarily due to the amplification of small variations in $\mathrm{d}C/\mathrm{d}a$ by the Paris exponent, $n$. Furthermore, the phase-field simulation shows a slight upward crack deviation not present in the other models, which increases the total crack length and affects the life prediction. Consequently, the skeleton-corrected phase-field results align more closely with the LEFM and elasticity solutions than the Bourdin-corrected results.

\subsection{Specimens with holes}
\begin{figure}[h!]
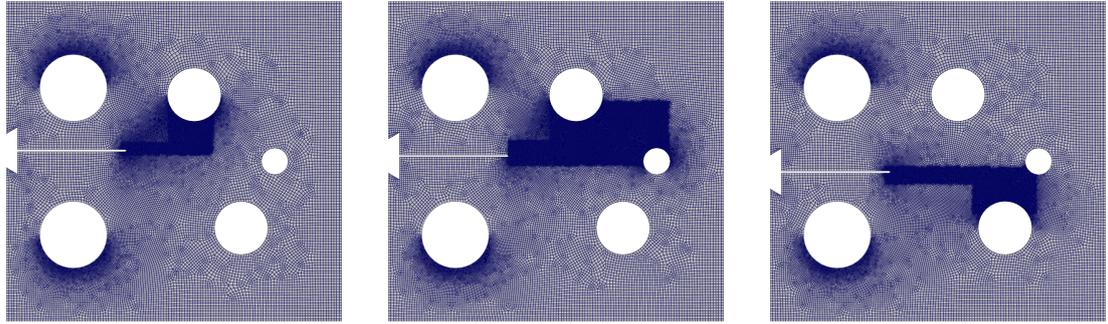

   \centering
   \subfigure[Specimen 2 ($H = 0.56\SpecimenWidth= 22.4$ mm)]{
      \figpath{examples/GmshGeoFiles/Compact_specimen/specimen_2_H16}[width=0.30\linewidth]
      \label{fig:mesh_specimen2}
   }
   \hfill
   \subfigure[Specimen 3 ($H = 0.58\SpecimenWidth= 23.2$ mm)]{
      \figpath{examples/GmshGeoFiles/Compact_specimen/specimen_3_H08}[width=0.30\linewidth]
      \label{fig:mesh_specimen3}
   }
   \hfill
   \subfigure[Specimen 4 ($H = 0.64\SpecimenWidth= 25.6$ mm)]{
      \figpath{examples/GmshGeoFiles/Compact_specimen/specimen_4_Hminus16}[width=0.30\linewidth]
      \label{fig:mesh_specimen4}
   }
   \caption{Finite element meshes for compact tension specimens: (a) Specimen 2, (b) Specimen 3, (c) Specimen 4. Meshes use quadrilateral elements with $h = 0.001 \SpecimenWidth = 0.04$ mm in crack regions.}
   \label{fig:compact_specimen__with_holes_meshes}
\end{figure}

This section presents the results of multiple simulations of compact specimens with varying initial crack positions, as outlined in Table~\ref{tab:compact_specimen_simulation_configurations}, and their comparison with experimental results of the literature. The material parameters, boundary conditions, mesh refinement, and non-variational solver settings are identical to those used in the phase-field simulation of Specimen 1. Figure~\ref{fig:compact_specimen__with_holes_meshes} illustrates the finite element meshes for specimens 2, 3, and 4. These meshes are carefully generated to provide sufficient resolution in the regions where crack initiation and propagation are expected.

\begin{figure}[h!]
   \centering

   \subfigure[Simulated phase-field result: final crack path predicted by the model.]{
      \figpath{examples/Phase_Field_Compact_Specimen/results_specimen_2_H16/paraview_phi}[width=0.45\textwidth, height=0.3\textheight, keepaspectratio=false]
      \label{fig:compact_specimen_2_phase_field}
   }
   \hfill
   \subfigure[Experimental result: observed crack path from reference data.]{
      \resizebox{0.45\textwidth}{0.3\textheight}{%
         \begin{tikzpicture}[scale=1.0]  
         \def\b{40}
         \def\hg{0.6*\b}         
         \def\a{0.2*\b}          
         \def\hone{0.275*\b}     
         \def\c{0.25*\b}         
         \def\D{0.25*\b}         
         \def\H{1.6}             
         
         \def\Da{0.2*\b}         
         \def\xa{0.45*\b}        
         \def\ha{0.25*\b}        
         
         \def\Db{0.1*\b}         
         \def\xb{0.75*\b}        
         \def\hb{0}              
         
         \def\Dc{0.2*\b}         
         \def\xc{0.625*\b}       
         \def\hc{-0.25*\b}       
         
         \def\f{0.06375*\b}      
         \def\j{0.08875*\b}      
         \def\k{0.04250*\b}      
         \def\l{0.00375*\b}      
         \def\g{\a - \l/0.57735}           
               
         \draw[thick] (-\c,-\hg) -- (\b,-\hg);
         \draw[thick] ( \b,-\hg) -- (\b, \hg);
         \draw[thick] (-\c, \hg) -- (\b, \hg);

         \draw[thick] (-\c,\hg) -- (-\c,\H+\f);
         \draw[thick] (-\c,\H+\f) -- (-\c+\k,\H+\j);
         \draw[thick] (-\c+\k,\H+\j) -- (-\c+\k,\H+\l);
         \draw[thick] (-\c+\k,\H+\l) -- (\g,\H+\l);
         \draw[thick] (\g,\H+\l) -- (\a,\H);
         \draw[thick] (\a,\H) -- (\g,\H-\l);
         \draw[thick] (\g,\H-\l) -- (-\c+\k,\H-\l);
         \draw[thick] (-\c+\k,\H-\l) -- (-\c+\k,\H-\j);
         \draw[thick] (-\c+\k,\H-\j) -- (-\c,\H-\f);
         \draw[thick] (-\c,\H-\f) -- (-\c,-\hg);
         
         \draw[thick] (0,\hone) circle (\D/2);
         \draw[thick] (0,-\hone) circle (\D/2);
         
         \draw[thick] (\xa,\ha) circle (\Da/2);
         \draw[thick] (\xb,\hb) circle (\Db/2);
         \draw[thick] (\xc,\hc) circle (\Dc/2);
         
          \ifuseflattened
            \draw[thick, red, line width=2pt] plot[shift={(0,0)}] file {\figdir examples-F-Papers_Data-F-Wagner_phd-F-fig510-F-experimental.txt};
          \else
            \draw[thick, red, line width=2pt] plot[shift={(0,0)}] file {examples/Papers_Data/Wagner_phd/fig510/experimental.txt};
          \fi

         \end{tikzpicture}%
      }
      \label{fig:simulation_2_geometry}
   }
   \caption{Comparison of simulated and experimental crack paths for specimen 2. (a) Simulated phase-field result showing the predicted crack trajectory. (b) Experimental observation from reference data, illustrating the actual crack path. The simulation shows excellent agreement with the experiment, with both cracks propagating toward hole 'a'.}
   \label{fig:simulation_H16_complete}
\end{figure}
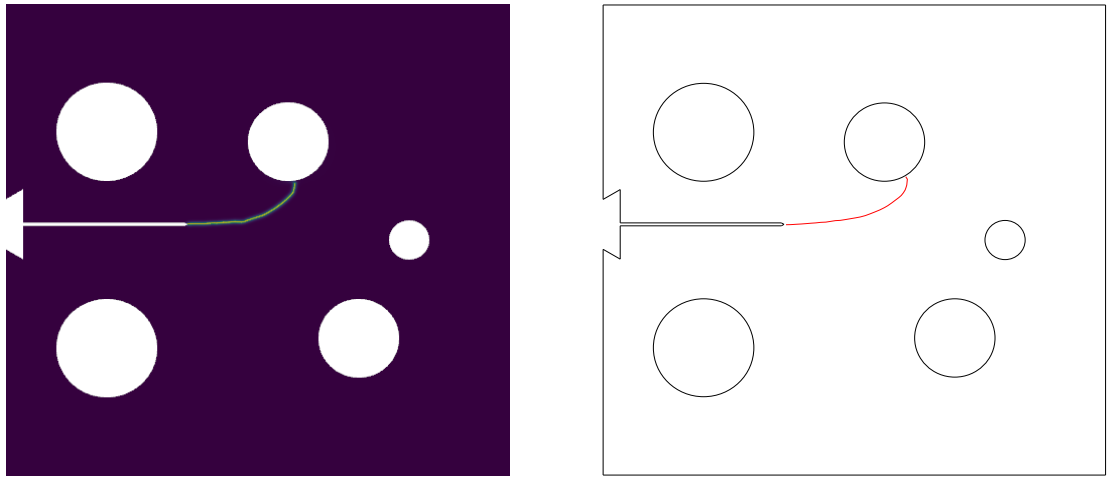

First, specimen 2 is analyzed, where the initial crack is positioned at $H = 0.56 \SpecimenWidth = 22.4$ mm. Figure \ref{fig:simulation_H16_complete} illustrates the final phase-field simulation results and compares the predicted crack path with the experimental data from \cite{example_Wagner2018_phd_thesis}. The predicted crack path shows excellent agreement with the experimental results; in both cases, the crack propagates toward and terminates at hole 'a'.

\begin{figure}[h!]
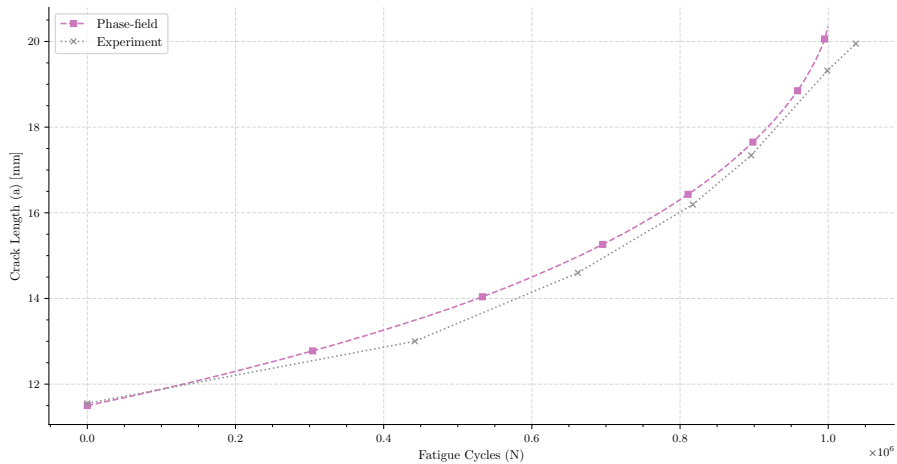

   \centering
   \figpath{examples/Compare_Compact_Specimen/compact_phase_field/compare_cycles_vs_crack_length_paper2}[width=0.8\textwidth]
   \caption{Fatigue life prediction for specimen 2 using Paris' law, comparing the phase-field method with experimental data from \cite{example_Wagner2018_phd_thesis}.}
   \label{fig:fatigue_life_specimen_2}
\end{figure}

Once the crack evolution and the compliance derivative $\mathrm{d}C/\mathrm{d}a$ are determined, the fatigue life is calculated using Eq.~\eqref{eq:total_number_of_cycles_dCda_experiment}. The predicted fatigue life is then compared with the experimental values reported in~\cite{example_Wagner2018_phd_thesis}. As shown in Figure~\ref{fig:fatigue_life_specimen_2}, the predicted curve of crack length versus number of cycles is in excellent agreement with the experimental data throughout the entire analysis. Notably, the prediction error for the final fatigue life is less than 4\%. Considering the high sensitivity of fatigue life predictions to simulation details and Paris law parameters, achieving such close agreement for both the crack path and the full life curve is a very promising result.

\begin{figure}[h!]
   \centering

   \subfigure[Simulated phase-field result: final crack path predicted by the model.]{
      \figpath{examples/Phase_Field_Compact_Specimen/results_specimen_3_H08/paraview_phi}[width=0.45\textwidth, height=0.3\textheight, keepaspectratio=false]
      \label{fig:compact_specimen_3_phase_field}
   }
   \hfill
   \subfigure[Experimental result: observed crack path from reference data.]{
      \resizebox{0.45\textwidth}{0.3\textheight}{%
         \begin{tikzpicture}[scale=1.0]  
         \def\b{40}
         \def\hg{0.6*\b}         
         \def\a{0.2*\b}          
         \def\hone{0.275*\b}     
         \def\c{0.25*\b}         
         \def\D{0.25*\b}         
         \def\H{0.8}             
         
         \def\Da{0.2*\b}         
         \def\xa{0.45*\b}        
         \def\ha{0.25*\b}        
         
         \def\Db{0.1*\b}         
         \def\xb{0.75*\b}        
         \def\hb{0}              
         
         \def\Dc{0.2*\b}         
         \def\xc{0.625*\b}       
         \def\hc{-0.25*\b}       
         
         \def\f{0.06375*\b}      
         \def\j{0.08875*\b}      
         \def\k{0.04250*\b}      
         \def\l{0.00375*\b}      
         \def\g{\a - \l/0.57735}             
               
         \draw[thick] (-\c,-\hg) -- (\b,-\hg);
         \draw[thick] ( \b,-\hg) -- (\b, \hg);
         \draw[thick] (-\c, \hg) -- (\b, \hg);

         \draw[thick] (-\c,\hg) -- (-\c,\H+\f);
         \draw[thick] (-\c,\H+\f) -- (-\c+\k,\H+\j);
         \draw[thick] (-\c+\k,\H+\j) -- (-\c+\k,\H+\l);
         \draw[thick] (-\c+\k,\H+\l) -- (\g,\H+\l);
         \draw[thick] (\g,\H+\l) -- (\a,\H);
         \draw[thick] (\a,\H) -- (\g,\H-\l);
         \draw[thick] (\g,\H-\l) -- (-\c+\k,\H-\l);
         \draw[thick] (-\c+\k,\H-\l) -- (-\c+\k,\H-\j);
         \draw[thick] (-\c+\k,\H-\j) -- (-\c,\H-\f);
         \draw[thick] (-\c,\H-\f) -- (-\c,-\hg);
         
         \draw[thick] (0,\hone) circle (\D/2);
         \draw[thick] (0,-\hone) circle (\D/2);
         
         \draw[thick] (\xa,\ha) circle (\Da/2);
         \draw[thick] (\xb,\hb) circle (\Db/2);
         \draw[thick] (\xc,\hc) circle (\Dc/2);
         
          \ifuseflattened
            \draw[thick, red, line width=2pt] plot[shift={(0,0)}] file {\figdir examples-F-Papers_Data-F-Wagner_phd-F-fig512a-F-experimental.txt};
          \else
            \draw[thick, red, line width=2pt] plot[shift={(0,0)}] file {examples/Papers_Data/Wagner_phd/fig512a/experimental.txt};
          \fi

         \end{tikzpicture}%
      }
      \label{fig:simulation_3_geometry}
   }
   \caption{Comparison of simulated and experimental crack paths for specimen 3. (a) Simulated phase-field result showing the predicted crack trajectory. (b) Experimental observation from reference data, illustrating the actual crack path. The simulation does not agree with the experiment, as the predicted crack propagates toward hole 'a' while the experimental crack propagates toward hole 'b'.}  
   \label{fig:simulation_3_complete}
\end{figure}
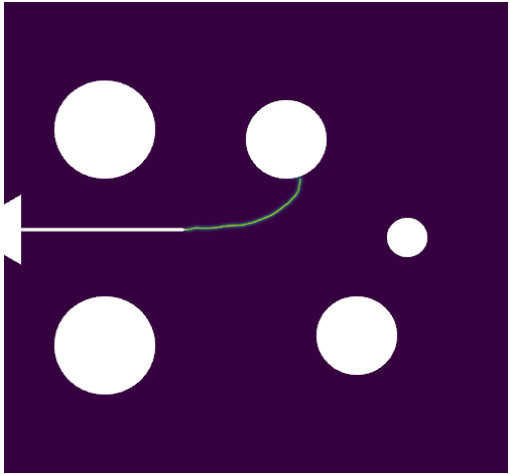
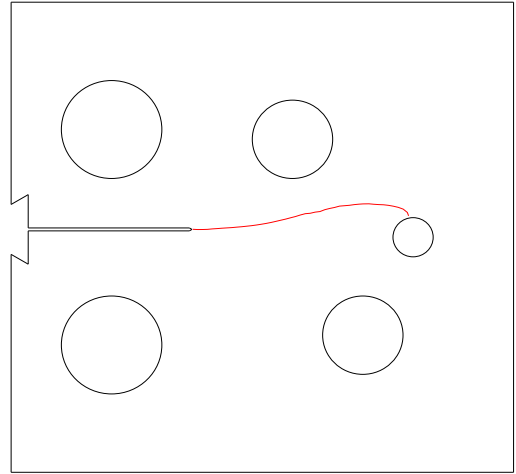

\begin{figure}[h!]
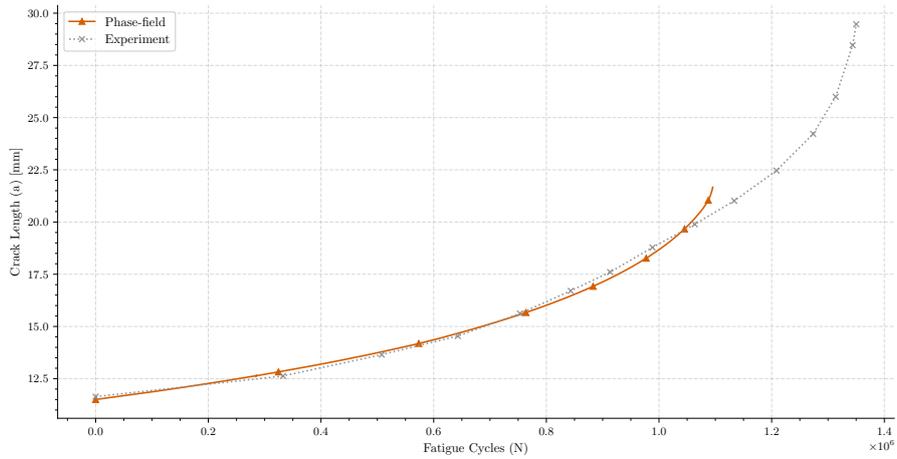

   \centering
   \figpath{examples/Compare_Compact_Specimen/compact_phase_field/compare_cycles_vs_crack_length_paper3}[width=0.8\textwidth]
   \caption{Fatigue life prediction for specimen 3 using Paris' law, comparing the phase-field method with experimental data from \cite{example_Wagner2018_phd_thesis}.}
   \label{fig:fatigue_life_specimen_3}
\end{figure}

Next, Specimen 3 is analyzed, where the initial crack is positioned closer to the specimen's mid-plane at $H = 0.58\SpecimenWidth = 23.2$~mm. Figure~\ref{fig:compact_specimen_3_phase_field} presents the results for this configuration. In this case, the crack path predicted by the phase-field simulation does not match the experimental results reported in~\cite{example_Wagner2018_phd_thesis}, where the crack terminates at hole 'b'. Instead, our simulation predicts a crack path similar to that observed for Specimen 2, propagating toward hole 'a'. As a results significant differences in fatigue life compared to the experimental observations are observed (Figure~\ref{fig:fatigue_life_specimen_3}).  

The discrepancy of the model with this particular experiment is probably due to the instability of crack path in this configuration. In our simulations it is observed that non-negligible path changes are achieved by changing the FE mesh or the way boundary conditions are imposed. Similar observations are reported in \cite{example_boundary_conditions}. In addition some details of the PFF model might influence in the results, such as the choice between isotropic and anisotropic models and the type of energy decomposition assumed. It is possible that other particular election of all these parameters would lead to better agreement but, since the objective is to evaluate the model predictive capacity under equal conditions, we have not made any parametric study to find optimal conditions to reproduce this mesh path.

\begin{figure}[h!]
   \centering

   \subfigure[Simulated phase-field result: final crack path predicted by the model.]{
      \figpath{examples/Phase_Field_Compact_Specimen/results_specimen_4_Hminus16/paraview_phi}[width=0.45\textwidth, height=0.3\textheight, keepaspectratio=false]
      \label{fig:simulation_4_results}
   }
   \hfill
   \subfigure[Experimental result: observed crack path from reference data.]{
      \resizebox{0.45\textwidth}{0.3\textheight}{%
         \begin{tikzpicture}[scale=1.0]  
         \def\b{40}
         \def\hg{0.6*\b}         
         \def\a{0.2*\b}          
         \def\hone{0.275*\b}     
         \def\c{0.25*\b}         
         \def\D{0.25*\b}         
         \def\H{-1.6}             
         
         \def\Da{0.2*\b}         
         \def\xa{0.45*\b}        
         \def\ha{0.25*\b}        
         
         \def\Db{0.1*\b}         
         \def\xb{0.75*\b}        
         \def\hb{0}              
         
         \def\Dc{0.2*\b}         
         \def\xc{0.625*\b}       
         \def\hc{-0.25*\b}       
         
         \def\f{0.06375*\b}      
         \def\j{0.08875*\b}      
         \def\k{0.04250*\b}      
         \def\l{0.00375*\b}      
         \def\g{\a - \l/0.57735}              
               
         \draw[thick] (-\c,-\hg) -- (\b,-\hg);
         \draw[thick] ( \b,-\hg) -- (\b, \hg);
         \draw[thick] (-\c, \hg) -- (\b, \hg);

         \draw[thick] (-\c,\hg) -- (-\c,\H+\f);
         \draw[thick] (-\c,\H+\f) -- (-\c+\k,\H+\j);
         \draw[thick] (-\c+\k,\H+\j) -- (-\c+\k,\H+\l);
         \draw[thick] (-\c+\k,\H+\l) -- (\g,\H+\l);
         \draw[thick] (\g,\H+\l) -- (\a,\H);
         \draw[thick] (\a,\H) -- (\g,\H-\l);
         \draw[thick] (\g,\H-\l) -- (-\c+\k,\H-\l);
         \draw[thick] (-\c+\k,\H-\l) -- (-\c+\k,\H-\j);
         \draw[thick] (-\c+\k,\H-\j) -- (-\c,\H-\f);
         \draw[thick] (-\c,\H-\f) -- (-\c,-\hg);
         
         \draw[thick] (0,\hone) circle (\D/2);
         \draw[thick] (0,-\hone) circle (\D/2);
         
         \draw[thick] (\xa,\ha) circle (\Da/2);
         \draw[thick] (\xb,\hb) circle (\Db/2);
         \draw[thick] (\xc,\hc) circle (\Dc/2);
           

         \ifuseflattened
            \draw[thick, red, line width=2pt] plot[shift={(0,0)}] file {\figdir examples-F-Papers_Data-F-Wagner_phd-F-fig511-F-experimental.txt};
         \else
            \draw[thick, red, line width=2pt] plot[shift={(0,0)}] file {examples/Papers_Data/Wagner_phd/fig511/experimental.txt};
         \fi

         \end{tikzpicture}%
      }
      \label{fig:simulation_4_geometry}
   }
   \caption{Comparison of simulated and experimental crack paths for specimen 4. (a) Simulated phase-field result showing the predicted crack trajectory. (b) Experimental observation from reference data, illustrating the actual crack path. The simulation shows excellent agreement with the experiment, with both cracks propagating toward hole 'c'.}
   \label{fig:simulation_H_minus16_complete}
\end{figure}
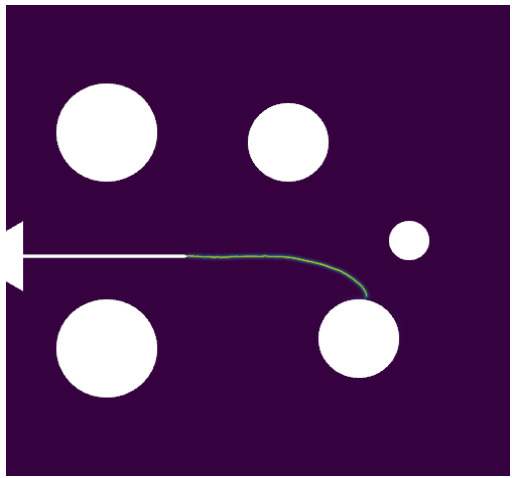
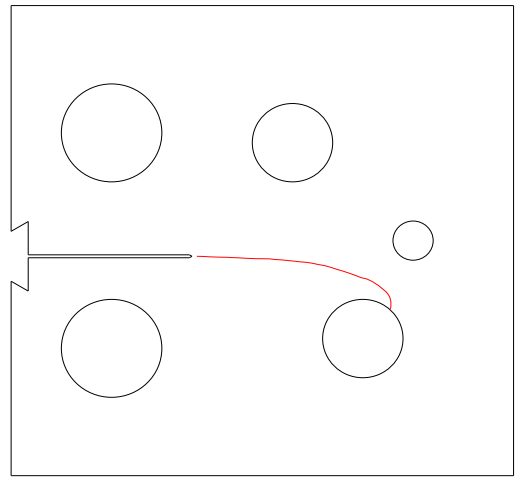

\begin{figure}[h!]
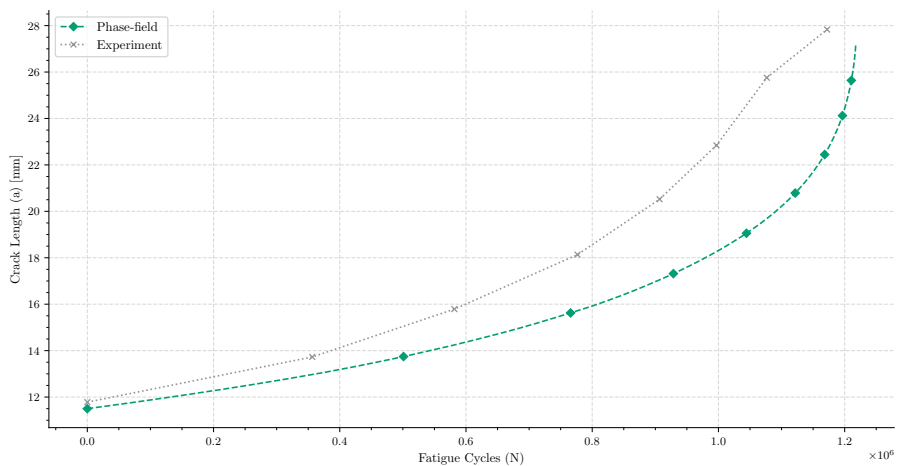

   \centering
   \figpath{examples/Compare_Compact_Specimen/compact_phase_field/compare_cycles_vs_crack_length_paper4}[width=0.8\textwidth]
   \caption{Fatigue life prediction for specimen 3 using Paris' law, comparing the phase-field method with experimental data from \cite{example_Wagner2018_phd_thesis}.}
   \label{fig:fatigue_life_specimen_4}
\end{figure}

Finally, Specimen 4 is analyzed, with the initial crack positioned at $H = 0.64\SpecimenWidth = 25.6$~mm. The simulation results, presented in Figure~\ref{fig:simulation_H_minus16_complete}, show the crack propagating downward and terminating at hole 'c', a path that aligns well with experimental observations~\cite{example_Wagner2018_phd_thesis}. This outcome confirms the model's robustness in handling varied asymmetric configurations. Following the established methodology, the fatigue life is calculated and compared with experimental data in Figure~\ref{fig:fatigue_life_specimen_4}, revealing a prediction error in the final number of cycles of less than 4\%. Notably, for both Specimen 2 and Specimen 4 ---the two cases where the crack path was correctly predicted—-- the fatigue life errors are of the same percentage. This consistency across different geometries provides strong evidence for the validity of the selected Paris law parameters for this material and these conditions.

As a summary, our model successfully predicted the crack path and fatigue life in 3 of 4 cases, with an error in the final number of cycles below 4\% for these three cases. These results are notable, considering the model's simplicity and the complex stress states and mode mixture present in the experiments. Moreover, these outcomes were achieved using a fixed and simple PFF configuration (isotropic damage AT2 model). It is expected that adapting the PFF strategy with different damage functions or energy decompositions could further improve these results.

It is noteworthy that the challenges in predicting the crack path for Specimen 3 are not unique to our model. The simulations reported in~\cite{example_Wagner2018_phd_thesis, example_Wagner2019_article}, which used the hypercomplex finite element method (ZFEM) and FRANC3D~\cite{example_franc3d}, also correctly predicted the crack paths for Specimens 2 and 4 but failed to do so for Specimen 3. In that case, the experimental result shows the crack terminating at hole 'b'. In contrast, our simulation predicts termination at hole 'a', while the ZFEM and FRANC3D simulations show the crack passing between holes 'a' and 'b' without terminating in either. This highlights that the high sensitivity of the crack propagation phenomenon is also observed with other simulation methods.

\begin{figure}[h!]
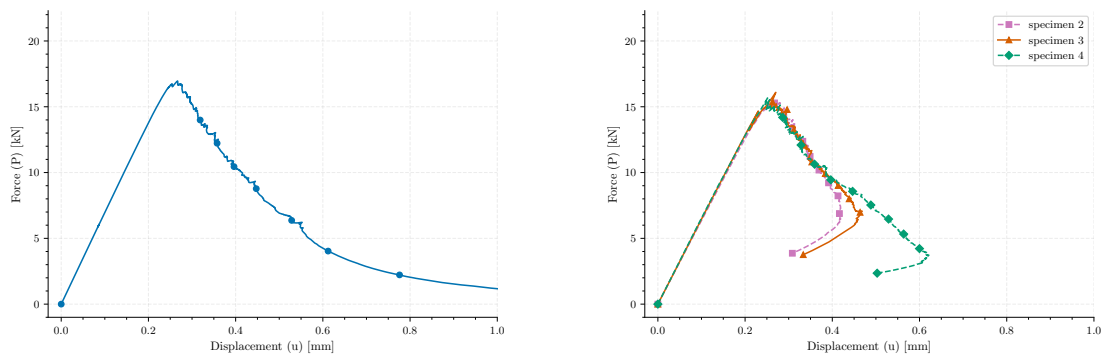

   \centering
   \subfigure[Force-displacement curve for the standard compact tension specimen without holes.]{
      \figpath{examples/Compare_Compact_Specimen/compact_phase_field/force_displacement_specimen_1}[width=0.45\textwidth]
      \label{fig:compact_force_displacent_specimen_1}
   }
   \hfill
   \subfigure[Force-displacement curves for the compact tension specimens with additional holes.]{
      \figpath{examples/Compare_Compact_Specimen/compact_phase_field/compare_force_displacement}[width=0.45\textwidth]
      \label{fig:compact_force_displacent_specimen_holes}
   }
   \caption{Force-displacement curves for the compact tension specimens. (a) Response of the standard specimen without holes. (b) Response of the specimens with holes, showing snap-back behavior as the crack approaches a hole.}
   \label{fig:compact_tension_force_displacement}
\end{figure}

\begin{figure}[h!]
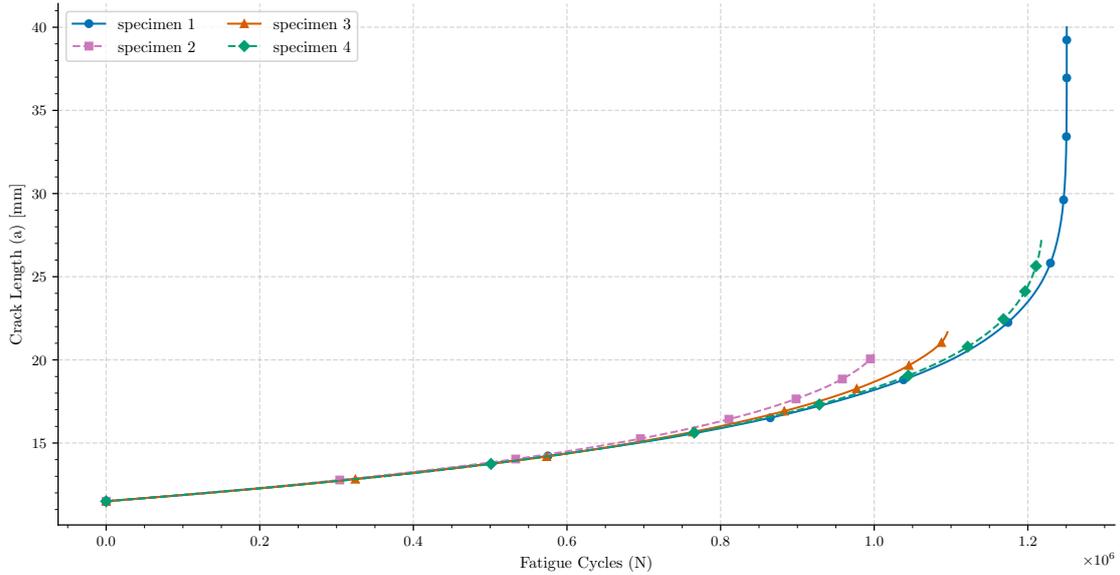

   \centering
   \figpath{examples/Compare_Compact_Specimen/compact_phase_field/paper_compare_cycles_vs_crack_length}[width=1.0\textwidth, keepaspectratio=true]
   \caption{Comparison of fatigue life predictions for all compact tension specimens. The plot shows the number of cycles versus crack length for the standard specimen (without holes) and the three modified specimens.}
   \label{fig:cycles_vs_crack_length_all}
\end{figure}

To further exploit the results of the proposed technique, the static force-displacement curve obtained under a crack-controlled experiments are computed and represented in Figure~\ref{fig:compact_force_displacent_specimen_1} for the specimen without holes and in Fig. \ref{fig:compact_force_displacent_specimen_holes} for the other cases. The results in both figures incorporate the skeleton correction. The specimens with holes exhibit snap-back behavior in the final part of their curves, which occurs as the crack approaches a hole. This snap-back implies a strong reduction on the energy dissipated during monotonic crack propagation, due to the reduction of the crack length. To ascertain if this snap-back response and its corresponding reduction of energy dissipation correlates with the fatigue performance, the fatigue crack evolution is represented in Figure~\ref{fig:cycles_vs_crack_length_all}. As expected, the specimen without holes exhibits the longest fatigue life. Specimen 2 has the shortest fatigue life, followed by Specimen 3, which shows a similar life despite its crack path differing from experimental results. Specimen 4, where the simulated crack terminates at the bottom hole, has the longest fatigue life among the specimens with holes. These results establish a clear qualitative correlation between the fatigue life of the specimen and the crack length (or total dissipated energy) under monotonic crack growth in a static experiment.

\section{Conclusions}
\label{sec:conclusions}

In this work, a novel approach for the simulation of fatigue crack propagation based the phase-field framework has been presented. This methodology is founded on a unified and computationally efficient framework that combines the principles of Linear Elastic Fracture Mechanics (LEFM) and Paris' law with phase-field fracture static simulations. The core of the methodology lies in the development of novel energy (crack length)-controlled solvers that robustly trace the complete quasi-static equilibrium path of a cracked body, including complex snap-back instabilities, in a single simulation. 

From an experimental view point the new approach does not require a specific parametrization, as other PFF based fatigue models, since it relies directly on the Paris' law parameters fit in standard experiments. From a numerical and computational perspective, the current approach does not rely on cycle-by-cycle simulation, as other PF fatigue approaches, making it possible to simulate HCF of UHCF cases with same cost as LCF. Secondly, in comparison with classic methods for crack propagation, remeshing or element enhancing on-the-fly is not necessary and the computation reduces to a single monotonic simulation.

The main contributions and findings of this study can be summarized as follows:
\begin{enumerate}
    \item Two equivalent energy-controlled schemes were developed. It was shown that both methods are capable of handling complex structural instabilities with the addition of only a single Lagrange multiplier. The choice between them relates to computational aspects, such as the symmetry of the variational solver in a Newton-Raphson approach versus obtaining physical parameters directly.
    \item The proposed framework leverages the energy-controlled simulation to directly compute the evolution of the specimen's compliance and its derivative with respect to the crack area, $\mathrm{d}C/\mathrm{d}a$. This avoids the need for computationally expensive cycle-by-cycle simulations, as the fatigue life can be directly integrated using Paris' law from the compliance data of a single quasi-static analysis. This significantly reduces computational cost compared to cycle-by-cycle methods and avoids extensive calibration work required by other phase-field fatigue models.
    \item An important limitation to consider is the typical overestimation of the crack length when measured via the $\gamma$ functional, which also leads to an effective critical energy release rate greater than the nominal one. A systematic procedure for correcting this inherent overestimation was implemented and validated. By applying correction factors based on either mesh parameters or image skeletonization, the simulation results show excellent agreement with analytical LEFM predictions for standard benchmarks.
    \item The framework's predictive power was demonstrated on a complex modified compact tension specimen with multiple holes, where the crack path is not known a priori. The model successfully predicted intricate, curvilinear crack trajectories in most cases, which were in good agreement with experimental data from the literature, showcasing its utility for analyzing realistic engineering components. However, one case presented a completely different crack path than that reported in other simulations and experiments. This discrepancy could be attributed to mesh dependency, the need for a smaller length-scale parameter, or the possibility that the isotropic energy degradation model was insufficient for that specific case.
    \item This methodology enables a rigorous and transparent assessment of its limitations. It allows for a clear evaluation of factors such as the non-linear trend in initial stiffness due to damage accumulation, overestimation of crack length, differences in crack path arising from isotropic versus anisotropic models, mesh dependency, and the influence of the length scale parameter. Although these limitations exist, it is shown that their influence can be studied in detail to account for them comprehensively.
\end{enumerate}

Future work could involve the straightforward extension of the framework to 3D problems. It is also noted that this approach can be adapted to other fatigue models, such as the Priddle equation, or to consider different stress ratios by applying the relevant LEFM relationships.

In conclusion, the proposed methodology offers a powerful and practical tool for fracture and fatigue analysis. By combining the descriptive power of phase-field models with the foundational principles of LEFM, it provides a computationally tractable approach for life prediction in complex engineering structures.

\section{Code Availability}
\label{sec:code_availability}
To ensure full reproducibility and promote open science, all source code, simulation scripts, mesh files, and post-processing routines used to generate the results in this paper are publicly available. The materials are hosted in a dedicated GitHub repository \cite{code_availability} and permanently archived on Zenodo \cite{code_availability_zenodo}.

The simulations were performed using `PhaseFieldX`~\cite{code_phasefieldx}, a custom open-source finite element library developed by one of the authors. This library, built upon the FEniCSx project~\cite{code_BarattaEtal2023}, implements the novel energy-controlled solvers, elasticity models, and crack measurement algorithms presented in this work. The public availability of both the specific simulation scripts and the underlying library provides a robust platform for validating our findings and fostering further research in the field.

The datasets generated and analyzed during the current study are entirely reproducible using the source code provided. No external or proprietary data sources were used.

\section{Acknowledgements}

This work was supported by the Spanish Ministry of Science and Innovation through the FPI grant PRE2020-092051 (MCIN/AEI/10.13039/501100011033). M. Castillón gratefully acknowledges Dr. Jørgen S. Dokken for his invaluable technical assistance with the implementation of Lagrange multiplier methods in the FEniCSx computational framework. His guidance proved essential for the development of the energy-controlled solvers presented in this work and their subsequent integration into the `PhaseFieldX` library.

\section{Declaration of generative AI and AI-assisted technologies in the writing process}
During the preparation of this work, the author(s) used Gemini 2.5 Pro to improve the readability and language of the manuscript. After using this tool, the author(s) reviewed and edited the content as needed and take full responsibility for the content of the published article.

\appendix
\section{Numerical implementation details}
\label{sec:appendix_fem}
This appendix summarizes the finite element discretization of the coupled phase-field fracture problem, detailing the construction of the residual vectors and tangent (Jacobian) matrices for the reference, variational, and non-variational energy-controlled schemes.

At each load or control increment, the nonlinear system is solved using a Newton-Raphson iterative scheme. In this approach, the unknowns (displacement $\boldsymbol{u}$, phase-field $\phi$, and, if present, the Lagrange multiplier $\lambda$) are updated by solving a linearized system of the form:
\begin{equation}
    \mathbf{K} \, \Delta \mathbf{x} = \mathbf{f}
\end{equation}
where $\mathbf{K}$ is the global tangent (Jacobian) matrix, $\Delta \mathbf{x}$ is the vector of unknown increments, and $\mathbf{f}$ is the global residual vector. The structure of $\mathbf{K}$ and $\mathbf{f}$ depends on the formulation, as detailed below.

\subsection{Reference (Standard) Phase-Field Formulation}

For the standard phase-field formulation, the coupled system for the increments $(\Delta \boldsymbol{u}, \Delta \phi)$ at each Newton-Raphson iteration is:
\begin{equation}
    \begin{bmatrix}
        K_{uu} & K_{u\phi} \\
        K_{\phi u} & K_{\phi\phi}
    \end{bmatrix}
    \begin{bmatrix}
        \Delta \boldsymbol{u} \\
        \Delta \phi
    \end{bmatrix}
    =
    \begin{bmatrix}
      \boldsymbol{f}_u \\
      \boldsymbol{f}_\phi \\
    \end{bmatrix}
\end{equation}

Given the finite element interpolation (see Section~\ref{sec:numerical_aspects}), the elemental residuals for each node $a \in N$ are defined as follows:

\begin{align}
    \boldsymbol{R}^{\boldsymbol{u}}_a &= \int_{\Omega} \left[ g(\phi_h)\, \boldsymbol{\sigma}_a(\boldsymbol{\epsilon}(\boldsymbol{u}_h)) + \boldsymbol{\sigma}_b(\boldsymbol{\epsilon}(\boldsymbol{u}_h)) \right] : \boldsymbol{\epsilon}(\Nax) \, \mathrm{d}\Omega
    - \int_{\partial_N\Omega} \boldsymbol{t} \cdot \Nax \, \mathrm{d}S, \\
    R^{\phi }_a &= \int_{\Omega} g'(\phi_h) \Nax \psi_a(\boldsymbol{\epsilon}(\boldsymbol{u}_h)) \, \mathrm{d}\Omega
    + G_c \int_{\Omega} \left( \frac{1}{l} \phi_h \Nax + l \nabla\phi_h \cdot \nabla \Nax \right) \mathrm{d}\Omega.
\end{align}
where $\boldsymbol{\sigma}_a$ and $\boldsymbol{\sigma}_b$ are the stress tensors derived from the strain energy densities: $\boldsymbol{\sigma}_a = \frac{\partial \psi_a}{\partial \boldsymbol{\epsilon}}$ and $\boldsymbol{\sigma}_b = \frac{\partial \psi_b}{\partial \boldsymbol{\epsilon}}$.
The explicit tangent (Jacobian) matrix blocks are:
\begin{align}
    K^{\boldsymbol{u},\boldsymbol{u}}_{a,b} &= \int_{\Omega} [g(\phi_h) \mathbb{C}_a(\boldsymbol{\epsilon}(\boldsymbol{u}_h)) + \mathbb{C}_b(\boldsymbol{\epsilon}(\boldsymbol{u}_h))] : \boldsymbol{\epsilon}(\Nbx) \, \mathrm{d}\Omega, \\
    K^{\boldsymbol{u},\phi}_{a,b} &= \int_{\Omega} g''(\phi_h) \Nax \Nbx \psi_a(\boldsymbol{\epsilon}(\boldsymbol{u}_h)) \, \mathrm{d}\Omega, \\
    K^{\phi,\boldsymbol{u}}_{a,b} &= \int_{\Omega_e} g'(\phi_h) N^\phi_a \sigma_a(\boldsymbol{\epsilon}(\boldsymbol{u}_h)) : \boldsymbol{\epsilon}(N^{\boldsymbol u}_b) \, \mathrm{d}\Omega, \\
    K^{\phi,\phi}_{a,b} &= \int_{\Omega_e} g''(\phi_h) N^\phi_a N^\phi_b \psi_a(\boldsymbol{\epsilon}(\boldsymbol{u}_h)) \, \mathrm{d}\Omega + G_c \int_{\Omega} \left( \frac{1}{l}N^\phi_a N^\phi_b + l \nabla N^\phi_a \cdot \nabla N^\phi_b \right) \, \mathrm{d}\Omega.
\end{align}
where $\mathbb{C}_a$ and $\mathbb{C}_b$ are the fourth-order elasticity tensors, representing the second derivatives of the strain energy densities with respect to the strain tensor: $\mathbb{C}_a = \frac{\partial^2 \psi_a}{\partial \boldsymbol{\epsilon}^2}$ and $\mathbb{C}_b = \frac{\partial^2 \psi_b}{\partial \boldsymbol{\epsilon}^2}$.

\subsection{Variational Energy-Controlled Scheme}

For the variational energy-controlled scheme, the unknowns are $(\Delta \boldsymbol{u}, \Delta \phi, \Delta \lambda)$, and the Newton-Raphson system at each increment is:
\begin{equation}
    \begin{bmatrix}
        K_{uu} & K_{u\phi} & K_{u\lambda} \\
        K_{\phi u} & K_{\phi\phi} & K_{\phi\lambda} \\
        K_{\lambda u} & K_{\lambda\phi} & 0
    \end{bmatrix}
    \begin{bmatrix}
        \Delta \boldsymbol{u} \\
        \Delta \phi \\
        \Delta \lambda
    \end{bmatrix}
    =
    \begin{bmatrix}
        \boldsymbol{f}_u \\
        \boldsymbol{f}_\phi \\
        \boldsymbol{f}_\lambda
    \end{bmatrix}
\end{equation}

The variational scheme introduces a Lagrange multiplier $\lambda$ and augments the residuals as follows:
\begin{align}
    R^{\boldsymbol{u}e,\text{var}}_a &= R^{\boldsymbol{u}e}_a + \lambda\, c_u \int_{\partial_N\Omega_e} \boldsymbol{t} \cdot N_a \, \mathrm{d}S, \\
    R^{\phi e,\text{var}}_a &= R^{\phi e}_a + \lambda\, c_\phi \int_{\Omega_e} \left( \frac{1}{l} \phi_h N_a + l \nabla\phi_h \cdot \nabla N_a \right) \mathrm{d}\Omega, \\
    R^{\lambda e}_a &= c_\phi \int_{\Omega_e} \left( \frac{1}{2l} \phi_h^2 + \frac{l}{2} |\nabla\phi_h|^2 \right) \mathrm{d}\Omega
    + c_u \int_{\partial_N\Omega_e} \boldsymbol{t} \cdot \boldsymbol{u}_h \, \mathrm{d}S - \tau(t).
\end{align}

The tangent blocks for the variational scheme are:
\begin{align}
    K^{\boldsymbol{u}\boldsymbol{u},\text{var}}_{ab} &= K^{\boldsymbol{u}\boldsymbol{u}}_{ab}, \\
    K^{\boldsymbol{u}\phi,\text{var}}_{ab} &= K^{\boldsymbol{u}\phi}_{ab}, \\
    K^{\boldsymbol{u}\lambda,\text{var}}_{ab} &= c_u \int_{\partial_N\Omega_e} \boldsymbol{t} \cdot N_a \, \mathrm{d}S, \\
    K^{\phi\boldsymbol{u},\text{var}}_{ab} &= K^{\phi\boldsymbol{u}}_{ab}, \\
    K^{\phi\phi,\text{var}}_{ab} &= K^{\phi\phi}_{ab} + \lambda c_\phi \int_{\Omega_e} \left( \frac{1}{l}N^\phi_a N^\phi_b + l \nabla N^\phi_a \cdot \nabla N^\phi_b \right) \, \mathrm{d}\Omega, \\
    K^{\phi\lambda,\text{var}}_{ab} &= c_\phi \int_{\Omega_e} \left( \frac{1}{l}\phi_h N^\phi_a + l \nabla\phi_h \cdot \nabla N^\phi_a \right) \, \mathrm{d}\Omega, \\
    K^{\lambda\boldsymbol{u},\text{var}}_{ab} &= c_u \int_{\partial_N\Omega_e} \boldsymbol{t} \cdot N_b \, \mathrm{d}S, \\
    K^{\lambda\phi,\text{var}}_{ab} &= c_\phi \int_{\Omega_e} \left( \frac{1}{l}\phi_h N^\phi_b + l \nabla\phi_h \cdot \nabla N^\phi_b \right) \, \mathrm{d}\Omega, \\
    K^{\lambda\lambda,\text{var}}_{ab} &= 0.
\end{align}

All terms $K^{\cdot\cdot,\text{var}}$ reference the standard tangent blocks defined in the reference case above.

\subsection{Non-Variational Energy-Controlled Scheme}

For the non-variational energy-controlled scheme, the Newton-Raphson system for $(\Delta \boldsymbol{u}, \Delta \phi, \Delta \lambda)$ is:
\begin{equation}
    \begin{bmatrix}
        K_{uu} & K_{u\phi} & K_{u\lambda} \\
        K_{\phi u} & K_{\phi\phi} & 0 \\
        K_{\lambda u} & K_{\lambda\phi} & 0
    \end{bmatrix}
    \begin{bmatrix}
        \Delta \boldsymbol{u} \\
        \Delta \phi \\
        \Delta \lambda
    \end{bmatrix}
    =
    \begin{bmatrix}
        \boldsymbol{f}_u \\
        \boldsymbol{f}_\phi \\
        \boldsymbol{f}_\lambda
    \end{bmatrix}
\end{equation}

For the non-variational scheme, the phase-field residual and its tangent do not depend on $\lambda$:
\begin{align}
    R^{\boldsymbol{u}e,\text{nv}}_a &= R^{\boldsymbol{u}e}_a + \lambda\, c_u \int_{\partial_N\Omega_e} \boldsymbol{t} \cdot N_a \, \mathrm{d}S, \\
    R^{\phi e,\text{nv}}_a &= R^{\phi e}_a, \\
    R^{\lambda e}_a &= c_\phi \int_{\Omega_e} \left( \frac{1}{2l} \phi_h^2 + \frac{l}{2} |\nabla\phi_h|^2 \right) \mathrm{d}\Omega
    + c_u \int_{\partial_N\Omega_e} \boldsymbol{t} \cdot \boldsymbol{u}_h \, \mathrm{d}S - \tau(t).
\end{align}

The tangent blocks for the non-variational scheme are:
\begin{align}
    K^{\boldsymbol{u}\boldsymbol{u},\text{nv}}_{ab} &= K^{\boldsymbol{u}\boldsymbol{u}}_{ab}, \\
    K^{\boldsymbol{u}\phi,\text{nv}}_{ab} &= K^{\boldsymbol{u}\phi}_{ab}, \\
    K^{\boldsymbol{u}\lambda,\text{nv}}_{ab} &= c_u \int_{\partial_N\Omega_e} \boldsymbol{t} \cdot N_a \, \mathrm{d}S, \\
    K^{\phi\boldsymbol{u},\text{nv}}_{ab} &= K^{\phi\boldsymbol{u}}_{ab}, \\
    K^{\phi\phi,\text{nv}}_{ab} &= K^{\phi\phi}_{ab}, \\
    K^{\phi\lambda,\text{nv}}_{ab} &= 0, \\
    K^{\lambda\boldsymbol{u},\text{nv}}_{ab} &= c_u \int_{\partial_N\Omega_e} \boldsymbol{t} \cdot N_b \, \mathrm{d}S, \\
    K^{\lambda\phi,\text{nv}}_{ab} &= c_\phi \int_{\Omega_e} \left( \frac{1}{l}\phi_h N^\phi_b + l \nabla\phi_h \cdot \nabla N^\phi_b \right) \, \mathrm{d}\Omega, \\
    K^{\lambda\lambda,\text{nv}}_{ab} &= 0.
\end{align}

Again, all terms reference the standard tangent blocks defined in the reference case.

\section{Scaling relationships for different critical energy release rates}
\label{sec:scaling_relationships}
The critical energy release rate $G_c$ does not influence the structural compliance of the specimen, which depends solely on geometry and elastic properties. This fundamental property enables the establishment of scaling relationships between simulations with different $G_c$ values but identical geometric and elastic properties.

Consider two problems with different critical energy release rates $G_{c1}$ and $G_{c2}$. From the Griffith criterion (Eq.~\eqref{eq:energy_release_rate_compliance}), we have:
\begin{align}
   G_{c1} &= \frac{P_1^2}{2B} \frac{dC}{da} \label{eq:griffith_1} \\
   G_{c2} &= \frac{P_2^2}{2B} \frac{dC}{da} \label{eq:griffith_2}
\end{align}
Since the compliance derivative $dC/da$ is identical for both cases (same geometry and elastic properties), the scaling factor $\beta = \sqrt{\frac{G_{c2}}{G_{c1}}}$ yields the following relationships:
\begin{align}
   P_2 &= \beta P_1 \label{eq:force_scaling_final} \\
   u_2 &= \beta u_1 \label{eq:displacement_scaling_final} \\
   \psi_2 &= \beta^2 \psi_1  \label{eq:energy_scaling_final}
\end{align}
where the displacement scaling follows from the linear relationship $u = C \cdot P$ and the strain energy scaling from $\psi = \frac{1}{2} P u$. These relationships enable efficient transformation of mechanical responses between simulations with different fracture toughness values, facilitating parametric studies and model validation.

\section{Crack Length Measurement via Image Post-Processing and Skeletonization Algorithms}
\label{sec:appendix_crack_measurement}
The crack area obtained directly through the $\gamma(\phi)$ functional systematically overestimates the true crack area. To address this limitation, an automated crack length measurement procedure is implemented using image post-processing and skeletonization algorithms from the scikit-image package \cite{code_scikit_image}, specifically employing the algorithm described in \cite{code_skeleton_algorithm}. This skeletonization algorithm operates by iteratively removing pixels from object boundaries until a single-pixel-wide centerline representation of the crack is obtained.

\begin{figure}[h!]
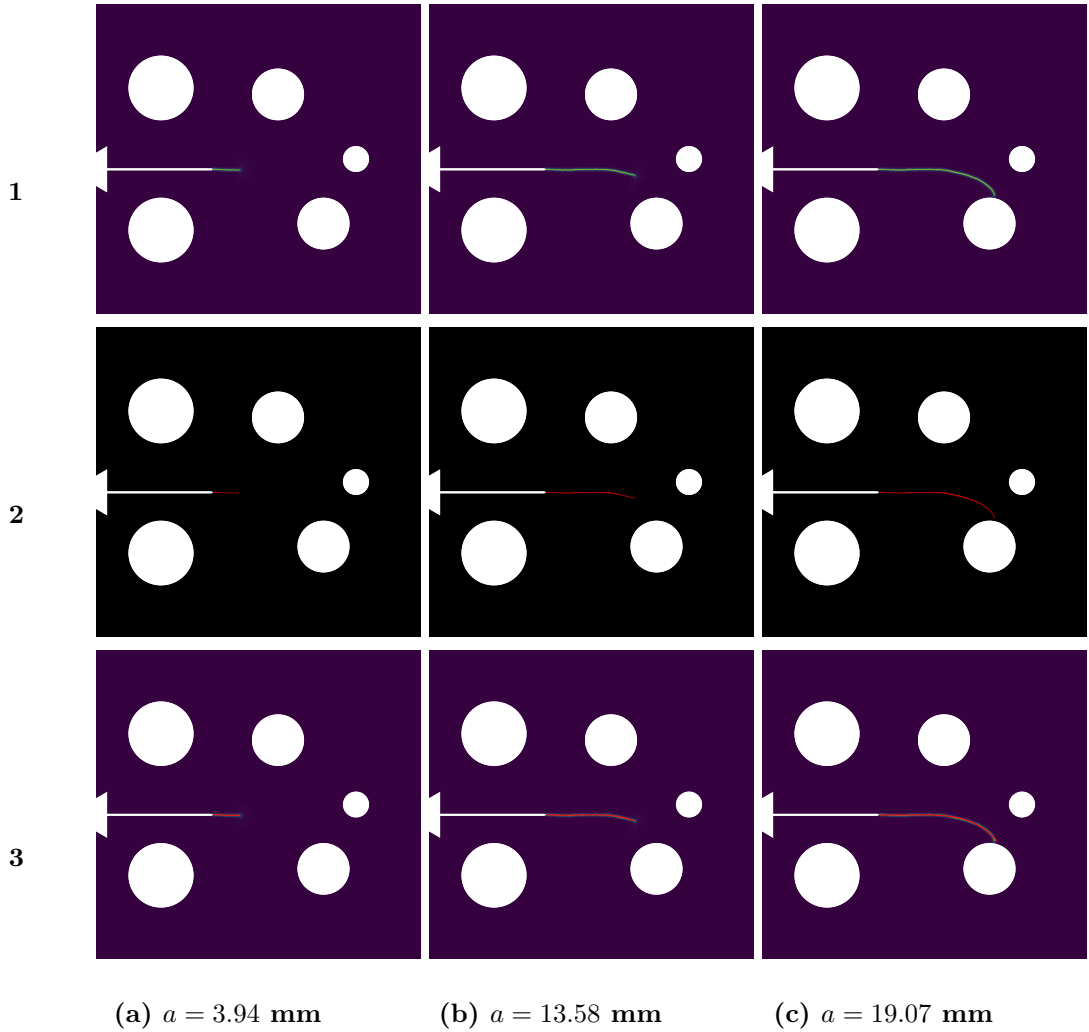

   \centering
   \begin{minipage}[b]{0.07\linewidth}
      \vspace{0.5cm}
      \raisebox{1.5cm}{\textbf{1}}
   \end{minipage}
   \begin{minipage}[b]{0.29\linewidth}
      \figpath{examples/Phase_Field_Compact_Specimen/results_specimen_4_Hminus16/crack_measurement/phi_images/phasefieldx_p0_000050}[width=\linewidth]
   \end{minipage}
   \begin{minipage}[b]{0.29\linewidth}
      \figpath{examples/Phase_Field_Compact_Specimen/results_specimen_4_Hminus16/crack_measurement/phi_images/phasefieldx_p0_000175}[width=\linewidth]
   \end{minipage}
   \begin{minipage}[b]{0.29\linewidth}
      \figpath{examples/Phase_Field_Compact_Specimen/results_specimen_4_Hminus16/crack_measurement/phi_images/phasefieldx_p0_000247}[width=\linewidth]
   \end{minipage}
    
   \vspace{1ex}
   \begin{minipage}[b]{0.07\linewidth}
      \vspace{0.5cm}
      \raisebox{1.5cm}{\textbf{2}}
   \end{minipage}
   \begin{minipage}[b]{0.29\linewidth}
      \figpath{examples/Phase_Field_Compact_Specimen/results_specimen_4_Hminus16/crack_measurement/phi_threshold_images/phasefieldx_p0_000050}[width=\linewidth]
   \end{minipage}
   \begin{minipage}[b]{0.29\linewidth}
      \figpath{examples/Phase_Field_Compact_Specimen/results_specimen_4_Hminus16/crack_measurement/phi_threshold_images/phasefieldx_p0_000175}[width=\linewidth]
   \end{minipage}
   \begin{minipage}[b]{0.29\linewidth}
      \figpath{examples/Phase_Field_Compact_Specimen/results_specimen_4_Hminus16/crack_measurement/phi_threshold_images/phasefieldx_p0_000247}[width=\linewidth]
   \end{minipage}
    
   \vspace{1ex}
   \begin{minipage}[b]{0.07\linewidth}
      \vspace{0.5cm}
      \raisebox{1.5cm}{\textbf{3}}
   \end{minipage}
   \begin{minipage}[b]{0.29\linewidth}
      \figpath{examples/Phase_Field_Compact_Specimen/results_specimen_4_Hminus16/crack_measurement/phi_images_splines/phasefieldx_p0_000050}[width=\linewidth]
      \\[-1ex]
      \centering \textbf{}
      \label{fig:spline_specimen2}
   \end{minipage}
   \begin{minipage}[b]{0.29\linewidth}
      \figpath{examples/Phase_Field_Compact_Specimen/results_specimen_4_Hminus16/crack_measurement/phi_images_splines/phasefieldx_p0_000175}[width=\linewidth]
      \\[-1ex]
      \centering \textbf{}
      \label{fig:spline_specimen3}
   \end{minipage}
   \begin{minipage}[b]{0.29\linewidth}
      \figpath{examples/Phase_Field_Compact_Specimen/results_specimen_4_Hminus16/crack_measurement/phi_images_splines/phasefieldx_p0_000247}[width=\linewidth]
      \\[-1ex]
      \centering \textbf{}
      \label{fig:spline_specimen4}
   \end{minipage}
   \\[1ex]
   \begin{minipage}[b]{0.07\linewidth}
   \end{minipage}
   \begin{minipage}[b]{0.29\linewidth}
      \centering \textbf{(a) $a=3.94$ mm}
   \end{minipage}
   \begin{minipage}[b]{0.29\linewidth}
      \centering \textbf{(b) $a=13.58$ mm}
   \end{minipage}
   \begin{minipage}[b]{0.29\linewidth}
      \centering \textbf{(c) $a=19.07$ mm}
   \end{minipage}
   \caption{Illustration of the automated crack length measurement procedure using image post-processing. The process is shown for three stages of crack growth (columns a, b, c). Row 1: The original phase-field solution ($\phi$). Row 2: A binary image created by applying a threshold ($\phi > 0.95$) to isolate the cracked region. Row 3: The skeletonization algorithm extracts the crack's centerline from the binary image. A spline is fitted to the resulting points and overlaid on the original phase-field solution to visualize the measured crack path.}
   \label{fig:compact_specimen_with_holes_results}
\end{figure}

The procedure automatically extracts and measures crack length from phase-field solutions stored in \texttt{.vtu} format through the following steps: (1) identification of cracked regions where $\phi > \phi_\text{th} = 0.95$; (2) generation of binary images with red zones indicating $\phi > \phi_\text{th}$ and black zones for $\phi < \phi_\text{th}$; (3) skeletonization to extract single-pixel-wide crack paths using \texttt{skimage.morphology.skeletonize}; (4) coordinate extraction from the single-pixel-wide path, where pixel coordinates are converted to real physical domain coordinates using the known pixel-to-domain mapping; (5) spline curve fitting to improve measurement accuracy by avoiding length mismeasurement that occurs when cracks follow diagonal paths; and (6) calculation of crack length from the fitted spline curves. Since the phase-field variable is not saved at every simulation step, quadratic interpolation is applied to obtain crack length at all time steps. Figure~\ref{fig:compact_specimen_with_holes_results} illustrates this procedure.

This automated procedure significantly improves the accuracy of crack length measurements in phase-field simulations by eliminating overestimation inherent in the $\gamma(\phi)$ functional, applying image processing and skeletonization techniques to extract the true crack path, and enabling direct mapping of image data to physical coordinates for precise measurements.

The algorithm requires only five input parameters: (1) the phase-field solution files in \texttt{.vtu} format, (2) the characteristic dimension of the computational domain, (3) the physical length and width of the specimen, (4) the threshold value $\phi_\text{th}$ for crack identification, and (5) the smoothing factor for spline fitting. The procedure has been successfully validated through the examples presented in this work. The generated splines demonstrate excellent agreement with the phase-field crack paths, and the measured crack lengths show high accuracy compared to expected values. As shown in Figure~\ref{fig:compact_specimen_with_holes_results}, the fitted splines accurately capture the crack trajectory evolution throughout the simulation. This algorithm is specifically designed and validated for 2D simulations, providing a robust and reliable method for crack length measurement in phase-field fracture analysis.
\bibliographystyle{unsrt}
\bibliography{references.bib}

\end{document}